\documentclass[12pt,doublespace]{JHEP3}
\JHEPspecialurl{http://jhep.sissa.it/JOURNAL/JHEP3.tar.gz}

\usepackage{amsmath,amssymb,amsfonts,xspace}
\usepackage{epsfig,supertabular}

\DeclareMathOperator{\Li}{Li}
\DeclareMathOperator{\Td}{Td}
\DeclareMathOperator{\ch}{ch}
\DeclareMathOperator{\rk}{rk}
\renewcommand{\Im}{\imag}

\newcommand{\dwrt}[1]{\frac{\partial}{\partial#1}}

\newcommand{\expe}[1]{{\bf E}\!\left( #1\right)}
\newcommand{\be}{\begin{equation}}
\newcommand{\ee}{\end{equation}}
\newcommand{\beq}{\begin{eqnarray}}
\newcommand{\eeq}{\end{eqnarray}}
\newcommand{\bea}[2]{\be\label{#2}\begin{array}{#1}}
\newcommand{\eea}{\end{array}\ee}

\def\det{\,{\rm det}\, }

\def\Im{\,{\rm Im}\, }

\def\Re{\,{\rm Re}\, }

\def\({\left(}
\def\){\right)}
\def\[{\left[}
\def\]{\right]}

\def\11{1\!\! 1}

\def\hf{f}

\def\eps{\varepsilon}

   \def\CX {{\cal X}}

\def\bW{ \bar W}

\def\bF{\bar F}

\def\ba{\bar a}

\def\bz{\bar z}

\def\ttau{\tilde \tau}

\def\tleta{\tilde\eta}

\newcommand{\CP}{\IC P^1}

\renewcommand{\d}{\mathrm{d}}
\newcommand{\de}{\mathrm{d}}

\newcommand{\I}{\mathrm{i}}
\newcommand{\cA}{\mathcal{A}}
\newcommand{\cB}{\mathcal{B}}
\newcommand{\cL}{\mathcal{L}}
\newcommand{\cQ}{\mathcal{Q}}
\newcommand{\cH}{\mathcal{H}}

\newcommand{\half}{\frac{1}{2}}

\newcommand{\cF}{\mathcal{F}}

\newcommand{\cC}{\mathcal{C}}
\newcommand{\cS}{\mathcal{S}}
\newcommand{\cG}{\mathcal{G}}
\newcommand{\cK}{\mathcal{K}}
\newcommand{\cM}{\mathcal{M}}

\newcommand{\cN}{\mathcal{N}}

\newcommand{\cX}{\mathcal{X}}
\newcommand{\cY}{\mathcal{Y}}
\newcommand{\cO}{\mathcal{O}}

\newcommand{\cT}{\mathcal{T}}
\newcommand{\cJ}{\mathcal{J}}
\newcommand{\cU}{\mathcal{U}}
\newcommand{\cD}{\mathcal{D}}

\newcommand{\cZ}{\mathcal{Z}}
\newcommand{\IR}{\mathbb{R}}
\newcommand{\IC}{\mathbb{C}}
\newcommand{\IZ}{\mathbb{Z}}

\newcommand{\IQ}{\mathbb{Q}}
\newcommand{\sk}{\mathcal{SK}}

\def\Hij#1{H^{[#1]}}

\def\hHpij#1{\hHij{#1}_{\scriptscriptstyle{\smash{(1)}}}}

\def\xii#1{\xi_{[#1]}}

\def\txii#1{{\tilde\xi}^{[#1]}}
\def\ai#1{{\alpha}^{[#1]}}
\def\tai#1{{\tilde\alpha}^{[#1]}}

\def\ui#1{^{[#1]}}
\def\di#1{_{[#1]}}

\def\Wkl{W_{\gamma}}

\def\Wg#1{W_{\gamma_{#1}}}

\def\bWg#1{\bW_{\gamma_{#1}}}
\def\Thg#1{\Theta_{\gamma_{#1}}}

\def\Igg#1{\cJ_{\gamma_{#1}}}

\def\ellg#1{\ell_{#1}}

\def\Thkl{\Theta_{\gamma}}

\def\varpi{t}

\def\tlp{\tilde p}

\def\pa{\partial}

\newcommand{\tzeta}{{\tilde\zeta}}
\newcommand{\txi}{\tilde\xi}
\newcommand{\nn}{\nonumber}
\newcommand{\kahler}{{K\"ahler}\xspace}
\newcommand{\hk}{{hyperk\"ahler}\xspace}
\newcommand{\qk}{{quaternion-K\"ahler}\xspace}

\def\bse{\begin{subequations}}
\def\ese{\end{subequations}}

\def\ellg#1{\ell_{#1}}

\def\Xigi#1{\Xi_{\gamma_{#1}}}

\def\qli2{{\bf E}}

\newcommand{\hCX}{\mathcal{X}}

\def\hSij#1{S^{[#1]}}
\def\hHij#1{H^{[#1]}}
\def\hnkl{\Omega(\gamma)}
\def\thnkl{\overline{\Omega}(\gamma)}

\def\hgam{\hat \gamma}
\def\Gauss{{\rm G}}

\def\ZG#1{\cZ^{(#1)}_{\Gauss}}
\def\ZNS#1{\cZ^{(#1)}_{\rm NS5}}
\def\ZNSG#1{\cZ^{(#1)}_{\mbox{\scriptsize NS5-G}}}

\def\phid#1{\phi_{\text{D},#1}^{\hphantom{A}}}
\def\thetad#1{\theta_{\text{D}}^{#1}}

\numberwithin{equation}{section}
\title{Fivebrane instantons, topological wave functions and hypermultiplet moduli spaces}

\author{Sergei Alexandrov$^1$, Daniel Persson$^2$,
Boris Pioline$^{3}$,
\\
\\
$^1$ {\it Laboratoire de Physique Th\'eorique \&
Astroparticules, CNRS UMR 5207, \\
Universit\'e Montpellier II, 34095 Montpellier Cedex 05, France}\\

$^2$  {\it Institut f\"ur Theoretische Physik, ETH Z\"urich,}\\ {\it CH-8093 Z\"urich, Switzerland}\\

$^3$ {\it Laboratoire de Physique Th\'eorique et Hautes
Energies, CNRS UMR 7589, \\
Universit\'e Pierre et Marie Curie - Paris 6,
4 place Jussieu, \\
75252 Paris Cedex 05, France} \\

\vspace*{2mm} {\tt e-mail: \email{sergey.alexandrov@univ-montp2.fr},
\email{daniel.persson@itp.phys.ethz.ch},
\email{pioline@lpthe.jussieu.fr}
} \vspace*{-3mm}

}

\abstract{We  investigate  quantum corrections to the hypermultiplet moduli space
$\cM$ in Calabi-Yau compactifications of type II string theories, with particular emphasis
 on instanton effects from Euclidean NS5-branes. Based on the consistency
of D- and NS5-instanton corrections, we determine the topology of
the hypermultiplet moduli space at fixed string coupling,
as previewed in \cite{Alexandrov:2010np}.
On the type IIB side, we compute corrections from $(p,k)$-fivebrane
instantons to the metric on $\cM$ (specifically, the correction to the complex
contact structure on its twistor space $\cZ$) by applying S-duality
to the D-instanton sum. For fixed fivebrane charge $k$, the corrections
can be written as a non-Gaussian theta series, whose summand for $k=1$
reduces to the topological A-model amplitude. By mirror symmetry, instanton corrections
induced from the chiral type IIA NS5-brane are similarly governed by the wave function of the topological B-model.
In the course of this investigation we clarify charge quantization for coherent sheaves and find hitherto
unnoticed corrections to the Heisenberg,  monodromy and S-duality actions on $\cM$,
as well as to the mirror map for Ramond-Ramond fields and D-brane charges.
}

\begin{document}

\newpage

\section{Introduction and Summary}\label{sec_intro}

To date, the hypermultiplet (HM) moduli space $\cM$ in type II string theories compactified
on a Calabi-Yau (CY) threefold $\cX$ has remained poorly understood.
Yet, its importance can hardly
be overstated: quantum corrections to the moduli space metric encode important geometric
invariants of $\cX$, which are closely related to the degeneracies of
BPS black holes \cite{Gunaydin:2005mx}.
Moreover, this encoding naturally incorporates chamber dependence and duality invariance.
Obtaining the exact moduli space metric on $\cM$ would have far-reaching implications
for mirror symmetry, and provide incisive tests of string dualities.
It may also be relevant for model building, as many $\cN=1$ string vacua
can be obtained by gauging isometries on the hypermultiplet branch.

Quantum corrections to the tree-level metric on $\cM$ fall in three categories: perturbative,
D-brane instantons and NS5-brane instantons \cite{Becker:1995kb}.
Each of these corrections must preserve
the \qk property of the metric, as required for supersymmetry \cite{Bagger:1983tt}.
While the perturbative and
D-instanton corrections are by now fairly well understood, NS5-brane instantons have
remained elusive. One technical reason is that fivebrane instantons break all continuous isometries,
and their implementation consistently with supersymmetry necessitates the full
machinery of twistor techniques for \qk manifolds. Another deeper reason is that
the type IIA fivebrane supports a chiral worldvolume theory \cite{Callan:1991ky},
the partition function
of which is notoriously subtle to define.  On the type IIB side, the world-volume theory
of the NS5-brane is
non-chiral \cite{Callan:1991ky} and follows in principle  by S-duality from that of
the D5-brane, but S-duality is also notoriously subtle in vacua with $\cN=2$ supersymmetry.
In this paper,  we take steps towards constructing fivebrane instanton corrections
on both sides, using insights from
topological strings, mirror symmetry and S-duality. The remainder of the introduction
is devoted to an account of the relevant background for the paper, while concurrently providing a
concise summary of our main results.

\subsection{The hypermultiplet moduli space}

At tree level, the \qk metric on the HM moduli space $\cM$ in type IIA (respectively, type IIB)
string theory compactified on $\cX$ is obtained by the so-called $c$-map procedure from
the special \kahler metric on the vector multiplet (VM) moduli space of type IIB (respectively,
type IIA) string theory compactified on the same CY
threefold $\cX$  \cite{Cecotti:1988qn,Ferrara:1989ik}. It describes the vacuum
expectation values and couplings of the dilaton $e^\phi\sim 1/g_{(4)}^2$,
the complex structure (respectively,  complexified \kahler) moduli $z^a$, the
periods $(\zeta^\Lambda,\tzeta_\Lambda)$ of the Ramond-Ramond (RR)
fields on $\cX$, and notably the Neveu-Schwarz (NS)
axion $\sigma$, dual to the Kalb-Ramond field  $B$ in four
dimensions (the range and meaning of the indices $a$ and $\Lambda$ will be specified below).
At fixed, vanishingly small string coupling $g_s$ and ignoring the NS-axion (i.e., modding out
by translations along $\sigma$, which are symmetries of the perturbative theory),
the ``reduced'' HM moduli space in type IIA string theory is known to be topologically the
intermediate Jacobian\footnote{The intermediate
Jacobian usually refers to the torus $\cT$ itself, rather than the total space of the fiber
bundle over $\cM_c(\cX)$. However, since $\cX$ is really a family of CY
manifolds with varying complex structure, we find this language convenient.} $\cJ_c(\cX)$ of
the CY family $\cX$, which is a torus bundle over the complex structure moduli space $\cM_c(\cX)$
with fiber $\cT\equiv H^3(\cX,\IR)/H^3(\cX,\IZ)$, which
parametrizes the space of harmonic RR three-form fields $C$ over $\cX$ modulo
large gauge transformations \cite{Morrison:1995yi}. In type IIB, the ``reduced" HM moduli
space is similarly a
torus bundle over the complexified \kahler moduli space $\cM_K(\cX)$, the
fiber of which is the quotient $\cT=H^{\rm even}(\cX)/\Gamma$ where
$\Gamma$ is the lattice of D-brane charges. As in type IIA, $\cT$
parametrizes the space of harmonic RR fields over $\cX$, modulo
large gauge transformations. We shall refer to the total space of this torus bundle
as  the symplectic Jacobian $\cJ_K(\cX)$. In type IIA (respectively, IIB), the NS-axion $\sigma$ parametrizes a
circle bundle $\cC$ over $\cJ_c(\cX)$ (respectively, $\cJ_K(\cX)$), the topology of which
has remained hitherto obscure. One of our goals will be to clarify the topology of the circle
bundle $\cC$, by analyzing the breaking of translational isometries along $\sigma$ due to fivebrane
instantons.

The quantum corrected
\qk metric on $\cM$ is most  conveniently described in terms of the twistor space
$\cZ_\cM $ associated to $\cM$, a complex contact manifold locally of the form $\CP\times \cM$
(see e.g.  \cite{Rocek:2006xb,Neitzke:2007ke,
Alexandrov:2008ds,Alexandrov:2008nk}). By the LeBrun-Salamon theorem,
the \qk metric on $\cM$ can be recovered from the complex contact structure on
$\cZ$, and linear perturbations of $\cM$ respecting the \qk property are
encoded in holomorphic sections of the \v{C}ech cohomology group $H^1(\cZ_\cM,\cO(2))$.
The one-loop correction to the tree-level metric was computed and further studied in
\cite{Antoniadis:1997eg,Gunther:1998sc,Antoniadis:2003sw,Robles-Llana:2006ez, Alexandrov:2007ec}.
It is believed to exhaust the series of perturbative corrections, in the sense that higher
loop corrections are expected to be removable by field redefinitions.
The twistorial description of the one-loop correction
was found in \cite{Alexandrov:2008nk}, as a logarithmic singularity
of a certain canonical Darboux coordinate at the north and south poles of $\CP$.

The D-instanton corrections to the complex contact structure of $\cZ_\cM$ (and, therefore,
to the \qk
metric on $\cM$) were derived in \cite{Alexandrov:2008gh,Alexandrov:2009zh,
Alexandrov:2009qq},
building on earlier work \cite{RoblesLlana:2006is,Alexandrov:2007ec,Saueressig:2007dr,RoblesLlana:2007ae}.
Key ingredients in this derivation were S-duality,
electric-magnetic duality (i.e. monodromy invariance) and mirror symmetry.
In the type IIA language, these D-instanton corrections are parametrized by a
chamber-dependent ``instanton measure" $n_\gamma(z)$,
which counts the number of stable special Lagrangian (sLag) submanifolds of $\cX$
in the homology class $\gamma\in H_3(\cX,\IZ)$, on which  Euclidean D2-branes can be wrapped.
The resulting complex contact structure is closely analogous to the
complex symplectic structure found on the twistor space of the (\hk) HM
moduli space in $\cN=2$ super-Yang Mills theories on $\IR^3\times S^1$ \cite{Gaiotto:2008cd}.
Indeed, upon compactifying the type IIA string theory further on a circle and applying
T-duality, $\cM$ is mapped to the VM
moduli space of the dual type IIB string theory in 3 dimensions, thus providing
a string theory analog of the set-up studied in \cite{Gaiotto:2008cd}.

Similarly as in $\cN=2$ field theories \cite{Gaiotto:2008cd},
the global consistency of the complex contact structure on $\cZ_\cM$
requires that the instanton measure $n_\gamma(z)$ satisfies certain
wall-crossing constraints, identical to the constraints for the generalized
Donaldson-Thomas (DT) invariants established in \cite{ks,Joyce:2009xv}. This supports the
expectation that the instanton measure $n_\gamma(z)$ is in fact equal to the
generalized DT invariant $\Omega(\gamma,z)$ defined in \cite{ks,Joyce:2009xv}, and
also to the indexed degeneracy of 4D black holes in type IIB string theory on $\cX$
in a vacuum with
asymptotic values of the VM moduli determined by $z^a$.
In the following we shall therefore identify these objects and denote
them by the common symbol $\Omega(\gamma,z)$, sometimes omitting the dependence
on the moduli $z^a$. Note however that due to the exponential growth of the
degeneracies of BPS black holes,
D-instanton corrections only make sense as an asymptotic series. The ambiguity of this
asymptotic series can be argued to be of the order of
the corrections expected from Kaluza-Klein monopoles
in type IIB on $\cX\times S^1$, or after T-duality,
NS5-branes in type IIA on $\cX$ \cite{Pioline:2009ia}.

\subsection{Sum over $H$-flux and Gaussian theta series}\label{sec_chiralsum}

In contrast to D-instantons, fivebrane instantons
have remained largely mysterious.
The contribution of one Euclidean type IIA NS5-brane
wrapped on $\cX$ (or one M5-brane in the M-theory picture)
is expected to involve a sum over harmonic configurations
$H\in \Gamma=H^3(\cX,\IZ)$ of the 3-form flux
$H$ supported  on the fivebrane worldvolume. Equivalently, one expects contributions
from all D2-NS5 (or M2-M5) bound states, where the D2/M2-brane can wrap any sLag
submanifold\footnote{Up to torsion, the lattices  $\Gamma=H^3(\cX,\IZ)$
and $\Gamma^*=H_3(\cX,\IZ)$ are isomorphic by Poincar\'e duality.
In this work we shall ignore torsion and  identify $\Gamma=\Gamma^*$.}
$\gamma\in \Gamma$ of $\cX$.
However, the self-duality constraint on $H$ implies that fluxes on three-cycles $\gamma,\gamma'$
with non-zero intersection product $\langle \gamma,\gamma' \rangle$ cannot be measured (nor defined)
simultaneously, and so the sum should run over a Lagrangian sublattice
$\Gamma_e\subset \Gamma$ only \cite{Witten:1996hc}
(see also \cite{Dolan:1998qk,Henningson:1999dm,Henningson:2001wh,Dijkgraaf:2002ac,
Diaconescu:2003bm,Moore:2004jv,Belov:2006jd,Belov:2006xj}).
As we review in Section \ref{sec_zns5}, the partition function $\ZG{1}$ for
a Gaussian self-dual three-form on $\cX$ is then  a holomorphic
section of a certain line bundle $\cL_\Theta$ over the
intermediate
Jacobian torus, which we shall call the ``theta line bundle''.
The first Chern class of $\cL_\Theta$ is known to be equal to
the \kahler class of $\cT$  \cite{Witten:1996hc}:
\be
\label{c1NS5T}
c_1(\cL_\Theta)\vert_\cT = \omega_\cT\ .
\ee

To specify the line bundle $\cL_\Theta$ uniquely,  one must choose a set
of characteristics\footnote{The
characteristics may in general depend on the metric on $\cX$. They can in principle
be computed in M-theory, see \cite{Diaconescu:2003bm}.}
$\Theta$ (more precisely, a quadratic refinement
$\sigma_\Theta$ of the intersection form on $H^3(\cX,\IZ)$) which
determine the periodicity properties of the sections of $\cL_\Theta$
under translations by $\Gamma$, i.e. under large gauge transformations
\cite{Witten:1996hc,Moore:2004jv,Belov:2006jd}. By the Kodaira vanishing
theorem, $\cL_{\Theta}$ admits a unique holomorphic
section $\vartheta_\Theta(\bar\cN,C)$, which is proportional
to the level 1/2 Siegel theta series
$\vartheta_{\rm Siegel} \[ {} ^\theta_\phi\](\bar\cN, \omega_\Lambda)$ of rank $b_3(\cX)/2$ with
characteristics $\Theta=(\theta^\Lambda,\phi_\Lambda)$, where $\cN$ is the (Weil) period matrix of $\cX$
and $\omega_\Lambda$ are complex coordinates on the torus $\cT$. Thus,
the Gaussian flux partition function $\ZG{1}$ for $k=1$
is given, up to a metric-dependent normalization factor, by the Siegel theta series
$\vartheta_\Theta(\bar\cN,C)$.

 In general, for $k>1$ fivebranes  wrapped on $\cX$,
the partition sum $\ZG{k}$ over fluxes for the ``diagonal" Gaussian self-dual
three-form (related by supersymmetry to the scalar fields describing  overall transverse
fluctuations of the stack of fivebranes) is expected to be a holomorphic section
of $\cL_\Theta^k$. By the same vanishing theorem, $\cL_\Theta^k$ admits $k^{b_3(\cX)/2}$
linearly independent holomorphic sections, given by level $k/2$ generalizations of
the Siegel theta series, and thus $\ZG{k}$ is expected to be a linear combination
of these theta series with coefficients depending on the metric on $\cX$.

\subsection{NS5-brane partition function and non-Gaussian theta series\label{sec_intsusy}}

Although it is a closely related object, the partition sum $\ZG{k}$
of a Gaussian self-dual three-form reviewed above is not yet the fivebrane partition
function $\ZNS{k}$ relevant for instanton corrections to
the hypermultiplet moduli space. There are several reasons for this:

\begin{itemize}
\item[i)] Firstly, $\ZG{1}$ only captures the topological degrees of freedom in the three-form $H$.
However, for the purpose of computing NS5-instanton corrections to the HM moduli space
metric, we require the partition function  of the full (2,0)
supermultiplet on the brane world-volume. While the other fields in the multiplets do not couple
to the $C$-field, they do contribute to the metric-dependent normalization factor.

\item[ii)] Secondly, one should include an
insertion of $(-1)^{2J_3} (2J_3)^2$, where $J_3$ is a angular momentum operator
on the fivebrane, in order to absorb fermionic zero-modes. Since the $H$-flux
contributes to angular momentum, this may result in additional insertions in the
Siegel theta series.
\item[iii)]  Thirdly, the action for $H$  is  Gaussian only
in the limit where the flux $H$ is much smaller than the inverse string coupling $1/g_s$.
Thus, the Gaussian partition function $Z^{(1)}_{\Gauss}$
is only valid at $g_s=0$, after factoring out a
vanishingly small factor  $e^{-S_{\rm NS5}}$ corresponding to the classical action
$S_{\rm NS5}=4\pi e^\phi+\I\pi\sigma$ of the fivebrane. To compute the partition
function at non-zero $g_s$, we  require the full non-linear, kappa-symmetric
action for the (2,0) multiplet on $\cX$, topologically twisted as a result of the insertion
of $(-1)^{2J_3} (2J_3)^2$.

To address this last point in more detail, recall that two distinct kappa-symmetric
descriptions of the (2,0) multiplet have been developed
in the literature:
(a) The action constructed in \cite{Bandos:1997ui,Aganagic:1997zq,Howe:1997fb,Bandos:2000az},
of Born-Infeld type, contains an auxiliary one-form (the gradient of a scalar field),
whose equation of motion implies the self-duality condition;
this action is well suited when $\cX$ allows for globally well-defined one-forms, but
seems inadequate for our purposes since for CY-threefolds $b_1(\cX)=0$.
(b) In the second approach \cite{Cederwall:1997gg,Sezgin:1998tm}, the action is polynomial and
contains an auxiliary 5-form, the elimination of which again leads to a Born-Infeld type action;
the self-duality condition must however be imposed by hand
(the action of \cite{Cederwall:1997gg,Sezgin:1998tm} is then more properly a pseudo-action).
Using this approach, it may be possible to separate the self-dual and
anti-self-dual parts of $H$ after a suitable Poisson
resummation, generalizing the procedure in \cite{Witten:1996hc,Dolan:1998qk,Dijkgraaf:2002ac,Belov:2006jd},
and then perform the topological twist.

\item[iv)]  Finally, for $k>1$ fivebranes , the degrees of freedom on the fivebrane  worldvolume
are presumably no-longer field-theoretical, but should involve the full little string theory in
6 dimensions \cite{Berkooz:1997cq,Seiberg:1997zk}.
With our current understanding of this theory, there seems to be no way of
fixing the linear combinations of level $k/2$
theta series entering in $\ZG{k}$ from first principles.

\end{itemize}

While all the reasons above show that the Gaussian theta series $\ZG{k}$ and
the fivebrane  partition function $\ZNS{k}$ governing  instanton
corrections to the HM moduli space are distinct functions of the metric and of the $C$-field,
we nevertheless expect that both should transform in the same way under large
gauge transformations of the $C$-field. Indeed, the latter depends only on the
Wess-Zumino-type coupling between the $C$-field and the $H$-flux, which is
independent of the detailed dynamics on the fivebrane  world-volume. Thus
we expect that $\ZG{k}$ and $\ZNS{k}$
are both sections of the same circle bundle $\cC_{\Theta}=\cL_\Theta^\circ$
over the intermediate
Jacobian torus $\cT$, where $\cL_\Theta^\circ$ denotes the unit circle inside
the line bundle $\cL_\Theta$. In particular, $\ZNS{k}$ should be given
by a theta series with the same characteristics $\Theta$, albeit of non-Gaussian type.
Unlike $\ZG{k}$ however,
$\ZNS{k}$ does not need to extend to a holomorphic section
of $\cL_\Theta$, and may have  different metric dependence.

Supersymmetry further
requires that $\ZNS{k}$ should be independent of the  \kahler moduli
(since the latter are part of the VM moduli space in type IIA),
and that corrections induced from $\ZNS{k}$
should preserve the \qk property of the HM metric. Thus,
$\ZNS{k}$ should (in a sense which we shall make precise below) lift to
a holomorphic section on the twistor space $\cZ_\cM$ over $\cM$. Using twistorial
techniques as well as insights
from S-duality and mirror symmetry, we shall construct the non-Gaussian
theta series $\ZNS{k}$, and show that its zero-coupling limit is a
Gaussian theta series $\ZNSG{k}$. The latter is almost identical to the usual Gaussian
flux partition function $\ZG{k}$, but differs  in its normalization factor, and in having
an insertion of a power of $H$ in the sum over $H$-fluxes,
cf. Eq. \eqref{NS5partitionfunctioncorrect} below. In particular, for $k=1$ the
normalization factor of $\ZG{1}$ can
be expressed as a certain product of holomorphic Ray-Singer torsions, the same
product which appears in the one-loop amplitude of the topological
B-model \cite{Bershadsky:1994cx}.\footnote{This connection to the topological B-model
is not surprising in view of the fact that the BPS sector of the world-volume
dynamics of a single M5-brane is governed by a deformation of Kodaira-Spencer
theory \cite{Marino:1999af,Bao:2006ef,Bengtsson:2008sp}.
It is also generally expected from mirror symmetry and S-duality, as
we discuss in Section \ref{subsec_qmS}.} This meshes nicely with the topology
of the one-loop corrected HM moduli space, as we now discuss.

\subsection{Topology of the hypermultiplet moduli space and quantum corrections}
\label{subsec_topology}

Irrespective of their microscopic origin, instanton corrections from $k$
fivebranes wrapped on $\cX$ are characterized by their non-trivial dependence
on the  Neveu-Schwarz (NS) axion $\sigma$ through an overall phase factor
$e^{-\I k \pi \sigma}$.
Large gauge transformations of the $B$-field require that
$\sigma$ is a compact coordinate with periodicity $\sigma\mapsto\sigma+2$ (in our conventions),
consistent with the above exponential dependence of fivebrane instantons.
Thus, $e^{\I\pi\sigma}$, $0\leq \sigma<2$ parametrizes the fiber of a circle bundle $\cC$
over the intermediate Jacobian $\cJ_c(\cX)$. On the other hand,
fivebrane instanton corrections to the HM metric are expected to be of the form
\be
\delta \de s^2\vert_{\text{NS5}}\sim e^{-4\pi|k| e^\phi-\pi \I k\sigma} \ZNSG{k}\, .
\label{NS5coupling}
\ee
Indeed, as we show in Section \ref{instantoncorrections},
such a coupling reproduces the classical fivebrane action  computed from
the 4D effective supergravity description in \cite{deVroome:2006xu}. For the coupling \eqref{NS5coupling}
to be globally well-defined, the circle bundle $\cC$ should be isomorphic to
the circle bundle $\cC_{\text{NS5}}$ where the fivebrane partition function $\ZNS{1}$ is valued.

On the other hand, the perturbative moduli space metric \eqref{hypmetone} exhibits
a  one-loop correction to the kinetic term \eqref{Dsigone} of the NS-axion
proportional to the Euler number $\chi(\cX)$ of the threefold $\cX$.
As we explain in Section \ref{sec_pert}, the  origin of this correction
can be traced to the familiar $B\wedge I_8$ topological
coupling in 10 dimensions. This fact provides an alternative derivation of
the one-loop correction to the HM metric, related  by supersymmetry to the computation in
\cite{Antoniadis:1997eg,Antoniadis:2003sw}. In particular, the form of the
axion kinetic term implies that  the curvature of the connection on the circle bundle $\cC$ is
given by
\be
\label{c1C}
\de \left(\frac{D\sigma}{2}\right) = \omega_\cT + \frac{\chi(\cX)}{24}\, \omega_\sk\, ,
\ee
where $\omega_\cT$ and $\omega_\sk$ are the \kahler forms
on $\cT$ and $\cM_c$.
The first term in \eqref{c1C} has support on the intermediate
Jacobian torus $\cT$. It reflects the fact that translations
along the torus fiber $\cT$ of $\cJ_c(\cX)$ (i.e. large gauge transformations of the $C$-field)
commute only up to a translation along the circle fiber of $\cC$ (i.e. a large gauge transformation
of the $B$-field).  As a result, large gauge transformations obey
a Heisenberg group law, see Eq. \eqref{grouplaw}.
Comparing \eqref{c1NS5T} and \eqref{c1C}, we see that
$c_1(\cC)-c_1(\cC_{\rm NS5})$ has no support on  $\cT$.
Thus, the coupling \eqref{NS5coupling} is indeed invariant under large gauge transformations,
provided $e^{\I \pi \sigma}$  transforms in the same way as $\ZNSG{1}$, i.e. as a section of $\cL_\Theta$.

The  second term in \eqref{c1C}, on the other hand,  has support on the complex
structure moduli space $\cM_c(\cX)$, and implies that the NS-axion picks up
anomalous variations under monodromies in $\cM_c$. At this point it is worth
pointing out that  the $B\wedge I_8$ topological
coupling, which is responsible for the second term in \eqref{c1C}, is also
responsible for an anomaly in the phase of the fivebrane partition
function~\cite{Witten:1996hc}. Indeed, the two effects are related by the
anomaly inflow mechanism~\cite{Blum:1993yd,Duff:1995wd}, which ultimately ensures
that the fivebrane instanton coupling \eqref{NS5coupling} is well-defined.

Now, the second term  in \eqref{c1C},
is equal to  $\chi(\cX)/24$ times the curvature
of the Hodge line bundle $\cL$ over $\cM_c(\cX)$, in which the holomorphic
three-form $\Omega_{3,0}$ on $\cX$ is valued. This indicates that the restriction of the
circle bundle $\cC$ to $\cM_c(\cX)$ is $\cL^{\chi(\cX)/24}$. Unless
$\chi(\cX)$ is a multiple of 24 however, the definition of $\cL^{\chi(\cX)/24}$ requires
additional data, namely a homomorphism $M\mapsto e^{2\pi\I\kappa(M)}$ from the
monodromy group to the group of 24:th roots of unity -- put differently, $e^{\I\pi\sigma}$
is valued in a twisted circle bundle.\footnote{We are grateful
to G. Moore, A. Neitzke and D. Zagier for discussions which have helped shaping this
point of view.}
As a result, $\sigma$ transforms under monodromies according to
\be
\label{sigmonintro}
\sigma \mapsto \sigma + \frac{\chi(\cX)}{24\pi}  \Im f_M +2\kappa(M)\, ,
\ee
where $f_M$ is a local holomorphic function on $\cM_c(\cX)$ which determines
the rescaling of the holomorphic three-form under monodromy,
$\Omega_{3,0}\mapsto e^{f_M} \Omega_{3,0}$. (Here one must make a choice of
branch of the logarithm in defining $f_M$, correlated with the choice of branch
in defining $\kappa(M)$.) In order for the product \eqref{NS5coupling} to be well-defined,
the fivebrane partition function $\ZNSG{1}$ should transform in the opposite fashion.
Since, as mentioned at the end of the previous subsection,
$\ZNSG{k}$ is proportional to the holomorphic part $e^{f_1}$ of the
B-model one-loop amplitude, this entails
that $e^{2\pi \I \kappa(M)}$ also governs the modular properties of $e^{f_1}$ under monodromies in $\cM_c$.

\subsection{Quantum mirror symmetry, S-duality and $(p,k)$ fivebranes}
\label{subsec_qmS}

The NS5-brane instanton corrections to the HM metric may also
be obtained using mirror symmetry\footnote{For rigid Calabi-Yau threefolds,
mirror symmetry is not available. NS5-instanton effects in this case have been recently
addressed in \cite{Bao:2009fg,Bao:2010cc} by postulating automorphic symmetries.}
and S-duality. Indeed, according to the quantum mirror symmetry
conjecture \cite{Becker:1995kb} (see \cite{Saueressig:2007dr,RoblesLlana:2007ae,
Pioline:2009qt} for recent discussions), the
same moduli space $\cM$ must arise as the HM moduli space of type IIB string theory
compactified on the mirror threefold $\hat\cX$. At the perturbative level, this amounts
to the classical mirror symmetry statement $\cM_c(\cX)=\cM_K(\hat \cX)$ between
the complex structure moduli space of $\cX$ and the complexified \kahler moduli
space of $\hat \cX$, the one-loop correction being controlled by the same invariant
$\chi({\cX})=-\chi(\hat\cX)$ on both sides. At the non-perturbative level, the identity
between the two HM moduli spaces requires a matching between the D-instanton effects
on the type IIA side, corresponding to D2-branes wrapped on special Lagrangian submanifolds
of $\cX$, and D5-D3-D1-D(-1)-instantons  on the type IIB side, corresponding to
stable coherent sheaves on $\hat\cX$. In Section \ref{sec_mir},
we explain how the fractional charges predicted by the K-theory description of
D5-D3-D1-D(-1)-instantons are consistent with the fact that
D2-brane charges are classified\footnote{It would be interesting to formulate
mirror symmetry directly at the level of the K-theory groups $K(\hat\cX)$
and $K^1(\cX)$, but we shall
not attempt to do so here.} by $H_3(\cX,\IZ)$.  The resolution of this apparent paradox
hinges on the quadratic ambiguity $A_{\Lambda\Sigma}$ of the prepotential \eqref{lve}
in the large volume limit,
and in fact allows to derive constraints on the fractional part of the matrix $A_{\Lambda\Sigma}$,
some of which had been observed in the early days of classical mirror
symmetry \cite{Candelas:1990rm,Klemm:1992tx,Candelas:1993dm,Candelas:1994hw}.
Incorporating these effects leads to novel corrections to the  ``Ramond-Ramond mirror map"
and S-duality action derived in \cite{Gunther:1998sc,Alexandrov:2008gh,Alexandrov:2009qq},
which we exhibit in Eqs. \eqref{shze} and \eqref{SL2Z}.

Assuming that these D-instanton corrections do match on both sides
of the mirror map, quantum mirror symmetry
further requires that fivebrane corrections on the type IIA (or M-theory) side match NS5-brane
corrections, or more generally $(p,k)$5-brane corrections,
on the type IIB side. The latter are related by type IIB S-duality\footnote{As in
earlier investigations \cite{RoblesLlana:2006is,Alexandrov:2008gh,Alexandrov:2009qq},
we shall work under the assumption that the full $SL(2,\IZ)$ S-duality of ten-dimensional
type IIB string theory stays valid in vacua with $\cN=2$ supersymmetry.}
to D5-brane instantons, governed by the generalized DT invariants $  \hnkl  $.
Thus, by combining  S-duality and mirror symmetry, one expects that
the type IIA NS5-instanton contributions are expressible in terms of the
topological B-model amplitude on $\cX$  \cite{Dijkgraaf:2002ac,Kapustin:2004jm} for
$k=1$, and more generally in terms of the generalized
DT invariants of $\cX$.

To implement these ideas, we start from holomorphic
sections of $H^1(\cZ_\cM,\cO(2))$ describing D5-D3-D1-D(-1)
instanton configurations, with charges $\gamma=(p^0,p^a,q_a,q_0)$, and sum over their images
under S-duality.
This twistor construction ensures that all instanton corrections respect
the quaternion-K\"ahler property of the metric, as required by supersymmetry.
The resulting (local) holomorphic functions encode corrections to the contact structure from
$(p,k)$5-brane instantons with $\gcd(p,k)=p^0$, bound to D3-D1-D(-1) instantons,
and take the following schematic form
\be
H_{{\rm NS5}}^{(k)}(\xi,\txi,\tilde\alpha) \sim e^{-\I \pi k \tilde{\alpha}}\,
H^{(k)}_{\text{NS5}}(\xi,\tilde\xi),
\label{NS5couplingtwistorspace}
\ee
where the variables $( \xi, \tilde\xi,\tilde\alpha)$ are complex Darboux coordinates
on $\cZ_\cM$.
These functions form an invariant set under the Heisenberg action \eqref{heisext} and under
monodromies around the point of infinite volume  in $\cM_K(\hat\cX)$,
up to subtle phases which we compute in Appendix A. These phases
indicate some tension between S-duality, Heisenberg and monodromy invariance,
and suggest that it may be necessary to relax some of our assumptions about the
way these symmetries are realized. In the absence of a satisfactory resolution
and since, by and large,
 our results  seem to support our general assumptions,
we dauntlessly forge ahead.

Ignoring these subtle phases then, the holomorphic functions \eqref{NS5couplingtwistorspace}
can be combined into a formal Poincar\'e series, and rewritten as a non-Gaussian theta series
with kernel given in \eqref{HNS5}.
For $k=1$, the summand of the theta series can be recognized, using the
DT/GW relation between (ordinary, rank one) Donaldson-Thomas invariants
and Gromov-Witten invariants  \cite{gw-dt,gw-dt2}, as the topological amplitude
of the A-model on $\hat\cX$  (see Eq. \eqref{NS5DTrelation} below). Thus, the partition function
of a single type IIB NS5-brane is given by a theta series built on the
A-model topological string amplitude. Using mirror symmetry,
 the partition function
of a single type IIA NS5-brane is then also a theta series whose wave function coincides with that
 of the B-model on $\cX$. By similar arguments, the partition function of type IIB $(p,k)$5-branes is governed by rank $r=\gcd(p,k)$
Donaldson-Thomas invariants on $\hat\cX$.

\subsection{Fivebrane partition function and twistors}
\label{subsec_fivepftwist}

The object $H^{(k)}_{\text{NS5}}(\xi,\tilde\xi)$ in \eqref{NS5couplingtwistorspace}
is formally a holomorphic section of $H^{1}(\cZ_\cM,\cO(2))$, and provides, by
a standard (though cumbersome)
procedure described e.g. in \cite{Alexandrov:2008nk}, an infinitesimal
deformation of the metric tensor on $\cM$ consistent with the
\qk property. By construction, $H^{(k)}_{\text{NS5}}(\xi,\tilde\xi)$ and the
phase $e^{-\I \pi k \tilde{\alpha}}$ are separately invariant under rescaling
of the holomorphic three-form $\Omega_{3,0}$ (up to subtleties in the
definition of $\cL^{\chi(\cX)/24}$
mentioned at the end of Section \ref{subsec_topology}).
Moreover, similarly as in \eqref{NS5coupling},
the two factors in \eqref{NS5couplingtwistorspace} transform oppositely under
translations along the torus $\cT$. Hence the fivebrane correction
\eqref{NS5couplingtwistorspace} to the complex contact structure on $\cZ$
is well-defined under both large gauge transformations and monodromies
(up to subtle phases that we ignore).

While this gives a (partial) answer to the question of main interest in this paper, namely
to compute fivebrane  instanton corrections to the metric on $\cM$, one may ask a
different though related question: is it possible to construct a \emph{scalar-valued function}
on $\cM$ which would generalize the Gaussian fivebrane partition function
$\ZG{k}$ at finite coupling? A possible answer to this question is obtained by
viewing  \eqref{NS5couplingtwistorspace} instead as a formal holomorphic section of
$H^{1}(\cZ_\cM,\cO(-2))$ (barring global issues) and applying the standard Penrose transform,
which relates sections of $H^{1}(\cZ_\cM,\cO(-2))$ to functions on $\cM$ satisfying
a certain set of second-order partial differential equations (see e.g. \cite{Neitzke:2007ke}).
While this procedure is not rigorous in general, it has been made rigorous in the case
of four-dimensional  quaternion-K\"ahler manifolds (see \cite{Alexandrov:2009vj}), and can
also serve as a warm-up for the more intensive computations involved in extracting
the explicit metric on $\cM$. In either case, the answer involves contour integrals over the $\CP$ fiber
of the twistor space $\cZ$, and reduces to the same saddle points in certain classical limits.

Performing this computation in the large volume limit, we indeed find that the Penrose transform
of \eqref{NS5couplingtwistorspace} reduces to a non-Gaussian theta series, whose summand
displays the expected non-linear action for $(p,k)$5-branes. Moreover, in the zero coupling
limit, this reduces to a Gaussian partition function of the type discussed in Section \ref{sec_chiralsum},
with a normalization factor given by the same product of holomorphic Ray-Singer torsions
which appears in the one-loop amplitude of the topological B-model on $\cX$, and
an additional flux-dependent insertion in the sum.
As a side-product, we also find a twistorial interpretation
for the ``auxiliary parameter" $t$ introduced
by \cite{Dijkgraaf:2002ac} to relate the Griffiths and Weil complex structures
on the intermediate Jacobian of $\cX$.

\subsection{Outline}

The remainder of this article is organized as follows. $\bullet$
In Section \ref{sec_zns5}, we review the construction of
the partition function of a Gaussian self-dual 3-form by holomorphic
factorization, and we recall some useful facts about topological strings.
$\bullet$
In Section \ref{sec_pert}, we discuss the topology of the
hypermultiplet moduli space, using the qualitative form of D-instanton and
NS5-instanton corrections to identify the periodicities of the RR and NS axions.
$\bullet$
Then, in Section \ref{sec_IIB}, we describe the corresponding type IIB picture and,
using mirror symmetry and S-duality, we find the classical action for general
fivebrane instantons, and show that it reduces to the Gaussian flux action
in the weak coupling limit.
$\bullet$
In Section \ref{sec_ns5twi}, we review the twistorial
description of the perturbative and D-instanton corrected HM moduli space,
and obtain a candidate section of $H^1(\cZ_\cM,\cO(2))$
which should govern deformations of the complex contact structure
on twistor space $\cZ_\cM$ induced by $(p,k)$-fivebranes.
The partition function $\ZNS{k}$ of $k$ NS5-branes, including non-linear effects,
follows by Penrose transform from this holomorphic section.
$\bullet$ We conclude in Section \ref{sec-dis} with a summary and discussion.
$\bullet$ The transformation properties of the candidate fivebrane transition
functions  $H_{k,p,\hgam}$ under various actions are discussed in  Appendix \ref{apsec_Heis},
and a directory of notations is provided in Appendix B. Note in particular the
pervasive use of the notation $\expe{z}=\exp\[2\pi\I z\]$.

\section{Flux partition functions and topological wave functions\label{sec_zns5}}

In this section we begin by revisiting previous constructions of the fivebrane partition function,
originally spelled out in \cite{Witten:1996hc} and further discussed in
\cite{Henningson:1999dm,Dijkgraaf:2002ac,Diaconescu:2003bm,Moore:2004jv,Belov:2006jd}.
We also review some relevant background material on topological strings,
which will play an important role in Section \ref{sec_ns5twi}.

$\bullet$ In Section \ref{sec_prel}, we set up the problem and
introduce our conventions for the moduli space
of metrics and RR $C$-fields on $\cX$.
$\bullet$
In Section \ref{sec_factorize}, we obtain, up to an overall metric-dependent factor,
the partition function $\ZG{k}$ for a self-dual 3-form, by holomorphic factorization
of the partition function of a non-chiral Gaussian harmonic three-form.
$\bullet$ In this construction, a key role is played by a certain $\mathbb{Z}_2$-valued
function $\sigma_\Theta$ (a quadratic refinement of the intersection form on $H^{3}(\cX,\mathbb{Z})$ modulo 2).
As we discuss in Section \ref{sec_theta}, the quadratic refinement $\sigma_\Theta$
ensures that  $\ZG{k}$ is valued in a single line bundle $\cL_\Theta^k$ with characteristics
$\Theta\in (\mathbb{Z}/2\mathbb{Z})^{b_3}$.
$\bullet$ In Section \ref{subsec-nonG} we discuss non-Gaussian generalizations
of this partition function and their interpretation in terms of certain wave functions.
$\bullet$ Section \ref{sec_top} is then devoted
to a discussion of various aspects of the topological A- and B-model, with particular
emphasis on the wave function properties of their respective partition functions.
In particular, we recall some useful relations between topological wave functions
and the partition functions of Donaldson-Thomas and Gromov-Witten invariants.

Our exposition in  Sections \ref{sec_factorize}
and \ref{sec_theta} follows closely  \cite{Witten:1996hc,Belov:2006jd} but is specifically
geared towards the partition function of the NS5-brane on a Calabi-Yau threefold.
As was already stressed in Section \ref{sec_intro}, $\ZG{k}$ is
not the partition function of the NS5-brane, but only an approximation thereof.
In order to construct $\ZG{k}$ through holomorphic factorization,
one is forced to make some simplifying assumptions, which we find in Section \ref{sec_ns5twi}
are in fact not valid to accurately describe the NS5-brane. Nevertheless,
we find it useful and illuminating to go through the derivation of $\ZG{k}$
in some detail in our conventions in order to facilitate the comparison with the correct
NS5-partition function constructed in Section \ref{sec_ns5twi} using twistor techniques.
Moreover, $\ZG{k}$ does capture many of the key topological features which are expected
of the NS5-brane partition function.

\subsection{Setting the stage}\label{sec_prel}

Let us consider type IIA string theory compactified on a Calabi-Yau threefold $\cX$, with a
stack of $k$ Euclidean NS5-branes whose world-volume $W$ consists of $\cX$ itself.
For $k=1$,  the world-volume $W$ supports 5 scalar fields (describing
transverse fluctuations of $W$ in $\IR^{10}\times S^1$, where $S^1$ is
the M-theory circle), two symplectic Majorana-Weyl fermions and a 2-form
potential $\cB$, whose 3-form field strength $H=\de \cB$ must be imaginary
self-dual \cite{Callan:1991ky},
\be
\star_W H = \I  H\, .
\ee
The world-volume $W$ is a magnetic source for the NS 2-form potential $B$ propagating
in the 10-dimensional bulk of space-time, while the flux $H$ is an electric source
for the RR 3-form potential $C$, mimicking D2-branes bound to the NS5-brane.

For $k>1$ fivebranes wrapping $\cX$, the world-volume dynamics is poorly understood,
but presumably splits into ``twisted sectors" labeled by pairs of integers $(p,q)$ such
that $p q=k$ \cite{Belov:2006jd},
where the $k$ fivebranes on $\cX$ recombine into $q$ fivebranes
wrapping $p$ copies of $\cX$. The dynamics in the $(p,q)$  twisted sector should
be described by $q^3$ interacting self-dual two-forms $\cB_{ijk}$, $i,j,k=1\dots q$,
such that the $q$ two-forms $\cB_{iii}$ stay massless on the Coulomb
branch. We shall focus on the Abelian
2-form  $\cB=\sum_{i=1\dots q} \cB_{iii}$  describing the center of mass degrees of freedom.
Moreover, we restrict to a regime where
$g_s |H|\ll 1$, such that the flux $H$ contributes quadratically to the fivebrane action.

Our goal in this Section is to revisit the construction of the Gaussian topological
partition sum $\ZG{k}$ for the field $H$ together with its supersymmetric partners,
as a function of the moduli of the CY threefold $\cX$ and the periods of the background $C$-field on $\cX$.
By topological partition sum, we mean the partition sum weighted by $(-1)^{2J_3} (2J_3)^2$,
which is relevant for two-derivative couplings in the low energy effective action.
As already mentioned in Section \ref{sec_intsusy}, the assumption that $g_s |H|\ll 1$
eventually breaks down for large values of the flux, and
the Gaussian action should  be replaced by its non-linear, Born-Infeld type
completion \cite{Bandos:1997ui,Aganagic:1997zq,Howe:1997fb,Cederwall:1997gg,Bandos:2000az}
in order to obtain the physical NS5-brane partition function. We shall return to
this issue in Section \ref{sec_nonlin}.

As will become apparent shortly, $\ZG{k}$ is independent of the \kahler structure of $\cX$,
and depends only on its complex structure. The moduli space  $\cM_c(\cX)$
of complex structures can be parametrized as usual by the complex periods
\be
\label{defXF}
X^\Lambda=\int_{\mathcal{A}^\Lambda} \Omega_{3,0}\, , \qquad\qquad
F_\Lambda= \int_{\mathcal{B}_\Lambda} \Omega_{3,0}
\ee
of the holomorphic 3-form\footnote{We trust that the reader will not confuse
the holomorphic 3-form $\Omega$ with the generalized Donaldson-Thomas invariant $\Omega(\gamma)$.
The latter will always come with an argument.}
$\Omega$ along a symplectic  basis
$(\mathcal{A}^\Lambda,\mathcal{B}_\Lambda)$, $\Lambda=0,\dots, h_{2,1}(\cX)$,
of $H_3(\cX,\mathbb{Z})$ modulo holomorphic rescalings $\Omega_{3,0}\mapsto e^f \Omega_{3,0}$.
We shall denote $\Omega\equiv(X^\Lambda,F_\Lambda)^T$.
The vector $\Omega$ thus takes values in an homogeneous complex Lagrangian cone
$\cC_\cX$, generated locally by a prepotential $F(X^\Lambda)$ such that
$F_\Lambda=\partial_{X^\Lambda} F(X)$.
The complex structure moduli space $\cM_c(\cX)$ is the quotient $\cC_X/\mathbb{C}^\times$
by homogeneous rescalings, which we parametrize by complex
coordinates $z^a$. Conversely, the cone
$\cC_X$ provides a canonical complex line bundle over $\cM_c(\cX)$,
known as the Hodge line bundle, which we denote by $\cL$.
Recall that $\cM_c(\cX)$ carries a special \kahler metric, with  \kahler potential
\be
\cK=-\log[ \I ( \bar X^\Lambda F_\Lambda - X ^\Lambda \bar F_\Lambda)]
\ee
and \kahler class $\omega_\sk=-\frac{1}{2\pi} \de\cA_{K}$, where
\be
\cA_K=\frac{\I}{2}( \cK_a \de z^a -  \cK_{\ba} \de \bar z^{\ba})
\label{Kahlerconnection}
\ee
is the \kahler connection.\footnote{Note that our definition of $\cA_K$ differs by
a factor of 2 from that used in \cite{Alexandrov:2007ec,Alexandrov:2008nk,Alexandrov:2008gh}.}
Under a monodromy $M$ in $\cM_c(\cX)$, the symplectic basis $(\mathcal{A}^\Lambda,\mathcal{B}_\Lambda)$
transforms by an integer-valued symplectic rotation
$\rho(M)={\scriptsize \begin{pmatrix}
\cD & \cC \\ \cB & \cA \end{pmatrix}}
\in Sp(b_3(\mathcal{X}),\IZ)$ with $b_3(\mathcal{X})=2h_{2,1}+2$.
At the same time $\Omega_{3,0}\mapsto e^{f_M} \Omega_{3,0}$ is generically rescaled,
so the period vector $\Omega$  transforms into $e^{f_M} \rho(M) \Omega$.
In the process, $\cK$ and $\cA$ change respectively by \kahler transformations
and gauge transformations of the form
\be
\cK\ \longmapsto\ \cK-f_M-\bar f_M, \qquad\qquad \cA_K\ \longmapsto\ \cA_K + \de \Im(f_M).
\label{KahlerGaugetransf}
\ee

A topologically trivial real $C$-field on $\cX$ can be parametrized by its real periods
\be
\label{defzezet}
\zeta^{\Lambda}=\int_{\mathcal{A}^{\Lambda}} C, \qquad \qquad \tilde\zeta_\Lambda=\int_{\mathcal{B}_\Lambda}C\, .
\ee
Again, we shall abuse notation and write $C$ for the vector $(\zeta^\Lambda,\tzeta_\Lambda)^T$.
By charge quantization,  the entries of $C$ are periodic with unit periods,
and parametrize a point on the $b_3(\cX)$-dimensional real torus
\be
\label{Tfib}
\cT = H^{3}(\mathcal{X},\mathbb{R}) / H^{3}(\mathcal{X},\mathbb{Z}) \, ,
\ee
equipped with the canonical symplectic form\footnote{In this equation and \eqref{dsT} below,
$\de C$ denotes the variation of $C$, {\it not}  its exterior
derivative on $\cX$. }
\be
\omega_\cT = -\int_\cX \de C \wedge \de C = \de\tzeta_\Lambda  \wedge \de\zeta^\Lambda \, .
\label{KahlerformT}
\ee
The torus $\cT$ is fibered over the complex structure moduli space $\cM_c(\cX)$, such that
$(\zeta^{\Lambda},\tzeta_\Lambda)$ transforms by the symplectic rotation $\rho(M)$ under
a monodromy $M$.  We shall refer to the total space of this fibration,
\be
\label{intJac}
\begin{array}{ccc}
  H^{3}(\cX,\mathbb{R})/H^{3}(\cX,\mathbb{Z}) & \to & \cJ_c(\cX) \\
   &  & \downarrow \\
   &  & \cM_c(\cX),
\end{array}
\ee
as the intermediate Jacobian
$\mathcal{J}_c(\cX)$. The Hodge star $\star_\cX$ on $H^3(\cX,\mathbb{R})$
endows the fiber $\cT$ with a complex structure and a
positive-definite \kahler metric
\be
\label{dsT}
\de s^2_\cT =\frac12 \int_\cX \de C \wedge \star_\cX \de C= -\frac12
\de \omega_\Lambda  \Im \cN^{\Lambda\Sigma} \,\de\bar\omega_{\Sigma}\, ,
\ee
where the complex coordinates $\omega_\Lambda$ are given by
\be
\label{omjac}
\omega_\Lambda = \tzeta_\Lambda - \bar\cN_{\Lambda\Sigma} \zeta^\Sigma\, ,
\qquad
\bar \omega_\Lambda = \tzeta_\Lambda - \cN_{\Lambda\Sigma} \zeta^\Sigma\, .
\ee
The \kahler class of the metric on $\cT$ is the symplectic form $\omega_\cT$ in
\eqref{KahlerformT} above. Equipped with the complex structure just defined,
which we refer to as the Weil complex structure, $\cT$ is a principally polarized
Abelian variety.

The Weil period matrix $\mathcal{N}_{\Lambda\Sigma}$ is related to the Griffiths
period matrix $\tau_{\Lambda\Sigma}=\partial_\Lambda \partial_\Sigma F(X)$ via the usual
special geometry relation
\be
\label{defcN}
\cN_{\Lambda\Lambda'} =\bar \tau_{\Lambda\Lambda'} +2\I \frac{ [\Im\tau \cdot X]_\Lambda
[\Im \tau \cdot X]_{\Lambda'}}
{X^\Sigma \, \Im\tau_{\Sigma\Sigma'}X^{\Sigma'}}\, .
\ee
Recall that $\Im \cN_{\Lambda\Sigma}$ and its inverse $\Im \cN^{\Lambda\Sigma}$
are negative definite symmetric matrices, while $\Im \tau_{\Lambda\Sigma}$
and its inverse $\Im \tau^{\Lambda\Sigma}$ have signature $(1,b_3(\cX)-1)$.
Under monodromies in $\cM_c(\cX)$, both $\cN_{\Lambda\Lambda'}$ and
$\tau_{\Lambda\Lambda'} $ transform by fractional linear transformations
\be
\label{monoN}
\cN \mapsto (\cA \cN + \cB)\, (\cC\cN + \cD)^{-1}\, ,
\qquad
\tau \mapsto (\cA \tau+ \cB)\, (\cC\tau + \cD)^{-1}\, .
\ee
However, unlike the Griffiths period matrix $\tau_{\Lambda\Sigma}$, the Weil
period matrix $\cN_{\Lambda\Sigma}$ does not vary holomorphically
over $\cM_c(\cX)$.

As is well-known, the self-duality constraint on $H$ leads to several complications in defining
its partition function. A self-dual flux $H$ can only couple to the anti-self-dual
part of the background 3-form $C$, so one would naively expect $\ZG{k}$ to be a
holomorphic function
on $\cT$, independent of $\bar\omega_\Lambda$. Instead, as emphasized
in  \cite{Witten:1996hc} and reviewed in detail below,
$\ZG{k}$  is more properly viewed as a holomorphic section
of a certain line bundle over the torus $\cT$.
As in the case of two-dimensional chiral bosons \cite{AlvarezGaume:1986mi,
AlvarezGaume:1987vm}, it is convenient to obtain the
partition function of a chiral 2-form  by factorizing the partition function of a non-chiral two-form.

\subsection{Factorizing the non-chiral partition function}\label{sec_factorize}

To construct the partition function we first ignore the self-duality condition
on the self-dual 3-form, and restrict to the limit of small string coupling. Then
the action for the 3-form flux on the world-volume $p \cX$ of $q$ fivebranes
is given by \cite{Witten:1996hc,Belov:2006jd}
\be
S(H,C)= p\left[ \frac{\pi}{q}  \int_\cX (H-q C)  \wedge \star (H-q C)
- \I \pi \int_\cX C \wedge H \right] .
\label{GeneralFivebraneAction}
\ee
The overall factor of $p$ reflects the wrapping number of each fivebrane on $\cX$,
while the $1/q$ factor is the familiar fractionization effect, coming from expanding
$\sqrt{q^2/g_s^4+ H^2/g_s^2}-|q|/g_s^2$ at small coupling. The factor of $q$ in $H-qC$ arises by summing
$H_{iii}-C$ over $i=1\dots q$.

Flux quantization requires that $H\in H^3(\cX,\IZ)$, i.e. that the periods
\be
n^\Lambda=\int_{\mathcal{A}^{\Lambda}} H, \qquad \qquad m_{\Lambda}=\int_{\mathcal{B}_\Lambda}H
\label{defmn}
\ee
be integer valued. Using \eqref{dsT},
 the classical action (\ref{GeneralFivebraneAction}) then reads
\be
\label{Smn}
S(\cN, H,C)= -\frac{\pi p}{2 q}
\left( \tilde m_\Lambda - \cN_{\Lambda\Lambda'}
\tilde n^{\Lambda'}  \right) \Im \cN^{\Lambda\Sigma}
 \left( \tilde m_{\Sigma} -  \bar \cN_{\Sigma\Sigma'}
\tilde n^{\Sigma'} \right)
- \I \pi p \,(m_\Lambda \zeta^\Lambda - n^\Lambda \tzeta_\Lambda),
\ee
where we defined $\tilde m_\Lambda=m_\Lambda-q \tzeta_\Lambda,\
\tilde n^\Lambda=n^\Lambda-q \zeta^\Lambda$. We also indicated
the dependence on the period matrix $\cN$. As advertised above,
\eqref{Smn} is independent of the \kahler moduli of $\cX$. In terms of the
self-dual and anti-self-dual components of the flux $H$,
\be
R_\Lambda =  m_\Lambda - \bar\cN_{\Lambda\Sigma} n^\Sigma\, ,
\qquad
\bar R_\Lambda = m_\Lambda - \cN_{\Lambda\Sigma} n^\Sigma\, ,
\ee
and of the $C$-field, Eq. \eqref{omjac}, this is equivalently rewritten (in
agreement with \cite{Belov:2006jd}, Eq. 2.4) as
\be
\label{SHR}
S(\cN, H,C)= -\Im \cN^{\Lambda\Sigma} \left[ \frac{\pi p}{2 q}\, R_\Lambda \bar R_\Sigma
- \pi p \, \bar\omega_\Lambda R_\Sigma
+ \frac{\pi p q}{2}\,\omega_\Lambda \bar\omega_\Sigma \right] .
\ee
Thus, the self-dual part $\omega_\Lambda$ of $C$ decouples, save
for the last term in \eqref{SHR} which originates from the quadratic term
$C\wedge \star_\cX C$ in (\ref{GeneralFivebraneAction}).

The partition function for a Gaussian 3-form flux in the twisted
sector $(p,q)$, with $k=pq$, is now obtained as
\be
\cZ_\Gauss^{(p,q)}(\cN,C)
= \sum_{H\in H^3(\cX,\IZ)} r(\cN, H, C) \, e^{-S(\cN, H,C)}\, ,
\label{nonchiralpartitionfunction}
\ee
where the prefactor $r(\cN, H, C)$ corresponds to the fluctuation determinant
around the harmonic flux configuration $H=n^{\Lambda}\alpha_\Lambda-m_\Lambda \beta^{\Lambda}$,
where $(\alpha_\Lambda,\beta^{\Lambda})$ is the symplectic basis
on the lattice $\Gamma=H^{3}(\mathcal{X},\mathbb{Z})$ dual\footnote{Namely,
$\int_{\cA^\Lambda}\alpha_\Sigma=\delta^{\Lambda}_{\Sigma}$,
$\int_{\cB_\Lambda}\beta^\Sigma=-\delta_{\Lambda}^{\Sigma}$,
$\int_{\cA^\Lambda}\beta^\Sigma=\int_{\cB^\Lambda}\alpha_\Sigma=0$.}
 to $(\mathcal{A}^{\Lambda},
\mathcal{B}_\Lambda)$. In view of our ignorance of the non-Abelian dynamics of $q$
fivebranes, we do not know how to evaluate $r(\cN, H, C)$ from first principles. Nevertheless,
it was argued in \cite{Moore:2004jv,Belov:2006jd} on topological grounds that  $r(\cN, H, C)$
must take the form
\be
r(\cN, H, C) = |\cF|^2\, \left[\sigma(H)\right]^k\, ,
\label{r(H)}
\ee
where the normalization factor
$|\cF|^2$ depends only on the metric on $\cX$
(and possibly, on the string coupling), but not on $H$-flux nor
on the background $C$-field. We shall see in Section \ref{sec_nonlin}
that this assumption does not hold for the fivebrane instanton partition $\ZNSG{k}$
derived from twistor techniques, where the factor $|\cF|^2$ appears to depend on
$H$ and $C$, see \eqref{normalizationfactor} below.
This dependence however is unessential for the topological
properties that we wish to emphasize here, and we proceed
in the remainder of this section under the assumption that $|\cF|^2$
is independent of the flux $H$ and the background field $C$.

The second (essential) factor $\sigma(H)$ in \eqref{r(H)}
is a ``quadratic refinement of the intersection form"
on the lattice $\Gamma$. This is defined as a homomorphism
$\sigma: \Gamma\to U(1)$, i.e. a phase
assignment such that
\be
\label{qrifprop}
\sigma(H+H') = (-1)^{\langle H, H' \rangle}\, \sigma(H)\, \sigma(H') \, ,
\ee
where
\be
\langle H, H' \rangle=m_\Lambda n'^\Lambda-m'_\Lambda n^\Lambda.
\label{SymplecticPairing}
\ee
The most general solution of \eqref{qrifprop} is given by \cite{Belov:2006jd}
\be
\sigma_\Theta(H) = \expe{-\frac12  m_\Lambda n^\Lambda + m_\Lambda \theta^\Lambda
- n^\Lambda \phi_\Lambda}\, ,
\label{quadraticrefinement}
\ee
where $\Theta=(\theta^\Lambda,\phi_\Lambda)^T$  are independent of $H$
and defined modulo $\Gamma$.
As we shall see below, $\Theta$ determines the characteristics of the theta series
governing the chiral partition function. By analogy with the standard Jacobi theta
series, $\Theta$ is sometimes referred to as a ``generalized spin structure" on $\cX$.
$\Theta$ in general depends on the metric on $\cX$: under monodromies in
$\cM_c(\cX)$, $\Theta$ transforms as
\be
\label{sympchar}
\begin{pmatrix} \theta^\Lambda \\ \phi_\Lambda \end{pmatrix}
\mapsto
\rho(M)
\cdot \[
\begin{pmatrix} \theta^\Lambda \\ \phi_\Lambda  \end{pmatrix}
-\frac12
\begin{pmatrix} (\cA^T\cC)_d \\ (\cD^T\cB)_d  \end{pmatrix}
\],
\ee
where $(A)_d$ denotes the diagonal of a matrix $A$ \cite{MR2062673}.
Since the $H$-flux transforms as $H\mapsto \rho(M)\cdot H$, the
quadratic refinement $\sigma_\Theta(H)$ is invariant.
Moreover, while it is consistent to assume that $\sigma_\Theta^2(H)=1$,
$2\Theta\in\Gamma$
for the partition of a self-dual three-form \cite{Witten:1996hc,Belov:2006jd},
in general, this is not so for the full fivebrane partition function,
where $\sigma_\Theta(H)$ may be $U(1)$-valued. In this subsection, we
nevertheless assume that $\sigma_\Theta^2(H)=1$, which
simplifies the holomorphic factorization procedure. We shall relax
this assumption in Section \ref{sec_theta} onwards.

In order to decouple the self-dual and anti-self dual parts of $H$, we now split the
lattice $\Gamma=H^3(\cX,\mathbb{\IZ})$ into the sum of two Lagrangian lattices
$\Gamma_e\oplus \Gamma_m$ (where '$e$' and '$m$' stand for 'electric' and 'magnetic'),
given by
\be
\Gamma_e = \sum_\Lambda \IZ \, \beta^\Lambda\,  ,
\qquad
\Gamma_m = \sum_\Lambda \IZ \, \alpha_\Lambda\, ,
\ee
and perform a Poisson resummation on the integers $m_\Lambda\in \Gamma_e$.
Denoting by $r^\Lambda\in \Gamma^*_e=\Gamma_m$ the dual summation variable, we find
\be
\mathcal{Z}^{(p,q)}_{\Gauss}(\mathcal{N},C) =
 \mathcal{D}^{1/2}\, |\cF|^2\,
 \sum_{(n^\Lambda,r^\Lambda)\in \Gamma_m} e^{-\tilde S}\, ,
\label{nonchiralpartition}
\ee
where $\mathcal{D}\equiv \det(- \frac{2q}{p} \Im\cN)$. The dual action separates into
\be
\label{Sdual}
\begin{split}
\tilde S =&\,  {\I\pi k}  \left(p_R^{\Lambda} \,\cN_{\Lambda\Sigma}\, p_R^{\Sigma}
-p_L^{\Lambda} \, \bar\cN_{\Lambda\Sigma} \, p_L^{\Sigma}  \right) + 2 \pi \I k\, \bar\omega_\Lambda p_R^\Lambda
+2\pi\I k q \, \phi_\Lambda ( p_L^\Lambda+p_R^\Lambda)\\
& -  \frac{\pi k}{2}\, \bar\omega_{\Lambda}  \Im\cN^{\Lambda\Sigma}
(\omega_{\Sigma}-\bar \omega_{\Sigma}),
\end{split}
\ee
where the vectors $p_L,p_R\in\Gamma_m/(2|k|)$ are the following linear combinations
of $n^\Lambda$ and $r^\Lambda$,
\beq
p_R^\Lambda &=& \frac12 (q^{-1}+q)\, n^\Lambda -(p^{-1} r^\Lambda+q \,\theta^\Lambda) \, ,
\\
p_L^\Lambda &=& \frac12 (q^{-1}-q)\, n^\Lambda + (p^{-1} r^\Lambda+q\, \theta^\Lambda) \, .
\eeq
For later reference, we note that the last term in \eqref{Sdual} may be rewritten in terms of the
real periods \eqref{defzezet} of the $C$-field as
\be
\label{wwwb}
\frac{\pi k}{2}\, \bar\omega_{\Lambda} \Im\cN^{\Lambda\Sigma}  (\omega_{\Sigma}-\bar \omega_{\Sigma})
=\I \pi k \, \zeta^\Lambda \left( \tzeta_\Lambda - \cN_{\Lambda\Sigma} \zeta^\Sigma \right).
\ee

Expressing the sum over  $(n^\Lambda,r^\Lambda)$ as a sum over $p_L,p_R$ in
suitably shifted lattices, the total partition function then decomposes
as (\cite{Belov:2006jd}, Thm. E.1)\footnote{Eq. \eqref{sumblocks} holds
under the assumptions that $\Theta$ is half-integer, $pq=p\ ({\rm mod}\ 2)$
and $\gcd(p,q)=1$, in particular $q$ is odd.
More generally, the partition function can always be decomposed into
a sum of products of level $pq/2m^2$ theta series, where $m=\gcd(p,q)$.
The quadratic refinement $\sigma(H)$ in \eqref{r(H)}
is crucial in ensuring that a single characteristics $\Theta$
appears in the sum \cite{AlvarezGaume:1987vm}.}
\be
\label{sumblocks}
\cZ^{(p,q)}_{\Gauss}(\mathcal{N}, C) = \mathcal{D}^{1/2}\, \sum_
{\substack{\mu_p\in (\Gamma_m/p)/\Gamma_m\\
\mu_q\in (\Gamma_m/q)/\Gamma_m}}
\cZ_{\Theta,\mu_p+\mu_q}^{(k)} ( \cN , 0)\,
\overline {\cZ_{\Theta,\mu_p-\mu_q}^{(k)}( \cN , C )}\, ,
\ee
analogous to the sum of products of holomorphic and anti-holomorphic conformal blocks
appearing in the partition function of a compact scalar field on a two-torus.
The ``holomorphic conformal block" $\cZ_{\Theta,\mu}^{(k)}$ appearing
in \eqref{sumblocks} is given by\footnote{This matches Eq. E.9 in \cite{Belov:2006jd}
upon identifying $(T,a_1,a_2,\gamma,n^I,w^I)$ there with
$(\bar\cN,\zeta,\tzeta,\mu,n^\Lambda,r^\Lambda+p q n^\Lambda)$ here.
}
\be
\label{thpl}
\cZ_{\Theta,\mu}^{(k)} (\cN, C)
 = \,\cF\,
 \!\! \!\!\sum_{n \in \Gamma_m+\mu+\theta}\!\!
 {\bf E}\bigg( \frac{k}{2} \left(
  n^{\Lambda} \bar\cN_{\Lambda\Sigma} n^{\Sigma} +\theta^\Lambda \phi_\Lambda\right)
  +k\left(\omega_\Lambda- \phi_\Lambda \right)n^{\Lambda} +  \frac{\I k}{4}
    \omega_\Lambda \Im\cN^{\Lambda\Sigma}
 (\omega_{\Sigma}-\bar \omega_{\Sigma}) \bigg) \, .
\ee
In this expression, both the modulus and phase of the prefactor $\cF$ are unspecified at this point,
but depend only on the complex structure of $\cX$ and string coupling.
Up to this normalization ambiguity, $\ZG{k}\equiv \cZ_{\Theta,\mu}^{(k)}$
is then the sought-for partition function of a self-dual three-form
on the worldvolume $\cX$, in the weak coupling limit,
and in the topological sector with quadratic refinement $\Theta=(\theta,\phi)$.
The shift vector $\mu$ runs
over the $|k|^{b_3(\cX)}$ elements in $(\Gamma_m/|k|)/\Gamma_m$, so
for $|k|\geq 1$ the partition function is
vector-valued. While the derivation of \eqref{thpl} from holomorphic factorization
has assumed that $2\Theta\in\Gamma$, the result \eqref{thpl}
is well-defined for any $\Theta$, allowing us the relax the assumption
that  $\Theta$ was half-integer.
We reiterate that $\ZG{k}$ differs from the
fivebrane instanton partition function in the weak coupling limit $\ZNSG{k}$
to be computed in Section \ref{sec_ns5twi},
although the difference is inessential at the level of the present discussion.

\subsection{Chiral partition function and Gaussian theta series\label{sec_theta}}

For $k=1$, the partition function of a Gaussian self-dual three-form
\eqref{thpl} is recognized as a
Siegel theta series of rank $b_3(\cX)/2$ with  in general, real
characteristics $\Theta=(\theta^\Lambda,\phi_\Lambda)$,
evaluated at the period matrix $\bar\cN$ on the Siegel upper-half plane.
More precisely, defining the standard holomorphic Siegel theta series as
\be
\label{thsiegel}
\vartheta_{\rm Siegel} \[ {} ^\theta_\phi\](\bar\cN, \omega_\Lambda)
 = \sum_{n \in \Gamma_m+\theta}\!\!
 {\bf E}\bigg( \frac{1}{2}
  n^{\Lambda} \bar\cN_{\Lambda\Sigma} n^{\Sigma}
  +\left(\omega_\Lambda- \phi_\Lambda \right)n^{\Lambda} \bigg) ,
\ee
we have, independently of $\mu$,
\be
\label{thplsiegel}
\cZ_{\Theta,\mu}^{(1)} (\cN, C)
= \,\cF\,
{\bf E}\bigg( \frac{1}{2} \theta^\Lambda \phi_\Lambda
 +  \frac{\I }{4} \,   \omega_\Lambda\Im\cN^{\Lambda\Sigma}
(\omega_{\Sigma}-\bar \omega_{\Sigma}) \bigg)
\vartheta_{\rm Siegel} \[ {} ^\theta_\phi\](\bar\cN, \omega_\Lambda)\, .
\ee
More generally, for $|k|>1$ Eq. \eqref{thpl} is a level $k/2$ generalization
of the Siegel theta series.
Using \eqref{wwwb}, it
may be usefully rewritten as
\be
\label{thpl2}
\cZ_{\Theta,\mu}^{(k)}(\cN, C) =\cF
\!\! \!\!\sum_{n \in \Gamma_m+\mu+\theta}
\!\!\!\expe{\frac{k}{2} (\zeta^\Lambda-n^{\Lambda})\bar \cN_{\Lambda\Sigma}
(\zeta^\Sigma-n^{\Sigma})
+k (\tilde\zeta_{\Lambda}-\phi_\Lambda) n^{\Lambda}
+\frac{k}{2} ( \theta^{\Lambda}\phi_{\Lambda}-\zeta^{\Lambda}\tilde\zeta_{\Lambda}
) }\, .
\ee

Under translations by a vector
$H=(\eta^\Lambda,\tleta_\Lambda)\in\Gamma$ on the torus~$\cT$, i.e. large
gauge transformations of the $C$-field,
the partition function \eqref{thpl2} satisfies the twisted periodicity property\footnote{Note that
the defining property \eqref{qrifprop} of the quadratic refinement is crucial
in ensuring the consistency of this periodicity condition.}
\be
\label{thperiod}
\cZ_{\Theta,\mu}^{(k)} (\cN, C+H)
= (\sigma_\Theta(H))^k \, \expe{\frac{k}{2}(\eta^{\Lambda}\tilde\zeta_\Lambda-\tleta_\Lambda\zeta^{\Lambda})}
\cZ_{\Theta,\mu}^{(k)}  (\cN,C).
\ee
Thus, $\cZ_{\Theta,\mu}^{(k)}$
is not a function on $\cT$, but rather
a section of  $(\cC_\Theta)^k$,  where $\cC_\Theta$ is the circle bundle over $\cT$
with first Chern class equal  to the \kahler class $\omega_\cT$ in \eqref{KahlerformT},
and whose sections satisfy the twisted periodicity property
\eqref{thperiod} for $k=1$ \cite{Witten:1996hc}.
It is important to stress that $\cC_\Theta$ and $\cC_{\Theta'}$
are non-isomorphic bundles unless $\Theta-\Theta'\in \Gamma$. Yet, the theta series for different
values of the characteristics are related by a translation along the torus,
\be
\label{thschar}
\cZ_{\Theta,\mu}^{(k)}(\cN, C)=
\expe{\frac{k}{2} \left( \langle \Theta,\Theta'\rangle
+  \langle C,\Theta- \Theta'\rangle\right)}
\cZ_{\Theta',\mu}^{(k)}(\cN, C+\Theta'-\Theta).
\ee

Moreover, thanks to its Gaussian character,
the  flux partition function
is actually a holomorphic section of  $(\cL_\Theta)^k$,
where $\cL_\Theta$ is the line bundle with circle bundle
$\cL^\circ_\Theta=\cC_\Theta$. Indeed, $\cZ_{\Theta,\mu}^{(k)}$
is annihilated by the covariant anti-holomorphic derivative
\be
\left( \frac{\partial}{\pa \bar \omega_{\Lambda}} - \frac{\pi k}{2}
\Im\cN^{\Lambda\Sigma} \omega_{\Sigma} \right) \,
\cZ_{\Theta,\mu}^{(k)}(\cN, \omega_\Lambda,\bar \omega_\Lambda)=0.
\label{holomorphicderivative}
\ee
By the usual index theorem and Kodaira vanishing argument \cite{Witten:1996hc},
the line bundle $\cL^k_\Theta$ admits exactly $|k|^{b_3(\cX)/2}$ holomorphic sections,
corresponding to the possible values of $\mu\in (\Gamma_m/|k|)/\Gamma_m$. Physically,
this is the familiar degeneracy of the Landau levels for a particle moving on a
$b_3(\cX)$-dimensional torus with $k$ units of magnetic flux.

Having described the behavior of the flux partition function \eqref{thpl2} under
large gauge transformations, we now turn to its
behavior under
monodromies in complex structure moduli space $\cM_c(\cX)$. Under
a monodromy $M$, the period matrix transforms by fractional linear
transformations \eqref{monoN},  the $C$ field transforms as
$C\mapsto \rho(M) \cdot C$ and the characteristics $\Theta$ transform
as \eqref{sympchar}, leaving the non-chiral partition function $\cZ^{(p,q)}_\Gauss$ invariant.
In contrast, the modular properties of the Siegel theta series imply that
the chiral fivebrane partition function $\cZ_{\Theta,\mu}^{(k)}$
is mapped to a linear superposition of $\cZ_{\Theta',\mu'}^{(k)}$
with $\mu'$ ranging over $(\Gamma_m/|k|)/\Gamma_m$, and fixed characteristics
$\Theta'$. Thus, the bundle where
$\cZ_{\Theta,\mu}^{(k)}$ takes values is also non-trivially fibered over $\cM_c(\cX)$,
as will be further elaborated upon in Section \ref{sec_pert}.
Specifically, under symplectic transformations the
level $k/2$ theta series $\vartheta_{(\theta,\phi),\mu}^{(k)}\equiv
\cF^{-1}\cZ_{\Theta,\mu}^{(k)}$
 transforms as \cite{Belov:2006jd}:
\begin{enumerate}
\item for $\rho(M)={\scriptsize \begin{pmatrix}
\cA^{-T} & 0 \\ 0 & \cA \end{pmatrix}}$:
\be
\vartheta_{(\theta,\phi),\mu}^{(k)}(\cA \cN \cA^T, \zeta,\tzeta )=
\vartheta_{(\cA^T \theta, \cA^{-1} \phi),\cA^T \mu}^{(k)}(\cN, \cA^T \zeta,\cA^{-1}\tzeta)\, ,
\label{cAtrans}
\ee
\item  for $\rho(M)={\scriptsize \begin{pmatrix}
1 & 0 \\ \cB & 1 \end{pmatrix}}$, where $\cB$ is a symmetric integer matrix,
\beq
\label{vartheB}
Ê\vartheta_{(\theta,\phi),\mu}^{(k)}(\cN+\cB, \zeta,\tzeta ) &=&
\expe{-\frac{k}{4} \cB_{\Lambda\Lambda}(\theta^\Lambda+2\mu^\Lambda)
+\frac{k}{2}   \cB_{\Lambda\Sigma}\mu^\Lambda \mu^\Sigma}
\nonumber \\
& &  \times \vartheta_{( \theta^\Lambda, \phi_\Lambda-\cB_{\Lambda\Sigma}\theta^\Sigma
-\frac12 \cB_{\Lambda\Lambda}),\mu}^{(k)}(\cN, \zeta,\tzeta-\cB\zeta)\, ,
\eeq
\item for $\rho(M)={\scriptsize \begin{pmatrix}
0 & -1 \\ 1 & 0 \end{pmatrix}}$,
\be
\vartheta_{(\theta,\phi),\mu}^{(k)}(-\cN^{-1}, \zeta,\tzeta )=
\frac{\sqrt{\det(-\I \bar\cN)}}{ k^{b_3(\cX)/2}} \sum_{\mu'\in (\Gamma_e/|k|)/\Gamma_e}
\expe{- k \mu' \mu}\,
\vartheta_{(- \phi,\theta),\mu'}^{(k)}(\cN, -\tzeta,\zeta)\, .
\label{vartheS}
\ee
\end{enumerate}

\subsection{Non-Gaussian theta series and wave-functions}
\label{subsec-nonG}

Our discussion of the topological nature of the chiral partition function thus
far was based on the weak coupling result \eqref{thpl2}. However, as explained
in Section \ref{sec_intsusy}, we expect that
upon including the effects of non-Abelian dynamics and non-linearities on the
fivebrane world-volume, the exact NS5-brane partition function $\ZNS{k}$ will continue to
take values in the same bundle. With this in mind, we note that the most general solution to the periodicity
conditions  \eqref{thperiod}
can be written as a non-Gaussian
theta series
\be
\label{thpl2f}
\cZ_{\Theta,\mu}^{(k), {\rm NG}}(\cN, \zeta^\Lambda,\tzeta_\Lambda) =
\!\! \!\!\sum_{n\in \Gamma_m+\mu+\theta}\!\! \!
\Psi_\IR^{k,\mu}\left( \zeta^\Lambda - n^{\Lambda}\right)
\expe{  k (\tilde\zeta_{\Lambda}-\phi_\Lambda) n^{\Lambda}
+\frac{k}{2} (\theta^{\Lambda}\phi_{\Lambda}-\zeta^{\Lambda}\tilde\zeta_{\Lambda})} ,
\ee
where $\Psi_\IR^{k,\mu}(\zeta^\Lambda)$ will also depend on the complex structure moduli
and the string coupling. In general, unlike its Gaussian counterpart,
Eq. \eqref{thpl2f} is a section of the circle bundle
$\cC_\Theta^{k}=(\cL_\Theta^\circ)^k$ but not a holomorphic section of the line bundle
$\cL_\Theta^k$.

In order to satisfy the same transformation law as \eqref{thpl2} under monodromies
in $\cM_c(\cX)$, $\Psi_\IR^{k,\mu}\left( \zeta^\Lambda \right)$ should be invariant
under the combined action of fractional linear transformations \eqref{monoN} on
the period matrix $\cN$ and under monodromies acting in the metaplectic representation $\rho_m(M)$.
This implies that $\Psi_\IR^{k,\mu}\left( \zeta^\Lambda \right)$ should be viewed as
the wave function of a certain state $|\Psi^{k,\mu}\rangle$ in the Hilbert space $\mathcal{H}$ of
square-integrable functions on $H^{3}(\mathcal{X}, \mathbb{R})$.
To spell this out, consider first a specific state $| \Psi^{\Gamma_m,k,\mu} \rangle \in \cH$,
conveniently represented as follows in terms of its wave function
$\Psi_\IR^{\Gamma_m,k,\mu}(\zeta^\Lambda)$ in the $\zeta$-representation
(corresponding to the real polarization for the
topological wave function as further explained in the next subsection):
\be
\Psi_\IR^{\Gamma_m,k,\mu}(\zeta^\Lambda)
=e^{-\I\pi k \theta^\Lambda \phi_\Lambda}
\sum_{n \in \Gamma_m+\mu+\theta}\!\!
\delta\left(\zeta^\Lambda - n^\Lambda\right)
\, e^{2\pi \I k \, \phi_\Lambda n^\Lambda }.
\label{PsiRNS5G}
\ee
We further introduce two sets of operators,
$T^\Lambda$ and $\tilde T_\Lambda$, acting on $\cH$.
Their action on an arbitrary wave function in the real polarization reads
\be
\label{qalg}
T^\Lambda \cdot \Psi_\IR(\zeta^\Lambda) =2\pi k  \zeta^\Lambda  \Psi_\IR(\zeta^\Lambda)  ,
\qquad
\tilde T_\Lambda \cdot \Psi_\IR(\zeta^\Lambda) =  -\I  \pa_{\zeta^\Lambda} \Psi_\IR(\zeta^\Lambda),
\ee
so that they satisfy the Heisenberg algebra
\be
[T^\Lambda,\tilde T_\Sigma]
=2\pi \I k \delta^\Lambda_\Sigma.
\label{Heisenbergalgebra}
\ee
Our main observation is then that the non-Gaussian theta series \eqref{thpl2f}
can be rewritten as a matrix element involving the two states introduced above:
\be
\label{matel}
\cZ_{\Theta,\mu}^{(k),\text{NG}} (\cN, C)
=\langle \Psi^{\Gamma_m,k,\mu} |
\, e^{-\I(\zeta^\Lambda \tilde T_\Lambda -\tzeta_\Lambda T^\Lambda)}\ |
\Psi^{k,\mu} \rangle\, .
\ee
In particular, the Gaussian chiral partition function \eqref{thpl2}
is recovered by choosing the state $|\Psi^{k,\mu}\rangle$ as follows:
\be
\Psi_\IR^{k,\mu}(\zeta^\Lambda)
=\cF\, e^{\pi\I k \,\zeta^\Lambda \bar\cN_{\Lambda\Sigma}
\zeta^\Sigma}\, ,
\label{PsiGauss}
\ee
where $\cF$ is the normalization factor in \eqref{thpl2}.

In Section \ref{sec_ns5twi} we will find that the relation \eqref{matel}
indeed captures the correct partition function $\ZNS{k}$ of the NS5-brane,
provided we identify $|\Psi^{k,\mu}\rangle$ with a particular state associated with the B-model topological string,
for which the Gaussian wave function \eqref{PsiGauss} corresponds to the weak coupling approximation.
To pave the way for these developments, we will in the next section introduce
some relevant background material on topological strings.

\subsection{Topological wave functions in different polarizations \label{sec_top}}

In this section we review in some detail various important properties of topological string
wave functions that will play an important role in Section \ref{sec_ns5twi}.
We begin the analysis from the point of view of the topological B-model,
while towards the end discussing the map to the A-model.
A novel observation of this section is that
the chiral partition function $\cZ_{\Theta,\mu}^{(k)}$ in (\ref{thpl2}), restricted to $k=1$,
can be represented as the state \eqref{PsiRNS5G}
in the so called ``Weil polarization'' as shown in \eqref{NS5overlap}.

\subsubsection{Griffiths polarization}

Recall that in the topological B-model on $\cX$, the partition function is defined as the generating function
of the genus $g$ correlation functions $C^{(g)}_{a_1 \cdots a_n}$ of $n$ chiral fields
 \cite{Bershadsky:1994cx}:
\be
\label{BCOVW}
 \Psi_{\rm BCOV}(z,\bar z; \lambda,x) = \lambda^{\frac{\chi(\cX)}{24}-1}\,
 \exp\left( \sum_{g=0}^{\infty}
\sum_{n=0}^{\infty} \frac{1}{n!} \lambda^{2g-2}
\ C^{(g)}_{a_1 \cdots a_n}(z,\bar z)\,  x^{a_1}
\cdots x^{a_n} \right) .
\ee
The correlation functions $C^{(g)}_{a_1 \cdots a_n}$, which are taken to vanish
for $2g-2+n\leq 0$, are global sections of the
vector bundle $(T^*)^n\otimes \cL^{2-2g}$ over the complex structure moduli
space $\cM_c(\cX)$. Here, $T^*$ is the holomorphic cotangent bundle of $\cM_c(\cX)$
and $\cL$ is the line bundle over $\cM_c(\cX)$ in which  the holomorphic three-form $\Omega_{3,0}$
is valued. In particular, under rescalings
$\Omega_{3,0}\mapsto e^f\Omega_{3,0}$, the correlation functions transform
as $C^{(g)}_{a_1 \cdots a_n}\mapsto e^{(2-2g)f}C^{(g)}_{a_1 \cdots a_n}$. As a result,
$ \Psi_{\rm BCOV}$ is a section of the line bundle\footnote{As mentioned at the end of Section
\ref{subsec_topology}, the definition of fractional power $\cL^{\chi/24}$ requires a
homomorphism $M\mapsto \expe{\kappa(M)}$ from the monodromy group
to the group of 24:th roots of unity. In the present case, this homomorphism is
just the multiplier system of the one-loop amplitude $e^{-f_1}/\sqrt{J_{\rm G}}$,
where $f_1$ and $J_{\rm G}$ are defined in \eqref{F1} and below \eqref{RtoG}.}
$\cL^{\frac{\chi(\cX)}{24}-1}$, provided $\lambda$ transforms as a section of $\cL$.

Moreover, $ \Psi_{\rm BCOV}$
satisfies the holomorphic anomaly conditions obtained in \cite{Bershadsky:1994cx},
Eq. 3.17-18.
As explained in \cite{Verlinde:2004ck},
it is illuminating to rescale $x^a$ and $\Psi_{\rm BCOV}$ as\footnote{The index G refers to
the Griffiths complex structure, as will become clear momentarily.}
\be
\label{psiovergg}
\Psi_{\rm G}(z,\bar z;\lambda^{-1}, x) =e^{f_1(z)}
\Psi_{\rm BCOV}(z,\bar z;\lambda,\lambda x) \, ,
\ee
where $f_1$ is (locally) a holomorphic function
determined from the factorization of the one-loop vacuum amplitude $F_1$ by
\be
F_1=\log \Big[ e^{f_1(z)+\bar{f}_1(\bar{z})} / \sqrt{M(z,\bar{z})}\Big],
\label{F1}
\ee
where
\be
\label{F1M}
M(z,\bar{z})=|g| \, e^{-2\left( \frac{b_3}{4}
- \frac{\chi(\cX)}{24} + 1 \right) \cK}, \qquad \quad |g|=\det(g_{a\bar b}).
\ee
More accurately,
$e^{f_1(z)}$ is  a section of
$\cL^{1-\frac{\chi(\cX)}{24}+ \frac{b_3}{4}} \otimes K_{c}^{1/2}$, where $K_{c}$ is the
canonical bundle of $\cM_c$, locally trivialized by the section
$dz^1 \wedge \cdots \wedge dz^{h_{2,1}}$.
Thus, $f_1$ transforms under holomorphic change of coordinates
$z\mapsto z'(z)$ and rescaling $(\lambda^{-1},x^a)\mapsto e^{-f}
(\lambda^{-1},x^a)$ according to
\be
 f_1\mapsto f_1 + \left( \frac{b_3}{4}
- \frac{\chi(\cX)}{24} + 1 \right) f + \frac12 \log \det(\pa z/\pa z')\, .
\label{f1transf}
\ee
This implies that the rescaled amplitude $\Psi_{\rm G}$ and its covariant
derivative $\nabla_a \Psi_{\rm G}$ transform as
$\cL^{b_3 /4}$-valued holomorphic half-densities on $\cM_c(\cX)$, namely
\be
\label{trPsiG}
\Psi_{\rm G}\mapsto \sqrt{ \det(\pa z/\pa z')} \, e^{b_3 f/4} \, \Psi_{\rm G}\, .
\ee
In terms of $\Psi_{\rm G}$, the
holomorphic anomaly equations of \cite{Bershadsky:1994cx} take the
more appealing form \cite{Verlinde:2004ck,Gunaydin:2006bz}
\be
\begin{split}
\left[ \dwrt{\bar z^a} -
\half e^{2\cK} \bar C_{\bar a\bar b\bar c} g^{b\bar b} g^{c\bar c}
\frac{\pa^2}{\pa x^b\pa x^c}
- g_{\bar a b} x^j \frac{\pa}{\pa \lambda^{-1}} \right] \Psi_{\rm G}  = 0,
\label{Ver1b}
\\
\left[ \nabla_{a} - \Gamma_{ab}^c x^b \frac{\pa}{\pa x^c} -
\half \partial_{a} \log |g| -
\lambda^{-1}\frac{\pa}{\pa x^i} + \frac12 C_{abc} x^b x^c \right] \Psi_{\rm G} = 0,
\end{split}
\ee
where $\nabla_a$ is the covariant derivative
\be
\nabla_a = \dwrt{z^a} + \pa_a \cK
\left( x^b \frac{\pa}{\pa x^b} + \lambda^{-1} \frac{\pa}{\pa{\lambda^{-1}}}
+ \frac{b_3}{4} \right).
\ee
The holomorphic anomaly equations \eqref{Ver1b} then guarantee that
the inner product
\be
\label{vinnerg}
\begin{split}
\langle \Psi | \Psi \rangle=&
\int \de x^a ~\de \bar x^{\bar a}~  \de\lambda^{-1} ~\de \bar \lambda^{-1}~
\sqrt{|g|} ~  e^{-\frac{b_3}{4}\cK} \\
&\exp\left( -e^{-\cK} x^a g_{a\bar b} \bar x^{\bar b} + e^{-\cK}
\lambda^{-1}\bar\lambda^{-1} \right)
\Psi_{\rm G}^{*}(\bz,z;\bar\lambda^{-1},\bar x) ~ \Psi_{\rm G}(z,\bz;\lambda^{-1},x)
\end{split}
\ee
is well defined, and (at least formally) independent of the complex structure moduli.

As explained in \cite{Witten:1993ed,Verlinde:2004ck,Gunaydin:2006bz,Aganagic:2006wq},
the above equations may be interpreted as the fact that $\Psi_{\rm G}$ is the wave-function
for a particular state $|\Psi\rangle \equiv |\Psi^{\rm top}\rangle $
in the Hilbert space $\cH$ quantizing the symplectic space
$H^3(\cX,\IR)$ in a complex polarization determined by the Griffiths complex structure.
Indeed, for a given choice of the holomorphic $\Omega_{3,0}$ on $\cX$,
any 3-form $C\in H^3(\cX,\IR)$
admits the Hodge decomposition
\be
\label{CH3}
\sqrt{2\pi}\, C =  {\lambda}^{-1} \Omega_{3,0} + x^a   D_a \Omega_{3,0}+
\bar x^{\bar a}  D_{\bar a} \bar \Omega_{3,0}
+  {\bar\lambda}^{-1} \bar\Omega_{3,0}\, ,
\ee
where ${\lambda}^{-1}, x^a$ are complex coordinates for the Griffiths complex
structure on $H^3(\cX,\IC)$ and the normalization factor $\sqrt{2\pi}$ is inserted
for later convenience.
The solutions of \eqref{Ver1b}  may be written as the overlap
\be
\label{psioverg}
\Psi_{\rm G}(z,\bar z;\lambda^{-1}, x) =   \, _{(z, \bar{z})} \langle
\lambda^{-1} , x| \Psi \rangle,
\ee
where $\ _{(z, \bar{z})} \langle \lambda^{-1}, x |$ is a basis of coherent states diagonalizing
the action of the operators quantizing the components $ {\lambda}^{-1}$ and
$x^a$ in \eqref{CH3}. The holomorphic equations \eqref{Ver1b}
then reflect the unitary transformation undergone by the coherent
states under changes of complex structure.

\subsubsection{Weil polarization}

Alternatively, one may choose to diagonalize the operators quantizing $\bar\lambda^{-1}$
and $x^a$, which are complex coordinates
for the Weil complex structure on $H^3(\cX,\IC)$.
The new topological wave function in the ``Weil complex polarization"
\be
\label{psioverw}
\Psi_{\rm W}(z,\bar z;\bar\lambda^{-1},x) =\,  _{(z, \bar{z})} \langle
 \bar\lambda^{-1}, x | \Psi \rangle
\ee
is obtained from the topological wave function in the ``Griffiths complex polarization"
\eqref{psioverg} by Fourier transforming over $\lambda^{-1}$,
\be
\label{psigtov}
\Psi_{\rm W}(z,\bar z;\bar\lambda^{-1}, x) =
\int \de\lambda^{-1}\, \exp\left( \I e^{-\cK} (\lambda \bar\lambda)^{-1} \right)\,
\Psi_{G}(z,\bar z;\lambda^{-1}, x) \, ,
\ee
such that the inner product is now given by a positive definite Gaussian kernel,
\be
\label{vinnerw}
\begin{split}
\langle \Psi | \Psi \rangle&=\int \de x^a ~\de \bar x^{\bar a}~  \de\lambda^{-1} ~\de \bar \lambda^{-1}~
\sqrt{|g|} ~  e^{-\frac{b_3}{4}\cK} \\
&\exp\left( -e^{-\cK} x^a g_{a\bar b} \bar x^{\bar b} - e^{-\cK}
\lambda^{-1}\bar\lambda^{-1} \right)
\Psi_W^{*}(\bz,z;\lambda^{-1},\bar x) ~ \Psi_W(z,\bar z;\bar\lambda^{-1},x).
\end{split}
\ee
It is in principle straightforward to work out the analog of \eqref{Ver1b}
for this new polarization. Since  $\Psi_{\rm G}$ transformed as \eqref{trPsiG} under holomorphic change
of coordinates and rescalings, $\Psi_{\rm W}$ must now transform
as a  $\cL^{b_3/4-1}$-valued holomorphic half-density,
\be
\label{trPsiW}
\Psi_{\rm W}\mapsto \sqrt{ \det(\pa z/\pa z')} \, e^{(\frac{b_3}{4}-1) f} \, \Psi_{\rm W}\, .
\ee
Importantly, the relations \eqref{trPsiG}, \eqref{trPsiW}
are properties of the coherent state bases $ _{(z, \bar{z})} \langle
\lambda^{-1}, x |$ and  $_{(z, \bar{z})} \langle
\bar\lambda^{-1}, x |$, and hold for an arbitrary state $|\Psi\rangle\in \cH$, not only
$|\Psi^{\rm top}\rangle\in \cH$. Moreover, both $\Psi_{\rm G}$  and
$\Psi_{\rm W}$  depend on the complex structure
of $\cX$, but are independent of any choice of symplectic
basis on $H^3(\cX,\IR)$.

\subsubsection{Real polarization and intertwiners}

However, at the expense of fixing a symplectic basis $\alpha_\Lambda,\beta^\Lambda$ of
$H^3(\cX,\IR)$, it becomes possible to express the state $|\Psi\rangle $ in a ``background
independent" way, i.e. independent of the complex structure on $\cX$. Indeed,
one may expand $C\in H^3(\cX,\IR)$ as in \eqref{defzezet},
\be
C= \zeta^\Lambda \alpha_\Lambda - \tzeta_\Lambda \beta^\Lambda\, ,
\ee
where $\zeta^\Lambda$ and $\tzeta_\Lambda$ are real Darboux coordinates on
$H^3(\cX,\IR)$, and express $|\Psi\rangle $ on a basis of wave functions
diagonalizing the operators associated to $\zeta^\Lambda$, say. The resulting wave
function
\begin{equation}
\Psi_{\IR}(\zeta^\Lambda) = \langle \zeta^\Lambda | \Psi \rangle\, ,
\end{equation}
in the so called ``real polarization",
is locally independent of the complex structure of $\cX$, but transforms according to
the metaplectic representation $\rho_m$ under changes of symplectic basis,
in particular under monodromies in $\cM_c(\cX)$. On square-integrable functions
$\Psi_{\IR}(\zeta^{\Lambda})$, the metaplectic representation acts according to
\be
\label{metarep}
\begin{split}
\left[\rho_m\left( {\scriptsize \begin{pmatrix}
\cA^{-T} & 0 \\ 0 & \cA \end{pmatrix}} \right)\cdot \Psi_{\IR}\right] (\zeta^{\Lambda})
&= \Psi_{\IR} (\cA^T \zeta^{\Lambda})\ ,\\
\left[\rho_m\left( {\scriptsize \begin{pmatrix}
1 & 0 \\ \cB & 1 \end{pmatrix}} \right)\cdot \Psi_{\IR}\right] (\zeta^\Lambda)
&= \expe{\frac12\,\cB_{\Lambda\Sigma} \,\zeta^\Lambda \zeta^\Sigma}\, \Psi_{\IR} (\zeta^\Lambda)\, ,
\\
\left[\rho_m\left( {\scriptsize \begin{pmatrix}
0 & -1 \\ 1 & 0 \end{pmatrix}} \right)\cdot \Psi_{\IR}\right] (\zeta^\Lambda)
&=\int \expe{- k \, \zeta^\Lambda \tzeta_\Lambda}\, \Psi_{\IR}
( \tzeta_\Lambda)\, \de\tzeta_\Lambda\, .
\end{split}
\ee
We further note that the metaplectic representation
is unitary with respect to  the inner product
\be
\langle \Psi | \Psi \rangle = \int\, \Psi_\IR^*(\zeta^\Lambda)\,  \Psi_\IR(\zeta^\Lambda)\,
\de \zeta^\Lambda\, .
\ee

The intertwiner between the real polarization and the
Griffiths complex polarization  was discussed
in \cite{Verlinde:2004ck,Gunaydin:2006bz,Aganagic:2006wq,Schwarz:2006br}.
Since the Griffiths complex coordinates $\lambda^{-1}, x^a$
and the real coordinates $\zeta^\Lambda,\tzeta_\Lambda$ are related classically by
\be
\label{xtolx}
x_\Lambda\equiv \tzeta_\Lambda-\bar\tau_{\Lambda\Sigma}\zeta^\Sigma=
2\I\, \Im\tau_{\Lambda\Sigma} \left(
\lambda^{-1} X^\Sigma + x^{a} \, D_{a} X^\Sigma\right)/\sqrt{2\pi}\, ,
\ee
the intertwiner from the real polarization to the
Griffiths  complex  polarization is given by the Gaussian kernel
\be
\label{RtoG}
\Psi_{\rm G}(z,\bar z; \lambda^{-1}, x)=\frac{\sqrt{\det\Im\tau}}{\sqrt{J_{\rm G}}}\,
\int
e^{ \frac{\pi}{2}\, x_\Lambda \Im\tau^{\Lambda\Sigma} x_\Sigma - 2\pi \I x_\Lambda \zeta^\Lambda
-\I\pi\, \zeta^\Lambda \bar\tau_{\Lambda\Sigma} \zeta^\Sigma}
\, \Psi_{\IR}(\zeta^\Lambda)\, \de \zeta^{\Lambda}\, ,
\ee
where $J_{\rm G}=\pa (\lambda^{-1}, x^a)/\pa x^\Lambda$ is the Jacobian
from the ``large phase space" variables $x^\Lambda=\Im\tau^{\Lambda\Sigma} x_\Sigma$
to the ``small phase space" variables $\lambda^{-1}, x^a$. A simple computation shows
that
\be
|J_{\rm G}|^2 =  e^{\frac{b_3}{2} \cK} \det(\Im\tau)\, /\, |g|,
\ee
therefore $J_{\rm G}$ transforms as a section of $\cL^{-b_3/2}\otimes K_c^{-1}\otimes
\det_{\rm G}^{-1}$, where $\det_{\rm G}$ denotes the line bundle whose sections transform as
$w\mapsto  \det(\cC\tau+\cD) \, w$ under monodromies. On the other hand,
as $\Psi_\IR(\zeta^\Lambda)$ transforms according to the metaplectic representation under monodromies,
the integral on the r.h.s.
of \eqref{RtoG} is valued in $\overline{\det_{\rm G}}^{1/2}$. As a result,
\eqref{RtoG} indeed transforms as $\cL^{b_3/4}\otimes K_c^{1/2}$ under monodromies.
In the special gauge $X^0=1$, $X^a=z^a$,
$J_{\rm G}$ evaluates to one, thereby identifying the bundles $\cL^{-b_3/2}\otimes K_c^{-1}$
and $\det_{\rm G}$.  Since the quadratic form $\Im\tau$ has indefinite signature, the
intertwiner \eqref{RtoG} is however only valid formally.

In contrast, the Weil complex coordinates $(\bar\lambda^{-1}, x^a)$
are related to the real coordinates $(\zeta^\Lambda,\tzeta_\Lambda)$ by
\be
\omega_\Lambda \equiv \tzeta_\Lambda-\bar\cN_{\Lambda\Sigma}\zeta^\Sigma=
2\I\, \Im\cN_{\Lambda\Sigma} \left(
\lambda^{-1} X^\Sigma + \bar x^{\bar a} \, D_{\bar a} \bar X^\Sigma\right)/\sqrt{2\pi}\, .
\ee
The intertwiner between the real polarization and the
Weil  complex polarization is given by the Gaussian kernel \cite{Gunaydin:2006bz}
\be
\label{RtoW}
\Psi_{\rm W}(z,\bar z; \bar \lambda^{-1},x)=\frac{e^{\cK/2}\sqrt{\det(-\Im\cN)}}{\sqrt{J_{\rm W}}}\,
\int
e^{-\frac{\pi}{2}\, \bar \omega_\Lambda \Im {\cN}^{\Lambda\Sigma} \bar \omega_\Sigma
- 2\pi \I \bar \omega_\Lambda \zeta^\Lambda
-\I\pi\, \zeta^\Lambda \cN_{\Lambda\Sigma} \zeta^\Sigma}
\, \Psi_{\IR}(\zeta^\Lambda)\, \de \zeta^{\Lambda}\, ,
\ee
where  $J_{\rm W}=\pa (\bar\lambda^{-1}, x^a)/\pa \bar w^\Lambda$ is the Jacobian
from  $\bar w^\Lambda=\Im\cN^{\Lambda\Sigma} \bar w_\Sigma$
to  $\bar \lambda^{-1}, x^a$, and the quadratic term in the exponential is now negative definite.
Using the relations
\be
|J_{\rm W}|^2 =  e^{\frac{b_3}{2} \cK}  \, \det(-\Im\cN) / |g|\, ,\qquad
\frac{J_{\rm W}}{J_{\rm G}}=\frac{e^{-\cK}}{
(\bar X^\Lambda\Im\tau_{\Lambda\Sigma}\bar X^\Sigma)}\ ,
\ee
it is apparent
that $J_{\rm W}$ is a section of $\cL^{1-\frac{b_3}{2}} \otimes \bar\cL^{-1} \otimes K_c^{-1}
\otimes \overline{\det_{\rm W}}^{-1}$, where $ \det_{\rm W}$ denotes the line bundle whose sections
transform as $w\mapsto \det(\cC\cN+\cD)\, w$ under monodromies. Since the integral on the r.h.s.
of \eqref{RtoW} is valued in $\det_{\rm W}^{1/2}$,
\eqref{RtoW} indeed transforms as a section of
$\cL^{\frac{b_3}{4}-1}\otimes K_c^{1/2}$. For convenience, we summarize the
transformation properties of the various topological wave functions in Table \ref{tabtrans}
on page \pageref{tabtrans}.

We are now in a position to connect the present discussion to the Gaussian chiral
partition function. Let us choose $\Psi_{\mathbb{R}}(\zeta^{\Lambda})$
in \eqref{RtoW} to be the real-polarized wave function associated to the particular state
$| \Psi^{\Gamma_m,k,\mu} \rangle\in \mathcal{H}$ in \eqref{PsiRNS5G}.
Then (the complex conjugate of) the chiral partition function
\eqref{thpl} for $k=1$ can be written up to a Gaussian prefactor as the wave function of this state
in the Weil polarization
\be
\overline{Ê\cZ_{\Theta,\mu}^{(1)} (\cN, C)}=\bar{\cF}\,
e^{-\cK/2} \sqrt{\frac{J_{\rm W}}
{\det(-\Im \cN)}} \,
e^{\frac{\pi}{2}\omega_{\Lambda} \Im\cN^{\Lambda\Sigma}\bar \omega_{\Sigma} }\,
\Psi_{\rm W}^{\Gamma_m,1,\mu}\big(  \bar\omega_\Lambda\big)\, ,
\label{NS5overlap}
\ee
which is independent of $\mu$ when $k=1$.

\subsubsection{Holomorphic topological partition function}

Let us now discuss the relation between the topological wave function $\Psi_{\rm BCOV}$
and the so-called holomorphic topological partition function $F_{\rm hol}$. For this purpose,
we first take the holomorphic limit $\bar z^{\bar a}\to \infty$, while
keeping $z^a$ fixed. We further assume that in this limit, the Griffiths period matrix
$\bar\tau_{\Lambda\Sigma}\to \infty$. In this case, the
intertwiner \eqref{RtoG} reduces to a delta function with support on
$\zeta^\Lambda = -\bar\tau^{\Lambda\Sigma} x_\Sigma$, so that
\be
\Psi_{\rm G}(x_\Lambda)\sim J_{\rm G}^{-1/2} \Psi_\IR(\zeta^\Lambda)\, .
\ee
Let us consider the restriction of the left-hand side to the locus $x^a=0$, $\lambda$ fixed,
where $x^a, \lambda^{-1}$ are related to $x_\Lambda$ by \eqref{xtolx}.  On this locus,
$\zeta^\Lambda$ becomes complexified and equal to $X^\Lambda/(\lambda\sqrt{2\pi})$,
which we
denote by $\xi^\Lambda$.  The BCOV amplitude therefore
becomes proportional to the topological string wave function in the real
polarization $\Psi_{\mathbb{R}}^{\text{top}}$, analytically continued to the complex domain:
\be
\label{BCOVtoR}
\Psi_{\rm BCOV}\big(z,\bar z \to \infty;\lambda,x=0\big) \sim
e^{-f_1(z)}\, \Psi^{\rm top}_\IR(\xi^\Lambda)\ .
\ee
Thus, defining the holomorphic topological partition function as
\be
\label{defFhol}
\exp\left[F_{\rm hol}(z,\lambda)\right]=e^{f_1(z)}\,
\lim\limits_{\bar z\to \infty}
\[\lambda^{1-\frac{\chi(\cX)}{24}} \Psi_{\rm BCOV}(z,\bar z,\lambda)\]\, ,
\ee
(where the prefactor restores the one-loop vacuum amplitude which was absent from
\eqref{BCOVW}), we have, in the \kahler gauge $X^0=1$, the relation \cite{Schwarz:2006br}
\be
\label{psiholR}
\exp\left[F_{\rm hol}(z,\lambda)\right]= \left(\xi^0\right)^{\frac{\chi(\cX)}{24}-1}
\Psi_{\IR}^{\rm top}(\xi^\Lambda)\ ,\quad
\lambda = \frac{1}{\xi^0\sqrt{2\pi}}\ ,\quad z^a = \frac{\xi^a}{\xi^0}\, .
\ee
This relation will play an important role in Section \ref{sec_ns5top}, where the coordinates
$\xi^\Lambda$ will be interpreted as Darboux coordinates on twistor space $\cZ_\cM$.

\subsubsection{Topological A-model and Donaldson-Thomas invariants}

So far we have been discussing topological wave functions in the context of the B-model.
However, the properties of the topological wave functions discussed in this section hold
both for the topological B-model on a CY threefold $\cX$ and for the topological
A-model on a CY threefold $\hat\cX$, provided one replaces the Euler
number $\chi(\cX)$ by $-\chi(\hat\cX)$. Indeed, mirror symmetry $\cX\mapsto
\hat\cX$ exchanges the
two  topological wave functions \cite{Bershadsky:1994cx, Antoniadis:1994ze}. We shall now
proceed to discuss some key features of the A-model,
which will play an important role in Section \ref{sec_ns5twi}.

In contrast to the B-model, the A-model topological string encodes deformations of
the (complexified) K\"ahler structure of the mirror Calabi-Yau threefold $\hat \cX$.
The holomorphic wave function $F_{\rm hol}(z^a,\lambda)$ of the topological A-model
on $\hat\cX$ therefore depends on the K\"ahler moduli $z^{a}\in \cM_K(\hat\cX)$
together with the topological string coupling $\lambda$.
The A-model wave function $F_{\rm hol}(z^a,\lambda)$ is moreover related to
the partition function of Gromov-Witten (GW) invariants via
\be
\exp\left[F_{\rm hol}(z^a,\lambda)-F_{\rm pol}(z^a,\lambda)\right] =
Z_{\text{GW}}\, ,
\label{fgw}
\ee
where
\be
\label{Fpol}
F_{\rm pol}(z^a,\lambda)=-\frac{(2\pi \I)^3}{\lambda^2}
\left( \frac16 \kappa_{abc}z^{a}z^{b}z^{c}
-\frac12 A_{\Lambda\Sigma} {z}^{\Lambda} {z}^{\Sigma}  \right)
-\frac{2\pi\I}{24}\, c_{2,a} z^a
\ee
is the ``polar part'' of $F_{\rm hol}(z^a,\lambda)$, and $\kappa_{abc}$
is the intersection product on $H^3(\hat \cX,\IZ)$.
Here  we included the ``quadratic ambiguity''  $A_{\Lambda\Sigma}{z}^{\Lambda} {z}^{\Sigma} $
in the prepotential,
where $A_{\Lambda\Sigma}$ is a constant real symmetric matrix and
$z^{\Lambda}=(1,z^{a})$. This quadratic term does not
affect the metric on the K\"ahler moduli space but plays a crucial role
for charge quantization, as we demonstrate in Section \ref{sec_IIB}.
The right hand side of \eqref{fgw} involves the partition function
$Z_{\text{GW}}=Z_{\text{GW}}^{0}\, Z'_{\text{GW}}$, where $Z'_{\text{GW}}$
encodes the non-degenerate GW-invariants, while
\be
Z_{\text{GW}}^{0}=\lambda^{-\frac{\chi(\hat\cX)}{24} \epsilon_{\text{GW}}}\,
[M(e^{-\lambda})]^{\chi(\hat\cX)/2}
\label{degGW}
\ee
contains the contribution from degenerate GW-invariants.
This follows from the weak coupling expansion of the Mac-Mahon
function $M(q)=\prod (1-q^{n})^{-n}$ (see e.g. App. E in \cite{Dabholkar:2005dt}),
\be
\label{mmahonw}
M(e^{-\lambda}) \sim K\, \lambda^{\frac{1}{12}} \,
\exp\left( \frac{ \zeta(3) }{\lambda^2}+  \sum_{n=0}^\infty (-1)^n \lambda^{2n+2}
\frac{\vert B_{2n+4} \vert}{(2n+4)!} \frac{(2n+3)}{(2n+2)} B_{2n+2}
\right),
\ee
where $B_n$ are the Bernoulli numbers and $K$ is a numerical constant.
Upon choosing the parameter $\epsilon_{\text{GW}}$ in \eqref{degGW} equal to one, the
power of $\lambda$ in \eqref{degGW} cancels the similar power in \eqref{mmahonw}, as
usually assumed in the topological string theory literature. In \cite{Dabholkar:2005dt, Denef:2007vg},
it was however noticed that the choice $\epsilon_{\text{GW}}=0$ was necessary in
order to match the OSV conjecture. We do not take sides at this point
and leave the parameter $\epsilon_{\text{GW}}$ arbitrary.

We further note that $Z'_{\text{GW}}$ is related to the partition function of
the (ordinary, rank one) Donaldson-Thomas
invariants $N_{DT}(Q_a,2J)$ via the ``GW/DT relation" \cite{gw-dt,gw-dt2}
\be
\label{zpgw}
Z'_{\text{GW}} = [M(e^{-\lambda})]^{-\chi(\hat\cX)\epsilon_{\text{DT}}  }\,
Z_{\text{DT}} \, ,
\ee
where the DT partition function is defined as
\be
\label{ZDTdef}
Z_{\text{DT}} \equiv
\sum_{Q_a,J}\, (-1)^{2J}\,N_{DT}(Q_a,2J)\,
e^{-2\lambda J +2 \pi \I Q_a z^a}.
\ee
In the relation \eqref{zpgw} we have also allowed for an arbitrary parameter $\epsilon_{\text{DT}}  $,
which is usually taken to be equal to one in the literature. We shall nevertheless leave
it unspecified for now. Physically, the DT invariants count bound states of
one D6-brane with $2J$ D0-branes and $Q_a$ D2-branes wrapped along $\gamma^a\in
H_2({\hat \cX},\mathbb{Z})$. They are also expected to provide the instanton measure for
D5-D1-D(-1) Euclidean configurations.

We may now combine the above relations and express the holomorphic wave function
$e^{F_{\rm hol}(z, \lambda)}$ in terms of the DT partition function:
\be
\label{gvc2}
\begin{split}
e^{F_{\rm hol}(z,\lambda)} =
\lambda^{-\frac{\chi(\hat\cX)}{24}\epsilon_{\text{GW}}}\,[M(e^{-\lambda})]^{(\frac12-\epsilon_{\text{DT}})
\chi(\hat\cX)}\,
e^{F_{\rm pol}}  \sum_{Q_a,J} (-1)^{2J}N_{DT}(Q_a,2J)\,
e^{-2\lambda J +2 \pi \I Q_a z^a}.
\end{split}
\ee
This relation will play an important role in Section \ref{sec_ns5top}.

\section{Topology of the type IIA hypermultiplet moduli space \label{sec_pert}}

We shall now discuss quantum corrections to the hypermultiplet (HM) moduli space
$\cM=\cQ_c(\cX)$ in type IIA string theory compactified on a CY 3-fold $\cX$. By the usual
T-duality and mirror symmetry arguments reviewed e.g. in \cite{Alexandrov:2008gh,
Pioline:2009qt}, the same space $\cM$ also appears as  the HM moduli space
in M-theory on $\cX$,  as  the HM moduli space in type IIB on the mirror 3-fold $\hat \cX$,
as the VM moduli space in type IIA on $\hat \cX\times S^1$
or as the VM moduli space in type IIB theories on $\cX\times S^1$ (see Table~\ref{dico}).
Here using insights from the previous section, we clarify the qualitative structure of
NS5-instanton corrections to the hypermultiplet metric, a result which allows us
to specify the topology of the perturbative moduli space in the weak coupling limit.
Some of the results in this section were already announced in \cite{Alexandrov:2010np}.

\begin{table}
$$
\begin{array}{|c|c|c|c|c|}
\hline
&
\multicolumn{2}{|c|}{\text{IIA}/\cX \times S^1_R} &
\multicolumn{2}{c|}{\text{IIB}/\hat\cX \times S^1_{R'}}
\\
\hline
& \text{HM} & \text{VM} & \text{VM} & \text{HM}
\\
\hline
\cM & \cQ_c(\cX) & \cQ_K(\cX) & \cQ_c(\hat\cX) & \cQ_K(\hat\cX)
\\
r & 1/g_{(4)}^2 & (R/l_{(4)})^2 & (R/l_{(4)})^2 & 1/g_{(4)}^2
\\
\cO(e^{-\sqrt{r}}) & \text{D2} & \text{D0-D2-D4-D6} & \text{D3} & \text{D(-1)-D1-D3-D5}
\\
\cO(e^{-r})  & \text{NS5} & \text{KKM} & \text{KKM} & \text{NS5}
\\
\hline
\end{array}
$$
\caption{Dictionary between various realizations of $\cQ_c(\cX)=\cQ_K(\hat \cX)$.
\label{dico}}
\end{table}

\subsection{Perturbative HM moduli space}
\label{subsec_pertHM}

The HM moduli space  in type IIA string theory compactified on
a CY 3-fold $\cX$ is a \qk  manifold
$\cM=\cQ_c(\cX)$ of real dimension $2b_{3}(\mathcal{X})=4(h_{2,1}+1)$.
Here the subscript $c$ refers to the fact that $\cQ_c(\cX)$ encodes the
moduli space of complex structures  $\cM_c(\cX)$, as well as
the four-dimensional dilaton $r\equiv e^\phi\sim 1/g_{(4)}^2$, RR-field $C$
and NS axion $\sigma$. In the weak coupling limit $r\to \infty$,
the \qk metric on $\cM$ is given, to all orders in $1/r$, by
\cite{Gunther:1998sc, Robles-Llana:2006ez,Alexandrov:2007ec}
\be
ds_{\cM}^2=\frac{r+2c}{r^2(r+c)}\,\de r^2
+\frac{4(r+c)}{r}\, \de s^2_{\cS\cK}
+\frac{\de s^2_\cT }{r}
+ \frac{2\, c}{r^2}\, e^{\cK}\, | z^\Lambda \de \tzeta_\Lambda - F_{\Lambda} \de \zeta^\Lambda|^2
+ \frac{r+c}{16 r^2(r+2c)} D\sigma^2\,  .
\label{hypmetone}
\ee
Here, $( \zeta^\Lambda, \tzeta_\Lambda)$ are the real periods \eqref{defzezet} of the $C$-field
on a symplectic basis of $H^3(\cX,\IR)$, $\de s^2_\cT$ is the metric \eqref{dsT}
on the (torus fiber of the) intermediate Jacobian,
$\de s^2_{\cS\cK}= \cK_{a{\bar b}}\,\de z^a \de \bar z^{\bar b}$
is the special \kahler metric on $\cS\cK\equiv \cM_c(\cX)$,
with \kahler connection \eqref{Kahlerconnection},
$D\sigma$ is the one-form
\be
\label{Dsigone}
D\sigma = \de \sigma + \tzeta_\Lambda \de \zeta^\Lambda -  \zeta^\Lambda \de \tzeta_\Lambda
+ 8 c \, \cA_K\, ,
\ee
and $c$ is a deformation
parameter which encodes the one-loop correction,
\be
c=-\chi(\cX)/(192 \pi) \, .
\label{cchi}
\ee
In the special case where $\chi(\cX)=0$ (or to leading, tree-level approximation in the
weak coupling limit $r\to\infty$), the metric \eqref{hypmetone} follows from the
special \kahler metric on $\cS\cK$ by the $c$-map construction \cite{Cecotti:1988qn,
Ferrara:1989ik}. All higher loop corrections are presumed to vanish,
for reasons that will be recalled shortly. On the other hand, non-perturbative corrections
from D2-brane and NS5-brane instantons will
correct \eqref{hypmetone} at order $e^{-\sqrt{r}}$ and $e^{-r}$,
respectively \cite{Becker:1995kb}. In particular, such corrections
should resolve the curvature singularity of the perturbative
metric \eqref{hypmetone} which is present at $r=-2c$ when $\chi(\cX)>0$.

\subsubsection{Ten-dimensional origin of the one-loop correction}

It is interesting to note that the connection
term in \eqref{Dsigone} amounts, after dualizing the NS-axion $\sigma$ and
the RR-axions $\tzeta_\Lambda$ into two-form potentials $B$ and $B^\Lambda$,
to a topological coupling
\be
\label{S4}
\int\Bigl(
 \Re \cN_{\Lambda\Sigma}    (\de B^\Lambda+\zeta^\Lambda \de B) \wedge d\zeta^\Sigma
 - 8 c \, B \wedge \omega_\sk\Bigr)
\ee
in the 4D effective action.\footnote{To dualize the NS-axion, we
add a term $- \de B \wedge \frac{\de\sigma}{2}$ to the Lagrangian $ \de s_{\cM}^2$
and integrate out the one-form
$\de\sigma$, leading to $-\frac16 r^2 \de B \wedge *\de B + B \wedge (\omega_\cT
+\frac{\chi({\cX})}{24}\omega_\sk)$, where $\omega_\sk=-\frac{1}{2\pi}\de\cA_K$.
\label{foodualsig}}
This in turn follows by dimensional reduction of the
topological coupling in 10-dimensional type IIA supergravity
\be
\label{S10}
 \int_\cY  \left( \frac16\, B\wedge \de C\wedge \de C - B\wedge I_8(R) \right),
\ee
where $I_8(R)=(p_2-\frac14 p_1^2)/48$ on $\cX$ (or from the similar coupling in 11D supergravity
on $\cX\times S^1$). Indeed, using the standard relations between Pontryagin classes
$p_1=c_1^2-2 c_2$ and $p_2=c_2^2-2c_1c_3+2c_4$ (see e.g. \cite{Eguchi:1980jx}, section 6.4),
the second term in \eqref{S10} can be rewritten, on an arbitrary complex manifold $\cY$,  as
\be
\label{BI8}
B\wedge I_8 = \frac1{24} B\wedge \left[ c_4 - c_1 \( c_3+ \frac18\, c_1^3 -\frac12\, c_1 c_2\) \right].
\ee
Taking\footnote{Taking instead $\cY=\IR^2\times \cG$, where $\cG$ is a CY four-fold,
Eq. \eqref{BI8} reproduces the $\frac1{24}\chi_{\cG} B$ one-point function found in
\cite{Sethi:1996es}. When $\cG$ is a CY threefold fibered over $\CP$, \eqref{BI8}
should instead reduce to Eq (19) in \cite{Brunner:1996pk}.}
$\cY=\IR^4\times \cX$, where the complex structure of $\cX$ varies as a function
of the position in $\IR^4$, and integrating the term
in parenthesis on the CY threefold $\cX$
produces a coupling $- \frac{\chi(\cX)}{24}\, B\wedge c_1$ on $\IR^4$,
which by the arguments in \cite{Bershadsky:1994cx} is equal to $\frac{\chi(\cX)}{24}\, B\wedge \omega_\sk$.
This derivation of the one-loop correction to the metric can be viewed as
a supersymmetric counterpart of the derivation in \cite{Antoniadis:1997eg,
Antoniadis:2003sw}\footnote{Note that the analysis
of \cite{Antoniadis:2003sw} was restricted to the
``universal sector", where the  $B\wedge \omega_\sk$ coupling vanishes.
In particular, for rigid CY threefolds, the last term in \eqref{Dsigone} is absent. We
are grateful to R. Minasian and P. Vanhove for discussions on these issues.},
where corrections to the scale factors in the metric \eqref{hypmetone}
were obtained by reduction of the CP-even $R^4$-type couplings
in 10 dimensional type IIA supergravity.

\subsection{Topology and instanton corrections}
\label{subsec_topolinst}

While \eqref{hypmetone} describes the local, Riemannian geometry of $\cM$, it is of
prime importance to understand its topology.  Since the metric \eqref{hypmetone}
is only valid in the weak coupling regime, we shall restrict our discussion to the topology of the
hypersurfaces  $\cC(r)$, corresponding to fixed values of the dilaton $r\in\IR_+$,
which foliate the full moduli space $\cM$. The global structure of $\cC(r)$ (which, evidently,
must be independent of $r$) can be specified by providing a group of discrete identifications
of the coordinates $z^a, \zeta^\Lambda, \tzeta_\Lambda,\sigma$, acting isometrically
on $\cC(r)$. The precise subgroup can be identified by studying the allowed
instanton corrections to the metric. In the rest of this section,
we shall discuss instanton corrections to \eqref{hypmetone}
at a qualitative, semi-classical level, ignoring the tensorial nature of the metric.
A more precise discussion will be given in Section \ref{sec_ns5twi}
using twistor techniques.

\subsubsection{Instanton contributions to the HM moduli space}\label{instantoncorrections}

In type IIA string theory on $\cX$, these correspond to
D2-brane instantons wrapping a sLag submanifold  of $\cX$ (more
precisely, to stable objects in the Fukaya category of $\cX$).
Such instantons
produce corrections to the metric on $\cQ_c(\cX)$ of the form \cite{Alexandrov:2008gh}
\be
\label{d2quali}
\delta \de s^2\vert_{\text{D2}} \sim \exp\left( -8\pi e^{\phi/2} |Z_\gamma|
- 2\pi\I (q_\Lambda \zeta^\Lambda-p^\Lambda\tzeta_\Lambda)\right)\, ,
\ee
where $(p^\Lambda,q_\Lambda)$ are integers which label the homology class
$\gamma=q_\Lambda \cA^\Lambda- p^\Lambda \cB_\Lambda\in H_3(\cX,\IZ)$
of the sLag,
while the central charge $Z_\gamma\equiv e^{\cK/2} (q_\Lambda X^\Lambda-p^\Lambda
F_\Lambda)$  (or, in mathematical
parlance, the stability data) is the volume of the sLag, induced from the CY metric.
The expression \eqref{d2quali} is only valid in the classical, weak
coupling limit $r\to \infty$, up to
a one-loop determinant prefactor which we discuss in
Section \ref{subsubsec_monodr}. There are also multi-D-instanton
corrections to the metric, but the dependence on the axions $(\zeta,\tzeta,\sigma)$
is always of the form  \eqref{d2quali}, where $(p^\Lambda,q_\Lambda)$ is the total
charge carried by the multi-instanton configuration.

At subleading order $e^{-r}$, there are in addition corrections from
NS5-brane instantons in the homology class $k \cX$  \cite{Becker:1995kb}.
For $k>0$, these corrections are expected to be of the form
\be
\label{ns5quali}
\delta \de s^2\vert_{\text{NS5}}  \sim
e^{-4 \pi k e^{\phi}-\I\pi  k \sigma} \,
{\ZNSG{k}(\cN, \zeta^\Lambda,\tzeta_\Lambda)}\, ,
\ee
where $\ZNSG{k}$ is the NS5-partition function in the Gaussian, weak coupling
approximation,
while the prefactor in \eqref{ns5quali} incorporates the Euclidean action for $k$ NS5-branes.
For the purposes of reading off from \eqref{ns5quali} the classical NS5-brane action,
one can replace $\ZNSG{k}$ with the self-dual flux
partition function $\cZ^{(k)}_{\Theta, \mu}$ obtained in \eqref{thpl2}.
Then retaining one of the terms in the sum \eqref{thpl2} and disregarding normalization factors,
the classical action for the non-perturbative corrections
in \eqref{ns5quali} is found to be
\be
\label{SdV}
\begin{split}
S_{\rm NS5/D2}=&\,\pi k\left[ 4 \, e^{\phi}
- \I (n^{\Lambda}-\zeta^\Lambda)\bar \cN_{\Lambda\Sigma} (n^{\Sigma}-\zeta^\Sigma)\right]
\\
&+ \I\pi  k (\sigma+\zeta^\Lambda \tzeta_\Lambda)
-2\pi \I k \, (\tilde\zeta_{\Lambda}-\phi_\Lambda) n^{\Lambda} -\I\pi k \theta^\Lambda \phi_\Lambda\, .
\end{split}
\ee
The first line matches precisely the instanton action obtained in \cite{deVroome:2006xu}, Eq. 4.54,
based on a classical analysis  of instanton solutions in $\cN=2$ supergravity,
while the second line restores the appropriate axionic couplings (including characteristics).
The integers $kn^\Lambda$ may be interpreted as the flux on the fivebrane,
or as charges of a D2-brane wrapped on $kn^\Lambda \cB_\Lambda \in H_3(\cX,\mathbb{Z})$,
although this classical interpretation is  misleading since fluxes
or D2-brane charges are inherently quantum-mechanical on the fivebrane
worldvolume.

\subsubsection{Large gauge transformations \label{sec_lgt}}

The isometry group of $\cC(r)$ contains in particular the continuous translations
\be
\label{heis0}
T_{H,\kappa}:\quad
\bigl(\zeta^\Lambda,\ \tzeta_\Lambda,\ \sigma\bigr)\mapsto
\bigl(\zeta^\Lambda + \eta^\Lambda ,\
\tzeta_\Lambda+ \tleta_\Lambda ,\
\sigma + 2 \kappa - \tleta_\Lambda \zeta^\Lambda + \eta^\Lambda \tzeta_\Lambda
\bigr)\, ,
\ee
with $H=(\eta^\Lambda,\tleta_\Lambda) \in \IR^{b_3}$, $\kappa \in \IR$, satisfying
the Heisenberg group law
\be
T_{H_2,\kappa_2} T_{H_1,\kappa_1} =
T_{H_1+H_2,\kappa_1+\kappa_2+\frac{1}{2}\langle H_1, H_2\rangle}\, ,
\label{grouplaw}
\ee
where $\langle H, H' \rangle$ is the symplectic pairing \eqref{SymplecticPairing}.
While being a symmetry to all orders in $1/r$, the continuous shifts $T_{H,0}$
are generally expected to be broken  to a discrete subgroup
by D-instantons.
The form of the D-instanton corrections \eqref{d2quali} implies that
$\partial_\sigma$ continues to be an isometry of the HM metric  at order $e^{-\sqrt{r}}$,
but that invariance under continuous translations of the $C$-field is
broken to the discrete group
$\Gamma= H^3(\cX,\IZ)$, dual to the lattice $H_3(\cX,\IZ)$ of D2-brane
charges, acting by integer shifts on the RR axions
\be
\label{heisab}
 \bigl(\zeta^\Lambda,\ \tzeta_\Lambda \bigr)\mapsto
\bigl(\zeta^\Lambda + \eta^\Lambda ,\ \tzeta_\Lambda+ \tleta_\Lambda\bigr),
\ee
where $H=(\eta^\Lambda,\tleta_\Lambda) \in H^3(\cX,\IZ)$. In order to be
an isometry of the perturbative metric \eqref{hypmetone}, this must be accompanied
by a shift of the NS-axion, so that $\Gamma$ acts as the transformation
$T_{H,0}$ in \eqref{heis0}, possibly combined with a certain translation
$T_{0,c(H) }$ of $\sigma$ which, at
this order, remains undetermined. Thus, ignoring the
NS-axion $\sigma$ and keeping the complex structure of $\cX$ fixed,
the hypersurface $\cC(r)$ projects to the intermediate
Jacobian torus $\cT=H^3(\cX,\IR)/H^3(\cX,\IZ)$,
as stated in \cite{Morrison:1995yi}.

The NS5-brane corrections \eqref{ns5quali} further break the continuous translations in $\sigma$
to discrete identifications\footnote{Here we assume that $k$ is integer. This can be
argued by carefully dualizing $\sigma$ as in footnote \ref{foodualsig},
and will be confirmed shortly using S-duality.}
$T_{(0,\kappa)}:\sigma\mapsto \sigma+2\kappa,\ \kappa\in\IZ$.
This implies that the hypersurface $\cC(r)$ is a circle bundle over the intermediate Jacobian
$\cJ_c(\cX)$, the fiber of which is parametrized by $e^{\I \pi \sigma},\ 0\leq \sigma <2$.
In addition, Eq. \eqref{ns5quali} implies that the restriction of $\cC(r)$ to the torus $\cT$
must be isomorphic to the circle bundle $\cC_\Theta$
where the NS5-brane partition function is valued. In particular,
the shift $T_{0,c(H)}$ which must accompany the large gauge transformation
\eqref{heisab} follows from the transformation property
\eqref{thperiod} of the fivebrane partition function.
This shift is given by
\be
\label{propkappaH}
c(H)=- \frac12\, \eta^\Lambda \tleta_\Lambda + \tleta_\Lambda\theta^\Lambda
- \eta^\Lambda\phi_\Lambda \ \ \ {\rm mod}\   1\, ,
\ee
where $\Theta=(\theta^{\Lambda}, \phi_\Lambda)$ are the characteristics governing
the chiral partition function \eqref{thpl2}.
In particular, $c(H)$ satisfies
\be
\label{propcH}
c(H_1+H_2) =c(H_1) + c(H_2) + \frac12  \langle H_1,H_2\rangle \mod 1\, ,
\ee
as a result of the quadratic refinement law \eqref{qrifprop} for $\sigma_\Theta(H)=(-1)^{2c(H)}$.
Thus, large gauge transformations
imply that the axions $(C,\sigma)$ take values in the quotient $(H^3(\cX,\IZ)\times \IR)/ \Gamma'$,
where $\Gamma'$ is the group generated by $T'_{H,\kappa}\equiv T_{H,\kappa+c(H)}$
acting via
\be
\begin{split}
\label{heisext}
T'_{H,\kappa}\ :\ \bigl(\zeta^\Lambda,\ \tzeta_\Lambda,\ \sigma\bigr)
\mapsto
\bigl(\zeta^\Lambda + \eta^\Lambda ,\
\tzeta_\Lambda+ \tleta_\Lambda ,\
\sigma + 2 \kappa
- \tleta_\Lambda \zeta^\Lambda+\eta^\Lambda \tzeta_\Lambda + 2 c(H)\bigr) ,
\end{split}
\ee
where $H=(\eta^\Lambda,\tleta_\Lambda) \in H^3(\cX,\IZ)$, $\kappa \in \IZ$ and $c(H) \in \IR$
is chosen such that \eqref{propkappaH} holds. It should be noted that
the extra shift\footnote{The shift in $\sigma$ induced by the quadratic refinement
was already observed in the context of rigid Calabi-Yau compactifications
upon assuming that the moduli space should be invariant under
a certain natural arithmetic group \cite{Bao:2010cc}. With hindsight, this additional
shift could also have been uncovered in the $SL(3,\mathbb{Z})$-invariant construction of
\cite{Pioline:2009qt}, had it not been obscured by a redefinition of $\kappa$.}
of $c(H) $ is crucial for the consistency of this action. Indeed, the composition
of two large gauge transformations
\be
T'_{H_2,\kappa_2} \, T'_{H_1,\kappa_1} =
T'_{H_1+H_2,\kappa_1+\kappa_2+ c(H_1)+c(H_2)+\frac12  \langle H_1,H_2\rangle-
c(H_1+H_2)}
\ee
is again a large gauge transformation $T'_{H_3,\kappa_3}$ with $\kappa_3\in \IZ$,
by virtue of \eqref{propcH}. Thus, upon acting on functions of the
form $F_k(\zeta^\Lambda,\tzeta_\Lambda,\sigma)\equiv F_k(\zeta^\Lambda,\tzeta_\Lambda) \,
e^{\I\pi k\sigma}$ with $k$ integer,
large gauge transformations $T'_{H,\kappa}$ are effectively Abelian.

To summarize, at a fixed point in  complex structure moduli space we have found that
the circle bundle  $\cC$ indeed restricts to the circle bundle $\cC_\Theta$ over
the torus $\cT=H^3(\cX,\IR)/H^3(\cX,\IZ)$ governing the fivebrane partition function.
We now discuss the fibration of  $\cC(r)$ over the complex structure
moduli space.

\label{page_sigma}

\subsubsection{Monodromies }
\label{subsubsec_monodr}

For discussing the fibration of the circle bundle $\cC(r)$ over $\cM_c(\cX)$,
and in particular its behavior under monodromies, it
becomes important to include the normalization factors  for the D-instanton
and NS5-instanton corrections  \eqref{d2quali} and \eqref{ns5quali}.

In the case of D-instantons, the normalization factor was determined in
\cite{Gaiotto:2008cd,Alexandrov:2008gh}. While the answer depends on the
metric component under consideration (see Section \ref{sec_Dtwi} for details),
the normalization factor universally involves
the product $\Omega(\gamma) \sigma_{\text{D}}(\gamma)$, where
$\Omega(\gamma)$ is the generalized DT invariant associated to the sLag
submanifold $\gamma=(p^\Lambda,q_\Lambda)$, and $\sigma_{\text{D}}(\gamma)$ is
again
 a quadratic refinement\footnote{It would be very
 interesting to obtain $\sigma_{\text{D}}(\gamma)$ from
the one-loop determinant for the topological
field theory on the D2-brane, along the lines of the superpotential
computation in \cite{Harvey:1999as}.} of the intersection form on $H^3(\cX,\IZ)$.
As in the case of the fivebrane partition function, $\sigma_{\text{D}}(\gamma)$
may be parametrized by characteristics $\Theta_{\text{D}}$,
\be
\sigma_{\text{D}}(\gamma) = \expe{-\frac12 q_\Lambda p^\Lambda
+ q_\Lambda \theta_{\text{D}}^\Lambda
- p^\Lambda \phi_{{\text{D}},\Lambda}}\equiv \sigma_{\Theta_{\text{D}}}(\gamma)\, .
\label{quadraticrefinementpq}
\ee
In order that the product $\Omega(\gamma) \sigma_{\text{D}}(\gamma)$ be invariant,
$\Theta_D$ must transform in the same way \eqref{sympchar}
as the fivebrane characteristics $\Theta$ under monodromies $C\mapsto \rho(M)\cdot C$.

While it seems natural to identify the two set of characteristics, $\Theta$ and $\Theta_{\text{D}}$,
we do not know for sure that this is a consistent choice.
On the one hand, it is desirable that physics
(in particular the HM moduli space $\cM$) be independent on any choice of quadratic refinement,
as it is the case for the class of $\cN=2$ field theories considered in  \cite{Gaiotto:2009hg}.
For what concerns the D-instanton corrections \eqref{d2quali},
a change $\Theta_{\text{D}}\mapsto \Theta'_{\text{D}}$ of the D-instanton characteristics
can be canceled by redefining the coordinate $C=(\zeta,\tzeta)$ into\footnote{The cost to
pay is that $C'$ no longer transforms as a symplectic vector under monodromies, as
observed in Section 4.1 of \cite{Gaiotto:2008cd}.}
\be
C'=C+\Theta'_{\text{D}}-\Theta_{\text{D}}.
\ee
Similarly, for what concerns the NS5-instanton corrections \eqref{ns5quali},
a change $\Theta\mapsto \Theta'$ of the
fivebrane characteristics can be canceled by redefining
\be
\hat C=C+\Theta'-\Theta\, ,
\qquad
\hat\sigma=\sigma+\langle \Theta-\Theta', C \rangle - \langle \Theta, \Theta' \rangle\ ,
\ee
where we have used \eqref{thschar}. In order for these two field redefinitions
to be consistent, we require that $\Theta_{\text{D}}$ and $\Theta$ should vary
in the same way. This suggests that the difference $\Theta-\Theta_{\text{D}}$
should be fixed, although it does not yet imply that it should vanish.
Now, it follows from \eqref{sympchar}
that the difference of characteristics
 transforms as a symplectic vector under monodromies,
modulo the addition of integers. In general, there is no non-zero vector invariant under
the full monodromy group, and so no canonical choice for the difference
$\Theta-\Theta_{\text{D}}$ except zero. On the other hand,  while type IIB S-duality
suggests that the two quadratic refinements should be related (as
one governs NS5-brane instantons while the other controls D5-brane instantons)
in Section \ref{sec_poinca} we shall find some tension between the equality of the two
characteristics and S-duality.
In view of this, we shall continue to distinguish the two quadratic refinements
in the sequel.

We now turn to the monodromy invariance of the fivebrane instanton  correction \eqref{ns5quali},
and to the topology of the NS-axion circle bundle $\cC$ over the complex structure moduli space
$\cM_c(\cX)$. As a first approach to this problem, let us evaluate
the curvature $\de (D\sigma/2)$ of the horizontal one-form \eqref{Dsigone},
suitably normalized to take into account the mod 2 periodicity of~$\sigma$:
\be
\label{c1L}
\de \left(\frac{D\sigma}{2}\right)= \omega_\cT + \frac{\chi(\cX)}{24}\, \omega_\sk \, .
\ee
The first term, equal to the \kahler class on the intermediate Jacobian torus $\cT$,
is recognized as the curvature of the circle bundle $\cC_{\Theta}$ discussed
in Section \ref{sec_theta}. The second term, proportional to the  \kahler class of the complex
structure moduli space $\cM_c(\cX)$, suggests that the restriction of $\cC$ to
$\cM_c(\cX)$ is isomorphic to $\cL^{\frac{\chi(\cX)}{24}}$, where $\cL$ is the
Hodge line bundle over $\cM_c(\cX)$. Unless $\chi(\cX)$ is divisible by 24,
there is no canonical definition of $\cL^{\frac{\chi(\cX)}{24}}$ (equivalently,
the curvature \eqref{c1L} is not an integer cohomology class). This suggests that
$\cC$ is a twisted circle bundle and that the coordinate $\sigma$ is in fact not globally
well-defined, as already observed in a slightly different context in \cite{Witten:1996bn}.

To address this point in more detail, observe that under local holomorphic
rescalings $\Omega_{3,0}\mapsto e^f \Omega_{3,0}$, the requirement that
the horizontal one-form \eqref{Dsigone} be invariant implies that
$\sigma$ must shift according to
\be
\sigma\mapsto \sigma+ \frac{\chi(\cX)}{24\pi}  \Im f + 2\kappa_f\,
\label{sigmaKahlershift}
\ee
(indeed, recall from \eqref{KahlerGaugetransf} that $\cA_K\mapsto \cA_K +\de \Im f$
under \kahler transformations $\cK\mapsto \cK-f-\bar f$). Here, $\kappa_f$ is an
undetermined constant modulo 1. More generally, under a monodromy transformation $M$ in
$\cM_c(\cX)$, $C$ and $\sigma$ must transform as
\be
\label{sigmon}
C\mapsto \rho(M)\cdot C\ ,\qquad
\sigma \mapsto \sigma + \frac{\chi(\cX)}{24\pi}  \Im f_M +2\kappa(M)\, ,
\ee
where  $f_M$ is a local
holomorphic function on $\cM_c(\cX)$ determined by the rescaling
$\Omega_{3,0}\mapsto e^{f_M} \Omega_{3,0}$ of the holomorphic 3-form
around the monodromy, and $\kappa(M)$ is again an undetermined constant
defined modulo 1. The consistency of the monodromy transformations \eqref{sigmon}
requires that $e^{2\pi\I\kappa(M)}$ be a unitary
character of the monodromy group. The combination
$s \equiv e^{-\frac{\chi(\cX)}{48} \cK+\I \pi \sigma}$ then transforms as
\be
\label{ansig}
s  \mapsto e^{2\pi\I\kappa(M)}\,
 e^{\frac{\chi}{24}\, f } \, s\, ,
\ee
which may be taken as the definition of the twisted bundle $\cL^{\frac{\chi(\cX)}{24}}$.
Since the bundle $\cL$ itself is well-defined,  the first factor must be
a $\chi(\cX)/24$:th root of unity, $\kappa(M)\in \chi(\cX)\IZ/24$.
Similarly, the additional term $2\kappa_f\in \chi(\cX)\IZ/12$ in \eqref{sigmaKahlershift} is
needed to escape the paradox raised in  \cite{Alexandrov:2010np}: in the absence
of this term, a trivial rescaling  $\Omega_{3,0}\mapsto e^{2\pi\I} \Omega_{3,0}$ would lead to
identifications $\sigma\equiv \sigma + \chi(\cX)/12$ in conflict with the periodicity
modulo 2 of $\sigma$. Let us further recall that the twisted bundle $\cL^{\frac{\chi(\cX)}{24}}$
from Section \ref{sec_top} also arises in the B-model topological
wave function, which is valued in $\cL^{ \frac{\chi(\cX)}{24}-1}$ \cite{Bershadsky:1994cx}.

In order to fully specify the topology of the circle bundle $\cC$, it is necessary to
determine the unitary character $e^{2\pi\I\kappa(M)}$, and more to the point, its
logarithm $\kappa(M)$. To appreciate the nature of this question, it is useful to
recall a similar problem which arises in the study of classical modular
forms\footnote{We are grateful to A. Neitzke and D. Zagier for suggesting this point of view.}:
determine the modular properties of the logarithm of the Dedekind $\eta$ function.
The answer to this question is well-known since the work of Dedekind  (whose sums
we shall encounter again in Section \ref{sec_pk}), but is most satisfactorily understood in
the language of determinant line bundles \cite{MR909232}. Transposing to the
present context, we expect that the role of the Dedekind $\eta$ function will
be played by the one-loop B-model topological amplitude $e^{f_1}/\sqrt{J_{\rm G}}$,
and that $\kappa(M)$ should be computable from the
fivebrane determinant line bundle along the lines of \cite{Belov:2006jd,Belov:2006xj,Monnier}.
Indeed, we shall find in Section \ref{sec_nonlin} that the metric-dependent normalization
factor of the fivebrane partition function in the weak coupling limit is proportional to the
one-loop B-model topological amplitude, consistently with the coupling \eqref{NS5coupling}.

Summarizing, we have found that the HM moduli space $\cM$, in the
weak coupling limit, is foliated by a family of hypersurfaces $\cC(r)$ which are
circle bundles\footnote{The dilaton factor $\IR^+_r$ may be used to
formally convert this circle bundle into a $\IC^{\times}$-bundle \cite{StienstraVandoren},
although the total space $\cM$ is not expected to carry a global complex structure.}
over the intermediate Jacobian \eqref{intJac},
\be
\cM \sim \IR^+_r \times \left(
\begin{array}{ccc}
  S^{1}_\sigma & \to & \cC(r)\\
   &  & \downarrow \\
   &  & \cJ_c(\cX)
\end{array}
\right),
\ee
whose curvature is given by \eqref{c1L}.
The perturbative moduli space  may be defined globally by the metric \eqref{hypmetone},
subject to  identifications \eqref{sigmon} under monodromies,
\eqref{heisext} under large gauge transformations,
and to possibly other identifications relating different leaves $\cC(r)$, as required
e.g. by S-duality.

\section{Mirror symmetry, S-duality and NS5-branes in type IIB}
\label{sec_IIB}

So far we have worked exclusively within the type IIA picture.
In this section we discuss the mirror map to type IIB where S-duality provides
a powerful tool to analyze NS5-brane effects. We recall that mirror symmetry identifies the HM
moduli space $\cQ_c(\cX)$ in type IIA string theory compactified on
a CY 3-fold $\cX$ with the HM moduli space $\cQ_K(\hat\cX)$
in type IIB compactified on the mirror CY 3-fold $\hat\cX$. While the former extends the
moduli space of complex structures $\cM_c(\cX)$ with data about the
dilaton, RR-fields and NS axion, the latter instead extends
the moduli space of (complexified) \kahler structures $\cM_K(\hat\cX)$
with similar data. In contrast to $\cM_c(\cX)$, the special \kahler metric on $\cM_K(\hat\cX)$
receives worldsheet  instanton corrections, and is most easily obtained from
$\cM_c(\cX)$ by mirror symmetry.

$\bullet$ In Section \ref{sec_mir} we review the mirror-dual description
of $\cM=\cQ_c(\cX)$ as the type IIB HM moduli space $\cM=\cQ_K(\hat\cX)$.
In the process, we discuss charge quantization, and reconcile the apparent
fractional charge shifts in the K-theory description of type IIB instantons (or type IIA
black holes) with the manifestly integer charges $\gamma\in H_3(\cX,\IZ)$
of type IIA instantons (or type IIB black holes).
$\bullet$
In Section \ref{sec_Sdual}, we then discuss the action of S-duality on the type IIB
hypermultiplet moduli space in the large volume limit and incorporate a
novel shift of the D3-brane axion $c_a$ needed to ensure consistency with
charge quantization.
$\bullet$
In Section \ref{sec_pk}, we use this action
to derive the classical action of $(p,k)$5-branes. In particular, we show that this
action is consistent with the Gaussian flux partition function
discussed in Section \ref{sec_zns5} in the weak coupling limit.

\subsection{Topology of the perturbative type IIB HM moduli space\label{sec_mir}}

Recall that in the weak coupling limit, the metric on $\cM\equiv \cQ_K(\hat\cX)$ is again
given by the one-loop corrected $c$-map metric \eqref{hypmetone}, where the special \kahler
manifold $\cS\cK=\cM_K(\hat \cX)$ is now the moduli space of complexified \kahler structures,
the torus $\cT$ parametrizes the periods of the RR potential
\be
A^{\rm even}=A^{(0)}+A^{(2)}+A^{(4)}+A^{(6)}\in H^{\rm even}(\hat\cX,\IR),
\ee
and the coordinates $r=e^\phi\sim 1/g_{(4)}^2$ and $\sigma$ still parametrize
the 4D string coupling and NS axion. The deformation parameter $c$ takes the
same value $c=\chi_{\hat \cX}/192\pi=-\chi({\cX})/192\pi$ as in \eqref{cchi}.
We refer to the total space of the torus $\cT$ over $\cM_K(\hat\cX)$ as the
symplectic Jacobian $\cJ_K(\hat\cX)$.

\subsubsection{\kahler moduli space}

To describe the geometry of $\cJ_K(\hat\cX)$ in more detail,
let us choose a basis $\gamma^a$,
$a=1,\dots, h^{1,1}(\hat\cX)$,  of 2-cycles (Poincar\'e dual to 4-forms $\omega^a$),
and a basis $\gamma_a$ of 4-cycles (Poincar\'e dual to 2-forms $\omega_a$), such that
\be\omega_a \wedge\omega_b = \kappa_{abc}\, \omega^c\, ,
\qquad
\omega_a \wedge \omega^b = \delta_a^b \omega_{\hat\cX}\, ,
\qquad
 \int_{\gamma^a}\omega_b= \int_{\gamma_b}\omega^a= \delta^a_b\, ,
\ee
where $\omega_{\hat\cX}$ is the volume form, normalized to $\int_{\hat\cX} \omega_{\hat\cX} =1$,
and $ \kappa_{abc}=\int_{\hat\cX} \omega_a \omega_b \omega_c
=\langle \gamma_a, \gamma_b, \gamma_c\rangle$
is the triple intersection product in $H_4({\hat\cX},\IZ)$.
The space $H^{\rm even}(\cX)$ carries a symplectic form given by
\be
\langle H, H' \rangle = \int_{\hat\cX} H \wedge \tau(H')\, ,
\ee
where $\tau$ flips the sign of the 2-form and 6-form part of $H'$.
It will be convenient  to let $\omega^\Lambda=(\omega_{\hat\cX},\omega^a),
\omega_\Lambda=(1,\omega_a)$ such that $(\omega^\Lambda,\omega_\Lambda)$
forms a symplectic basis of $H^{\rm even}(\cX,\IR)$.

The K\"ahler structure moduli space $\cM_K(\hat \cX)$
may then be parametrized by the periods $z^a\equiv b^a + \I t^a=\int_{\gamma^a}
(B+\I J)$ of the complexified \kahler form. In projective special coordinates $X^\Lambda=
(X^0,X^a)$ such that $z^a=X^a/X^0$, the prepotential governing $\cM_K(\hat\cX)$ is given
by\footnote{We restore the important
quadratic contribution $\frac12 A_{\Lambda\Sigma} X^\Lambda X^\Sigma$,
which was omitted in \cite{Alexandrov:2008gh}.
Equivalently, one may extend the range
of indices for the cubic form $\kappa_{abc}$ and define
$\kappa_{000}=-3A_{00},\ \kappa_{00a}=-2 A_{0a},\ \kappa_{0ab}=-A_{ab}$.}
\be
\label{lve}
F(X)=-\frac{N(X^a)}{X^0} + \frac12 A_{\Lambda\Sigma} X^\Lambda X^\Sigma +
\chi(\hat\cX)\,
\frac{\zeta(3)(X^0)^2}{2(2\pi\I)^3}
-\frac{(X^0)^2}{(2\pi\I)^3}{\sum_{k_a\gamma^a\in H_2^+(\hat \cX)}} n_{k_a}^{(0)}\, \Li_3\left[
\expe{ k_a \frac{X^a}{X^0}}\right],
\ee
where $N(X^a)\equiv \frac16 \kappa_{abc} X^a X^b X^c$, and $A_{\Lambda\Sigma}$
is a constant, real symmetric matrix.
The matrix $A_{\Lambda\Sigma}$ is ambiguous
up to integer shifts $A_{\Lambda\Sigma}\mapsto A_{\Lambda\Sigma}
+\cB_{\Lambda\Sigma}$,  $\cB_{\Lambda\Sigma}\in \IZ$, and does not
affect the \kahler potential $\cK=-\log[ \I ( \bar X^\Lambda F_\Lambda
- X ^\Lambda \bar F_\Lambda)]$ on $\cM_\cK(\hat\cX)$.
However, as we shall demonstrate shortly,  it is crucial for charge quantization.
In \eqref{lve}, the sum over effective divisors
represents the effect of genus 0 worldsheet  instantons, while the trilogarithm
sum $\Li_3(z)\equiv \sum_{n=1}^\infty z^n/n^3$ encodes multi-covering effects.

\subsubsection{Charge quantization \label{sec_cq}}

By the same reasoning as in the type IIA analysis of Section \ref{sec_prel},
the torus $\cT$ in the mirror type IIB can be written as
\be
\cT = H^{\rm even}(\hat\cX,\IR)/ \Gamma\, ,
\ee
where $\Gamma$ is the lattice dual to the charge lattice
classifying D5-D3-D1-D(-1)-instanton configurations.
To determine this lattice, we must invoke the
K-theory description of D-brane charges \cite{Minasian:1997mm}.

For non-vanishing D5-brane charge $p^0$, D5-D3-D1-D(-1)-instanton
configurations can be represented
as a  coherent sheaf $E$ on $\hat\cX$, of rank $\rk(E)=p^0$ and
generalized Mukai vector $\gamma'$, defined by\footnote{The prime anticipates the
fact that the vector $\gamma'$ lies in $H^{\rm even}(\hat\cX,\IQ)$. In Eq. \eqref{modifmukai}
we introduce a ``modified Mukai vector'' which lies in
$H^{\rm even}(\hat\cX,\IZ)$.}
\cite{Minasian:1997mm}
\be
\label{mukaimap}
\gamma' \equiv \ch(E)\sqrt{\Td(\hat \cX)} = p^0 + p^a \omega_a - q'_a \omega^a + q'_0 \omega_{\hat\cX}\, ,
\ee
where $\ch(E)$ and $\Td(\hat \cX)$ are the Chern character of $E$ and Todd class of
$T\hat\cX$. Using
\be
\begin{split}
\ch=&\rk+c_1+(\frac12 c_1^2-c_2)+
\frac12(c_3-c_1c_2+\frac13 c_1^3)+\dots
\\
\Td =& 1+\frac12 c_1+\frac1{12}(c_1^2+c_2)+\frac{1}{24}c_1 c_2\dots
=e^{c_1/2}\hat A,
\end{split}
\ee
where $\hat A$ is the Dirac genus, the coefficients $p^\Lambda,q'_\Lambda$
of $\gamma'$ on the symplectic basis $\omega_\Lambda, \omega^\Lambda$
are found to be~\cite{Douglas:2006jp}
\be
\begin{split}
& p^a =\int_{\gamma^a} c_1(E) \, ,
\qquad
q'_a =-\left( \int_{\gamma_a} \ch_2(E) +   \frac{p^0}{24}  c_{2,a}\right),
\\
& \qquad\quad q'_0 = \int_{\hat\cX}\left(\ch_3(E)+\frac{1}{24}\, c_1(E)\,c_2({\hat \cX})
\right) .
\end{split}
\label{ElectricChargesD5}
\ee
Due to the fractional coefficients appearing in \eqref{ElectricChargesD5} and in
$\ch_n(E)$,  the generalized Mukai vector $\gamma'=(p^\Lambda,q'_\Lambda)$
is not valued in $H^{\rm even}(\hat\cX,\IZ)$
but rather in $H^{\rm even}(\hat\cX,\IQ)$.
This appears to challenge the fact that the D2-brane charges in type IIA
compactified on the mirror 3-fold $\cX$ are classified
by $H_3(\cX,\IZ)$ modulo torsion.
To find the correct
integer-valued  charge vector $\gamma=(p^\Lambda,q_\Lambda)$,
it suffices to match the central charge
$ \int_{\hat \cX}  e^{-z^a \omega_a} \gamma'$ associated to
the sheaf $F$ with the holomorphic supergravity central charge
$e^{-\cK/2} Z_\gamma=q_\Lambda X^\Lambda -
p^\Lambda F_\Lambda$,
leading to
\be
q_\Lambda = q'_\Lambda + A_{\Lambda\Sigma} p^\Sigma\, .
\label{chargeshift}
\ee
Thus, $q_\Lambda$ and $q'_\Lambda$ differ by a (non-integer) symplectic transformation,
which removes the real quadratic terms in \eqref{lve}, transforming the prepotential $F$
into
\be
F'(X) = F(X) - \frac12 A_{\Lambda\Sigma} X^\Lambda X^\Sigma\, .
\ee

The magnetic charges $p^0$ and $p^a$ are manifestly integer.
To see that the electric charges $q_\Lambda$ are integer, it is convenient to rewrite
them  as
\be
\begin{split}
q_a=&\int_{\gamma_a} \text{c}_2(E)+p^0\left(A_{0a}-\frac{c_{2,a}}{24}\right)+A_{ab}p^b
-\frac12\kappa_{abc} p^b p^c,
\\
 q_0=&\left( \int_{\hat\cX}\text{ch}(E) \, \Td(\hat\cX) \right)
 +p^a\left(A_{0a}-\frac{c_{2,a}}{24}\right)+A_{00}p^{0} .
 \end{split}
\ee
Observing that the first term in $q_0$ is the index of the Dirac operator coupled to the
sheaf $F$, and therefore an integer,\footnote{We are grateful to R. Minasian for discussions
on related issues.} it follows that $q_\Lambda\in\IZ$ provided
the following additional constraints are imposed on the matrix $A_{\Lambda\Sigma}$:
\be
\begin{split}
i)&\ A_{ab}  p^b- \frac12\, \kappa_{abc} p^b p^c  \in \IZ\quad {\rm for}\ \forall p^a\in\IZ\, ,
\\
ii)&\ A_{0a}\in \frac{c_{2,a}}{24} + \IZ\, ,
\\
iii)&\ A_{00}\in \IZ\, .
\end{split}
\label{condA1}
\ee
The last condition shows that $A_{00}$ can be chosen to vanish.
The second condition was conjectured long ago in \cite{Hosono:1994av}
on the basis of a few examples, and is now
seen to follow from K-theory considerations. Without loss of generality we shall
choose $A_{0a}=c_{2a}/24$. The first condition, which appears to be novel,
requires that $2A_{ab}$ is integer.
It indeed holds true in the few examples where
$A_{\Lambda\Sigma}$ has actually been computed \cite{Candelas:1990rm,
Klemm:1992tx,Candelas:1993dm,Candelas:1994hw}.
For later reference, we note that the integrality of the Dirac operator
on $\hat\cX$ coupled to a rank one sheaf ($c_2(E)=c_3(E)=0$) further requires\footnote{
This condition also guarantees that the dimension of the moduli space of
the divisor Poincar\'e dual to $c_1(E)$ be even \cite{Maldacena:1997de}.}
\be
iv)\  N(p^a)+\frac1{12}\,c_{2,a}p^a\in\IZ \quad {\rm for}\ \forall p^a\in\IZ\, .
\label{condA2}
\ee
In view of $i)$ and $iv)$, it will be useful to define the integer-valued combinations
\be
\label{defL0La}
L_0(p^a)=  N(p^a)+\frac1{12}\,c_{2,a}\, p^a\, ,\qquad
L_a(p^a)= \frac12\, \kappa_{abc} p^b p^c - A_{ab}  p^b\, .
\ee

Assuming that these integrality conditions
are obeyed (as they must for mirror symmetry to hold), the charge vector
$\gamma=(p^\Lambda,q_\Lambda)$ is now valued in  $H^{\rm even}(\cX,\IZ)$,
and can be meaningfully identified with the homology class of a special Lagrangian submanifold on the mirror
type IIA side. On the other hand, the primed charges \eqref{ElectricChargesD5} satisfy
\be
q'_a\in \mathbb{Z} -\frac{p^{0}}{24}\, c_{2,a} - \frac12 \kappa_{abc} p^b p^c,
\qquad
q'_0\in \mathbb{Z}-\frac{1}{24}\, p^{a} c_{2,a}\, .
\label{fractionalshiftsD5}
\ee
For $p^0=0$, we recover the quantization conditions
$q'_a\in \mathbb{Z}  - \frac12 \kappa_{abc} p^b p^c$ and $q'_0\in \mathbb{Z}-\frac{1}{24}\, p^{a} c_{2,a}$
familiar from studies of
the D4-D2-D0 brane partition function \cite{Dabholkar:2005dt,deBoer:2006vg,Gaiotto:2006wm,Denef:2007vg}.
As we shall see in Section \ref{sec_Sdual}, the fractional shifts in
\eqref{fractionalshiftsD5} have important implications for S-duality.

\subsubsection{Symplectic Jacobian and mirror symmetry}

We are now finally ready to relate the variables $(\zeta^\Lambda,\tzeta_\Lambda)$
appearing in the metric \eqref{hypmetone} to the periods of the type IIB RR form $A^{\rm even}$,
and to specify the torus bundle $\cT\rightarrow \cJ_K(\hat\cX)\to \cM_K(\hat\cX)$.
For this purpose, recall that  the classical action of the D-instanton
associated to the sheaf $E$ is given by \cite{Marino:1999af,Aspinwall:2004jr}
\be
\label{Sbbrane}
S_{\gamma}= 8\pi e^{(\phi+\cK)/2} \left| \int_{\hat \cX}  e^{-\mathcal{J}} \gamma' \right|
+ 2\pi \I  \int_{\hat \cX}  \gamma' \wedge A^{\rm even}\, e^{-B} \, .
\ee
Identifying the periods of the RR potential $A^{\rm even}$ as
\be
\label{ABze}
A^{\rm even}\, e^{-B}
= \zeta^0 - \zeta^a \omega_a - \tzeta'_a \omega^a-\tzeta_0' \omega_{\hat \cX}
\ee
and defining the unprimed axions
\be
\label{shze}
\tzeta_\Lambda = \tzeta'_\Lambda + A_{\Lambda\Sigma} \zeta^\Sigma\, ,
\ee
we find that
the Euclidean action \eqref{Sbbrane} associated to the
D5-D3-D1-D(-1) bound state with integer charges $(p^\Lambda, q_\Lambda)$
becomes equal to the Euclidean action \eqref{d2quali} of a D2-brane wrapping a sLag
in the integer homology class $q_\Lambda \cA^\Lambda - p^\Lambda \cB_\Lambda\in H_3(\cX,\mathbb{Z})$
on the mirror threefold $\cX$. In particular, the coordinates $(\zeta^\Lambda,\tzeta_\Lambda)$
defined by \eqref{ABze}, \eqref{shze} have unit periodicity, and parametrize the torus
$\cT$.

Under a monodromy $M$ in the moduli space of \kahler structures $\cM_K(\hat\cX)$,
the torus transforms by a symplectic rotation $\rho(M)\in Sp(2b_2(\hat\cX)+2,\IZ)$
computable from the prepotential $F(X)$ (see Section \ref{sec_mon}).
Thus,  the symplectic Jacobian $\cJ_K(\hat\cX)$ is the total space of the torus bundle
\be
\label{intJac2}
\begin{array}{ccc}
  H^{\rm even}(\hat \cX,\IR)/\Gamma & \to & \cJ_K({\hat \cX})\\
   &  & \downarrow \\
   &  & \cM_K(\hat{\cX}),
\end{array}
\ee
where the lattice $\Gamma\subset H^{\rm even}(\hat \cX,\IZ)$,  dual to the lattice
of D5-D3-D1-D(-1) brane charges, is the image of the
K-theory lattice $K(\hat \cX)$ under the ``modified Mukai map"
\be
\label{modifmukai}
E\mapsto \gamma=e^{A} \ch(E) \sqrt{\Td({\hat{\cX}})}\ \in\ H^{\rm even}(\hat \cX,\IZ),
\ee
where $e^A$ represents the fractional symplectic transformation \eqref{chargeshift}.

Having specified the torus bundle $\cJ_K(\hat\cX)$, the topology of the
type IIB perturbative HM moduli space $\cM$ follows by the same reasoning as in
Section \ref{sec_pert}: $\cM$
is foliated by hypersurfaces $\cC(r)$ of constant coupling,
each of which is a circle bundle $\cC(r)$ over $\cJ_K(\hat\cX)$, with curvature given by
\be
\label{c1LB}
\text{d}\left(\frac{D\sigma}{2}\right)=\omega_\cT - \frac{\chi(\hat\cX)}{24}\, \omega_\sk \,  ,
\ee
where $\omega_\sk$ is the \kahler class of $\cM_K(\hat\cX)$,
and the characteristics $\Theta$ which remain to be specified.
This is in full agreement with the quantum mirror symmetry conjecture.
While classical mirror symmetry
identifies $\cM_K(\hat\cX)=\cM_c(\cX)$, semi-classical mirror symmetry
(in its weakest form) demands $\cJ_K(\hat\cX)=\cJ_c(\cX)$, quantum
mirror symmetry requires the identity of the full HM moduli spaces $\cQ_c(\cX)=\cQ_K(\hat\cX)$
as \qk manifolds.

\subsubsection{Monodromy around the large volume point \label{sec_mon}}

We mentioned previously that the torus $\cT$ is non-trivially fibered over $\cM_K(\hat\cX)$.
Here we consider the monodromy $M$:
\be
\label{monb}
b^a\mapsto b^a+\epsilon^a\ ,\qquad \epsilon^a\in \IZ
\ee
around the large volume point in $\cM_K(\hat\cX)$. This monodromy must be
accompanied by the symplectic rotation
\be
\label{bjacr}
\begin{split}
\zeta^a\mapsto \zeta^a + \epsilon^a \zeta^0\, ,
\qquad
\tzeta'_a\mapsto \tzeta'_a -\kappa_{abc}\zeta^b \epsilon^c
-\frac12\,\zeta^0\kappa_{abc} \epsilon^b \epsilon^c\, ,
\\
\tzeta'_0\mapsto \tzeta'_0 -\epsilon^a\tzeta'_a +\frac12\, \kappa_{abc}\zeta^a \epsilon^b \epsilon^c
+\frac16\, \zeta^0\kappa_{abc} \epsilon^a \epsilon^b \epsilon^c
\qquad
\end{split}
\ee
on the torus $\cT$, as follows from the definition \eqref{ABze}.
Upon transforming the charges according to
\be
\label{flow}
\begin{split}
p^0\mapsto p^0\, , &\qquad
p^a\mapsto p^a + \epsilon^a p^0\, ,
\qquad
q'_a\mapsto q'_a - \kappa_{abc} p^b \epsilon^c - \frac12\, p^0 \kappa_{abc} \epsilon^b
\epsilon^c\, ,
\\
&\quad q'_0\mapsto q'_0 - \epsilon^a q'_a  +\frac12\, \kappa_{abc} p^a \epsilon^b \epsilon^c
+ \frac16\, p^0 \kappa_{abc} \epsilon^a \epsilon^b
\epsilon^c\, ,
\end{split}
\ee
the instanton action \eqref{Sbbrane} remains invariant. The transformation
\eqref{flow} amounts to tensoring the sheaf $E$ with a line bundle $E'$ with first Chern class
$c_1(E')=\epsilon^a \omega_a$. We refer to the transformation \eqref{flow} as a ``spectral flow"
transformation with parameter $\epsilon^a$, and the transformed (unprimed)
charge vector as $\gamma[\epsilon]$. The spectral flow  \eqref{flow} can be rewritten as
an integral symplectic transformation of the integral charge vector $\gamma$,
\be
\label{monosymp}
\gamma[\epsilon]=\rho(M)\cdot\gamma\, ,
\qquad
\rho(M)=\(\begin{array}{cccc}
1\ & 0 & 0 & 0
\\
\epsilon^a & {\delta^a}_b & 0 & 0
\\
-L_a(\epsilon)\ & -\kappa_{abc}\epsilon^c & {\delta_a}^b & 0
\\
\ L_0(\epsilon) & L_b(\epsilon)+2A_{bc}\epsilon^c & \ -\epsilon^b\  & 1
\end{array}\) ,
\ee
which makes it clear that  the spectral flow
preserves the quantization conditions \eqref{fractionalshiftsD5}.

For $p^0\neq 0$, the combinations
\be
\label{qaq0hat}
\hat q_a=
q'_a + \frac12\, \kappa_{abc} \frac{p^b p^c}{p^0}
\, ,\qquad
\hat q_0 =
q'_0 +\frac{p^a q'_a}{p^0}+ \frac13\, \kappa_{abc} \frac{p^a p^b p^c}{(p^0)^2}\, ,
\ee
are invariant under the spectral flow \eqref{flow}. For a fixed $p^0$, we shall refer to
$\hgam=(p^a,\hat q_a,\hat q_0)$ as the reduced charge vector.
The invariant charges $\hat q_\Lambda$
may be expressed in terms of the Chern classes of $E$ as
\be
\label{qaq0hatc}
\begin{split}
\hat q_a&=
\int_{\gamma_a} \Big[c_2(E) +\frac{1-p^0}{2 p^0}\, c_1^2(E)\Big] - \frac{p^0}{24} \, c_{2,a}\, ,
\\
\hat q_0 & = \int_{\hat\cX} \left[ \ch_3(E)
 -\frac{1}{p^0}\, c_1(E)  \ch_2(E)
 + \frac{1}{3 (p^0)^2}\,c_1^3(E)
\right]\, .
\end{split}
\ee
These combinations coincide (up to the last term in $\hat q_a$) with
the homogeneous invariants of \cite{Douglas:2006jp}. In particular,
$\hat q_a+\frac{p^0}{24} c_{2,a}$ is recognized as the ``Bogomolov discriminant''.
It is also useful to observe that
\be
\label{defQJ}
Q_a\equiv p^0\left(  \hat q_a + \frac{p^0}{24} c_{2,a} \right),
\qquad
J\equiv \frac12 (p^0)^2 \hat q_0
\ee
are precisely (after matching conventions) the electric charge and
angular momentum of the 5D lift of the 4D D6-D4-D2-D0 black hole
with charges $(p^0,p^a,q_a',q_0')$ \cite{Gaiotto:2005gf,Gaiotto:2005xt,Gao:2008hw}.
Moreover, for $p^0=1$, $Q_a$ and $2J$ are integer.
These observations will play a role in Section \ref{sec_ns5top}.

While the combined action of the monodromy $M$ on the moduli via
\eqref{monb} and \eqref{bjacr} and on the charges via \eqref{flow} preserves
the D-instanton action and stability condition, and therefore the generalized
Donaldson-Thomas invariant $\Omega(\gamma,z^a)$, the latter is in general
not invariant under $\gamma\mapsto \gamma[\epsilon]$ at fixed values of the
moduli, due to wall-crossing phenomena. In the heuristic discussion of
Section \ref{sec_ns5twi}, we shall however ignore this
issue and assume that $\Omega(\gamma)$ is in fact invariant under spectral
flow, at fixed values of the moduli.

\subsection{S-duality\label{sec_Sdual}}

The type IIB description is more advantageous
for dealing with non-perturbative corrections as it provides an infinite discrete symmetry
mixing worldsheet  instantons and D-instantons, the 10D $SL(2,\IZ)$ S-duality
symmetry.\footnote{The same $SL(2,\IZ)$ action is also manifest geometrically
in the context of M-theory on $\hat\cX\times T^2$, whose VM moduli space is
given by the same space $\cQ_K(\hat\cX)$. As already pointed out in the introduction,
it is debatable whether the full $SL(2,\IZ)$ symmetry should stay
unbroken in vacua with $\cN=2$ supersymmetry, or whether it should be broken
to a finite index subgroup. At this point we assume that it does, though
our discussion can be adapted to accommodate the weaker option.}
In the zero coupling, infinite volume limit, where only the first,
cubic term in the prepotential \eqref{lve} is retained and the one-loop
correction can be ignored, the moduli space $\cQ_K(\hat\cX)$ admits
an isometric action of $SL(2,\IR)$ \cite{Gunther:1998sc,Bohm:1999uk,Alexandrov:2008gh},
corresponding to the S-duality of 10-dimensional type IIB string theory.
This action is most easily described using new coordinates $(\tau_1,c^a,c_a,c_0,\psi)$
related to the coordinates $(\zeta^\Lambda,\tzeta_\Lambda,\sigma)$ by
\be
\label{symptob}
\begin{split}
\zeta^0&=\tau_1\, ,
\qquad\qquad
\zeta^a = - (c^a - \tau_1 b^a)\, ,
\\
\tzeta'_a &=  c_a+ \frac{1}{2}\, \kappa_{abc} \,b^b (c^c - \tau_1 b^c)\, ,
\qquad
\tzeta'_0 =\, c_0-\frac{1}{6}\, \kappa_{abc} \,b^a b^b (c^c-\tau_1 b^c)\, ,
\\
\sigma &= -2 (\psi+\frac12  \tau_1 c_0) + c_a (c^a - \tau_1 b^a)
-\frac{1}{6}\,\kappa_{abc} \, b^a c^b (c^c - \tau_1 b^c)
\end{split}
\ee
and the ten-dimensional inverse coupling $\tau_2=1/g_{s}$ related to
the 4D dilaton by \cite{Alexandrov:2008gh}
\be
\label{phipertA}
e^{\phi}
= \frac{\tau_2^2}{16}\,e^{-\cK(z,\bz)}-\frac{\chi(\hat\cX)}{192\pi}\, .
\ee
Then, an element $\delta={\scriptsize \begin{pmatrix} a & b \\ c & d  \end{pmatrix}}
\in SL(2,\IR)$ acts by fractional linear transformations
of the axio-dilaton field $\tau\equiv \tau_1+\I \tau_2$, rescaling of the \kahler moduli
and by linear action on the RR and NS axions:
\be\label{SL2Z}
\begin{split}
&\quad \tau \mapsto \frac{a \tau +b}{c \tau + d} \, ,
\qquad
t^a \mapsto t^a |c\tau+d| \, ,
\qquad
c_a\mapsto c_a+\eps_a(\delta)\, ,
\\
&\quad
\begin{pmatrix} c^a \\ b^a \end{pmatrix} \mapsto
\begin{pmatrix} a & b \\ c & d  \end{pmatrix}
\begin{pmatrix} c^a \\ b^a \end{pmatrix}\, ,
\qquad
\begin{pmatrix} c_0 \\ \psi \end{pmatrix} \mapsto
\begin{pmatrix} d & -c \\ -b & a  \end{pmatrix}
\begin{pmatrix} c_0 \\ \psi \end{pmatrix}.
\end{split}
\ee
Here we deviate from \cite{Gunther:1998sc,Bohm:1999uk,Alexandrov:2008gh},
and allow for a shift in the D3-brane axion $c_a$ by an additive
character $\eps_a(\delta)$ of $SL(2,\IR)$,
i.e. such that $\eps_a(\delta\cdot \delta')=\eps_a(\delta)+\eps_a(\delta')$.

The continuous isometric action \eqref{SL2Z} is broken by the worldsheet  instanton
effects in \eqref{lve} and the one-loop correction $c$, but a discrete $SL(2,\IZ)$ subgroup can be
restored by including D(-1) and D1-brane instantons \cite{RoblesLlana:2006is}. This in
particular provides a common origin\footnote{In fact, this was used in \cite{RoblesLlana:2006is}
to confirm the normalization of $c$ in \eqref{cchi}.}
for the degenerate instanton contribution, proportional
to $\zeta(3)\chi(\hat\cX)$ in \eqref{lve} and the one-loop correction,
proportional to $\zeta(2)\chi(\hat\cX)$. The analysis in \cite{RoblesLlana:2006is}
did not probe the possibility of an additional character $\eps_a(\delta)$.

In the presence of D3-brane instantons, the character $\eps_a(\delta)$ is needed for the following
reason. Due to the fractional shift \eqref{fractionalshiftsD5} of the D(-1)-brane charge $q_0'$,
a D-instanton contribution proportional to
$\expe{p^\Lambda \tzeta_\Lambda-q_\Lambda \zeta^\Lambda}
=\expe{p^\Lambda \tzeta'_\Lambda-q'_\Lambda \zeta^\Lambda}$
transforms under an S-duality action
$\delta_b={\scriptsize \begin{pmatrix} 1 & b \\ 0 & 1  \end{pmatrix}}$
by a phase $\expe{\frac{p^a c_{2,a}}{24}\, b + p^a\eps_a(\delta_b)}$. Moreover,
this S-duality action differs from the Heisenberg shift \eqref{heisext}
with parameter $\eta^0=b$ by a fractional shift
$\tzeta_a\mapsto \tzeta_a+\eps_a(\delta_b)+\frac{ c_{2,a}}{24}\, b$.
Both these problems can be avoided by choosing the character $\eps_a(\delta)$
such that $\eps_a(\delta_b)=-\frac{ c_{2,a}}{24}\, b$. This is indeed the case
if $\eps_a$ is taken to be proportional to the multiplier system of the Dedekind eta function,
\be
\label{rhode}
\eps_a(\delta) = - c_{2,a}\, \eps(\delta)\, ,
\ee
where $\eps(\delta)$ is a fractional number defined, up to the addition of integers,
by the ratio\footnote{Strictly speaking, the two factors on the r.h.s. of
\eqref{multeta} both have sign ambiguities, and only their product is well-defined.
But since the condition \eqref{condA2} ensures that $c_{2,a}$ is even, one only
requires the value of $\eps(\delta)$ modulo 1/2.}
\be
\label{multeta}
\eta\left(\frac{a\tau+b}{c\tau+d}\right)\slash \eta(\tau)
=\expe{ \eps(\delta)}(c\tau+d)^{1/2}\, .
\ee
In particular, $24\eps(\delta)$ is integer, and may be obtained explicitly
as
\be
\label{defvareps}
\eps(\delta)=\left\{ \begin{array}{cc}
\frac{b}{24} \, {\rm sgn}(d)& (c=0)\, ,\\
\frac{a+d}{24 c}-\frac12\, s(d,c)-\frac18\, & (c > 0)\, ,\\
\end{array} \right.
\ee
where
\be
\label{Dedekindsum}
s(d,c)=\frac{1}{4|c|}\sum_{n=1}^{|c|-1}\cot\frac{\pi n}{c}\, \cot\frac{\pi n d}{c}
\ee
is the  Dedekind sum. The identification \eqref{rhode} is also
supported by the modular properties of the D4-D2-D0 partition
function \cite{deBoer:2006vg,Gaiotto:2006wm,Denef:2007vg,Manschot:2008zb,
Manschot:2009ia}, which
transforms with the multiplier system $\expe{-c_{2,a} p^a \eps(\delta)}$
~\cite{Denef:2007vg,Manschot:2008zb}.\footnote{We are grateful to J. Manschot for
discussions on this
issue.}

\subsection{Semi-classical $(p,k)$5-brane instantons\label{sec_pk}}

Having rectified the action of S-duality on the large volume, weak coupling HM moduli space,
we can now use it to obtain NS5-brane instantons from D-instantons. Indeed, from the
action on $(c_0,\psi)$ we see that S-duality maps D5-branes, with minimal
coupling to the D5 axion $c_0$, to NS5-branes coupled to the NS-axion $\psi$,
or more generally to $(p,k)$5-branes coupled to $p c_0+k\psi$. In particular,
since $c_0=\tzeta_0+\dots$ has unit periodicity and $\psi=-\frac12\, \sigma+\dots$,
we conclude that the NS-axion $\sigma$ must have periodicity 2, as anticipated
in Section \ref{sec_lgt}.

Let us now consider a configuration of D5-D3-D1-D(-1) instantons with charges $(p^0\neq 0$,$p^{a},q_a,q_0)$,
discussed in detail in the previous subsection.
The classical action associated to this charge configuration is
\be
\label{SD}
S_{\gamma} = 4\pi  | W_\gamma | + 2\pi\I ( q'_\Lambda \zeta^\Lambda - p^\Lambda
\tzeta'_\Lambda),
\ee
where
\be
\label{defWg}
W_\gamma = \frac{\tau_2}{2} \left( q'_\Lambda z^\Lambda - p^\Lambda
F'_\Lambda(z)\right)
\ee
is proportional to the  central charge $Z_\gamma=e^{\cK/2}
(q'_\Lambda X^\Lambda - p^\Lambda F'_\Lambda)$ of the D-instanton.
In the large volume limit where the cubic term in \eqref{lve} dominates,
\be
W_\gamma =\frac{\tau_2}{2}\left(  \frac{N( p^a - p^0 z^a)}{(p^0)^2} - \frac{\hat q_a(p^a-p^0 z^a)}{p^0}
+ \hat q_0\right)  .
\ee
For any two integers $(p,k)\neq (0,0)$, with greatest common divisor (gcd) $p^0$,
we now apply an S-duality transformation
\be
\label{Sdualde}
\delta = \begin{pmatrix} a & b \\ - k/p^0 & p/p^0 \end{pmatrix} \in SL(2,\IZ)\, ,
\ee
where the integers $(a,b)$, ambiguous up to the addition of $(k/p^0,-p/p^0)$,
are chosen such that $a p + b k = p^0$. This will map the
D5-brane configuration into a $(p,k)$5-brane.

Using \eqref{SL2Z}, one finds that the image of $W_\gamma$ under the map \eqref{Sdualde} is given by
\be
W_{k,p,\hgam}\equiv \delta\cdot W_\gamma=
\frac{\tau_2}{2|p-k\tau|^2} \[N\( \tilde p^a- i |p-k\tau| t^a \)- p^0 \hat q_a
\( \tlp^a- \I |p-k\tau| t^a \) + (p^0)^2 \hat q_0\],
\label{Wkg}
\ee
where we recall that.
As a result, the action \eqref{SD} is mapped to $S_{k,p,\hgam}$ which reads as follows
\beq
\label{ImSpk}
S_{k,p,\hgam}& =  &
4\pi \left| W_{k,p,\hgam} \right|+2\pi\I\biggl[
-(p c_0 + k \psi + p^a c_a)
+ \frac{p-k \tau_1 }{k |p-k\tau|^2} \left(N(\tilde p^a)-p^0\hat q_a \tlp^a+ (p^0)^2 \hat q_0  \right)
\biggr.
\\
&& + \frac{b^a}{2k}\(\frac13\,  \kappa_{abc} ( p b^b-k c^b) (3 \tilde p^c+p b^c - k c^c)
+\kappa_{abc}  \tilde p^b \tilde p^c -2p^0\hat q_a\)
 \biggl.
 -\frac{a }{k}\, q'_0 p^0 + c_{2,a}p^a \epsilon(\delta)\biggr] .\quad
\nonumber
\eeq
It should be stressed that the unusual additional terms in the imaginary part of $S_{k,p,\hgam}$ follow from the
ordinary axionic couplings in \eqref{SD}
by S-duality, and evade the no-go result of \cite{Chiodaroli:2009cz}. This is eventually
tied to the impossibility of defining instanton vacua with well defined mutually non-local D-instanton charges
in the presence of NS5-brane instantons.
Notice that the last two moduli-independent terms in \eqref{ImSpk} are the only ones depending explicitly
on the upper entries in the S-duality matrix \eqref{Sdualde}.
What is important however is that their combination is independent of
the choice of these entries so that the action is well defined.
Although some terms are singular at $k=0$, but the total sum
in \eqref{ImSpk} is regular at $k=0$, where it reduces to \eqref{SD}.
Moreover, for $\hat q_a=\hat q_0=0$, the action reduces to the answer found in
Eq. (3.72) of \cite{Pioline:2009qt}, based on the assumption that the S-duality group
is extended to $SL(3,\IZ)$.

From \eqref{ImSpk}, it is now apparent that $\psi$ is periodic with period 1, and therefore
$\sigma$ is periodic with period 2. Moreover, in the weak coupling limit
$\tau_2\to\infty$ and using the form of the period matrix $\cN$ for a cubic prepotential,
\be
\cN_{\Lambda\Sigma}=
\begin{pmatrix} -\frac13 (b b b) &\ \frac12\, (b b)_a \\
\frac12\, (b b)_a &\ - \kappa_{abc} b^c\end{pmatrix}
+\I
\begin{pmatrix} -\frac16 \, (t t t)
+ ( b b t) -\frac{1}{4V}\, (b t t)^2
&\ -(b t)_a+\frac{1}{4V}\, (b t t)(t t)_a
\\
-(b t)_a+\frac{1}{4V}\, (b t t)(t t)_a
&\ \kappa_{abc} t^c-\frac{1}{4V}\, (t t)_a(t t)_b\end{pmatrix},
\ee
where we introduced the notation $(xyz)=\kappa_{abc} x^a y^b z^c$, $(xy)_a=\kappa_{abc}x^b y^c$,
we find that the action for $k$ NS5-branes reduces to
\be
S_{k,p,\hgam} =2\pi |k| V \tau_2^2 +\pi\I k \left(  \sigma + \zeta^\Lambda \tzeta'_\Lambda
-2 n^\Lambda \tzeta'_\Lambda
- \bar\cN_{\Lambda\Sigma} (\zeta^\Lambda-n^\Lambda) (\zeta^\Sigma-n^\Sigma)\right)
-2\pi \I m_\Lambda z^\Lambda.
\label{NS5instact}
\ee
Here we defined
\be
n^0=p/k,
\qquad
n^a=p^a/k,
\qquad
m_a=p^0\hat q_a/k,
\qquad
m_0=a p^0 q'_0/k- c_{2,a}p^a \epsilon(\delta).
\label{defcarge}
\ee
For $m_\Lambda=0$ this reproduces \eqref{SdV}, and thus confirms the validity of \eqref{ns5quali}.

\section{NS5-instantons in twistor space \label{sec_ns5twi}}

In the previous sections we discussed instanton corrections to the HM moduli
space $\cM$ of type II string theory on a Calabi-Yau threefold at a qualitative level, ignoring
the tensorial nature of the metric on $\cM$ and jettisoning the constraints of supersymmetry.
The latter however require that all quantum corrections preserve the \qk property of the
metric on $\cM$  \cite{Bagger:1983tt}.
A convenient way to ensure this is to formulate the quantum corrections as deformations
of the complex contact structure on the twistor space $\cZ_\cM$,
a complex contact manifold locally of the form $\CP\times \cM$.

In this section, we take steps towards identifying the deformation of the complex contact structure
on $\cZ_\cM$ induced by NS5-brane instantons, by applying S-duality to the known twistorial
description of D-instantons. We work in the one-instanton approximation, i.e. treat
instanton corrections as an infinitesimal deformations of the complex contact structure on
$\cZ_{\rm pert}$, the twistor space of the perturbative HM moduli space discussed
in Section \ref{sec_pert}. Infinitesimal deformations are encoded in a holomorphic section
of $H^{1}(\cZ_{\rm pert},\cO(2))$. In practice it is most convenient to represent this section as
a set of local holomorphic functions defined
on the overlap of two patches in a suitable open covering of $\CP$. In general, this
set need only be invariant  under discrete symmetries up to local contact transformations.
Here we assume that this set is in fact simultaneously invariant under S-duality,
Heisenberg invariance and monodromy transformations, without the need for any compensating
contact transformation. This is the simplest approach to constructing a metric
with the required discrete symmetries. While we find strong indications that
this assumption is reasonable, we do encounter difficulties in realizing
all symmetries simultaneously, due to certain phases
spoiling exact invariance. As a result, Sections \ref{sec_poinca} onward
should be considered as exploratory, though we do believe that the
structure we uncover should be realized in a fully consistent treatment.

The outline of this section is as follows. $\bullet$  In Section \ref{sec_twitech}, we briefly recall
the twistorial description of general \qk manifolds,
following  \cite{Alexandrov:2008nk,Alexandrov:2008gh}.
$\bullet$  In Section \ref{sec_twimh}, we summarize the twistorial description of D-instanton
corrections established in \cite{Alexandrov:2008gh}, taking into account the
fractional shift \eqref{shze} and quadratic refinement which had been overlooked
in previous treatments. $\bullet$  In Section \ref{sec_poinca}, we obtain the twistorial
description of $(p,k)$5-brane instantons in type IIB string theory compactified on
$\hat\cX$, by applying S-duality on a general D5-D3-D1-D(-1) configuration.
We show that the corresponding deformation of the contact structure is
invariant under the Heisenberg shifts \eqref{heisext} and monodromy
around infinity \eqref{bjacr} (or rather their holomorphic lifts \eqref{bjac}, \eqref{heisalgz}
to the twistor space $\cZ_{\rm pert}$), up to subtle phases
which we do not understand.
$\bullet$
Under this caveat, we show in Section \ref{sec_ns5top} that
the total contribution of fixed NS5-brane charge $k$ can be expressed as a non-Gaussian theta
series with wave function $H_{k,\mu}$.
For a single NS5-brane, $k=1$, $H_{1,0}$ is recognized (up to certain prefactors)
as the A-model topological string wave function on $\hat\cX$, in the real polarization.
$\bullet$ Finally, in Section \ref{sec_nonlin}
we obtain the non-Gaussian chiral partition function $\ZNS{k}$ for $k$ fivebranes as the Penrose transform
of the holomorphic function encoding NS5-brane corrections to the contact structure.
In the saddle point approximation it reproduces the non-linear $(p,k)$-fivebrane action from Section \ref{sec_pk},
whereas the weak coupling limit of $\ZNS{k}$ reduces to a Gaussian partition function $\ZNSG{k}$ similar to
(but distinct from) the Gaussian flux partition function $\ZG{k}$ discussed in Section \ref{sec_zns5}.

\subsection{Twistor techniques \label{sec_twitech}}

Recall that a manifold $\cM$ of real dimension $4n>4$ is \qk if its holonomy group lies in
$USp(n)\times USp(1)\subset SO(4n)$, where $USp(1)=SU(2)$ (in dimension 4, this condition
must be replaced by the vanishing of the self-dual part of the Weyl curvature).
While $\cM$ admits locally a triplet of
almost-complex structures $\vec I$ satisfying the quaternion algebra, these are not
integrable. The space $\cM$ is nevertheless amenable to complex analysis by considering
its twistor space $\cZ_\cM$, a $\CP$ bundle over $\cM$ which carries a canonical, integrable
complex structure, \kahler-Einstein metric as well as a complex contact structure.
To explain how this comes about,
let $t$ be a complex stereographic coordinate on $\CP$, and $\vec p$ be the $SU(2)$ part
of the Levi-Civita connection on $\cM$. Under  $SU(2)$ frame rotations, $\vec p$
transforms as a $SU(2)$ gauge potential, while $t$ is acted upon by fractional linear
transformations.
The complex contact structure on $\cZ_\cM$ is defined by the kernel of the one-form
\be
\label{defDt}
Dt \equiv \de t + p_+ - \I p_3 t + p_- t^2 \ ,\qquad p_\pm = -\frac12 (p_1\mp \I p_2)\, .
\ee
Locally, on a patch
$\cU_i\subset \cZ_\cM$ there exists
a function $\Phi\di{i}$ on $\cZ_\cM$, holomorphic along the $\CP$-fiber (i.e. independent
of $\bar t$) and defined up to an additive holomorphic function on $\cU_i$, such that
the product
\be
\label{contact}
\CX\ui{i} = 2\, e^{\Phi\di{i}} \frac{D\varpi}{\I\varpi}\,
\ee
is a locally holomorphic one-form (i.e. $\bar\pa$-closed). We refer to $\Phi\di{i}$
as the contact potential. $\Phi\di{i}$ determines a
\kahler potential \cite{Alexandrov:2008nk}
\be
\label{Knuflat}
K_{\cZ_\cM}\ui{i} = \log\frac{1+\varpi\bar \varpi}{|\varpi|}
+ \Re\Phi\di{i}(x^\mu,\varpi)\,
\ee
for the canonical \kahler-Einstein  metric on $\cZ_\cM$
\begin{equation}\label{Z-metric}
\de s^2_{\cZ_\cM}=
\frac{|D\varpi|^2}{(1+\varpi{\bar \varpi})^2}+\frac{\nu}{4}\,\de s_{\cal M}^2\,
\end{equation}
(here $\nu$ is a normalization constant which determines the scalar curvature
of $\cM$). Moreover, $\cZ_\cM$ is endowed with a real structure acting as the
antipodal map on $\CP$, and preserving $\CX$.

According to theorems by Salamon and Lebrun \cite{MR664330,MR1001707},
the complex contact structure
and real structure on $\cZ_\cM$ uniquely specify the \qk metric on $\cM$. Locally,
one may always choose complex  Darboux coordinates
$\xii{i}^\Lambda, \txii{i}_\Lambda, \ai{i}$ ($\Lambda=0,\dots,n-1$) on $\cU_i\subset \cZ_\cM$
such that the contact one-form $\CX$ takes the Darboux form
\be
\label{contf}
\CX\ui{i}= \de\ai{i} + \xii{i}^\Lambda \, \de \txii{i}_\Lambda\, .
\ee
We shall find it convenient to define a variant of this coordinate system,
with $\tai{i}=-2\ai{i}- \xii{i}^\Lambda\txii{i}_\Lambda$
such that the contact one-form takes the symmetric Darboux form
\be
\label{contft}
{\cal X}^{[i]} = -\frac{1}{2}\left( \text{d} \tilde\alpha - \xi^{\Lambda}_{[i]}\,  \text{d} \tilde{\xi}^{[i]}_\Lambda + \tilde{\xi}^{[i]}_\Lambda\,  \text{d} \xi_{[i]}^{\Lambda} \right)\, .
\ee
On the overlap of two patches $\cU_i \cap \cU_j$, the Darboux coordinates are related
by complex contact transformations preserving $\CX\ui{i}$ up to holomorphic rescaling.
Thus, the complex contact structure on $\cZ_\cM$ may be specified globally by giving
generating functions $\hSij{ij}(\xii{i}^\Lambda,\txii{j}_\Lambda,\ai{j})$ of these
complex contact transformations. These must satisfy the obvious
reality conditions and compatibility constraints, and are themselves
defined up to independent complex contact transformations on each patch.

The \qk metric on $\cM$ can be obtained by solving the gluing conditions for the Darboux coordinates
on $\cU_i\cap \cU_j$ following from the generating functions $\hSij{ij}$,
\be\label{xitrafo}
\begin{split}
\xi\di{j}^\Lambda &=  \hf_{ij}^{-2} \, \pa_{\txii{j}_\Lambda} \hSij{ij}\, ,
\qquad\quad\quad\
\txii{i}_\Lambda = \pa_{\xii{i}^\Lambda} \hSij{ij}\, ,
\\
\ai{i} &= \hSij{ij} - \xii{i}^\Lambda \pa_{\xii{i}^\Lambda} \hSij{ij}\, ,
\qquad
e^{\Phi\di{i}} =  \hf_{ij}^{2} \,e^{\Phi\di{j}} \, ,
 \end{split}
\ee
where $\hf_{ij}^{2} = \pa_{\ai{j}} \hSij{ij}=\CX\ui{i}/\CX\ui{j}$ is the homogeneity
factor relating the contact one-forms. The parameter space $\{ x^\mu \}$ of solutions is $\cM$ itself,
while the solutions $\xi\di{i}^\Lambda(t; x^\mu)$, $\txi\ui{i}_\Lambda(t; x^\mu)$ and $\ai{i}(t; x^\mu)$
can be plugged in \eqref{contf} to read off the $SU(2)$ connection in  \eqref{defDt}. The
metric itself follows after some more steps explained in \cite{Alexandrov:2008nk}.

Furthermore, infinitesimal perturbations around a given \qk manifold
preserving the \qk structure are in one-to-one correspondence
with $H^1(\cZ_\cM,\cO(2))$  \cite{lebrun1994srp}.
They can be parametrized explicitly as a set of local holomorphic
functions $\hHpij{ij}$ on the overlap of two patches $\cU_i\cap \cU_j$,
corresponding to the variation of the quantity $\hHij{ij}$ entering in
the generating function $\hSij{ij}$ via \cite{Alexandrov:2008nk}
\be
\label{defH}
\hSij{ij}(\xii{i}^\Lambda,\txii{j}_\Lambda,\ai{j})
 = \ai{j}+ \xii{i}^\Lambda \,  \txii{j}_\Lambda -\Hij{ij}(\xii{i}^\Lambda,\txii{j}_\Lambda,\ai{j})\ .
\ee
The consistency conditions on $\hSij{ij}$ are equivalent to the cocycle conditions
on $\hHij{ij}$ at the linearized level. For perturbations around  a toric \qk
manifold, the variation of the Darboux coordinates and contact potential
can be obtained by certain contour integrals of $\hHij{ij}$, as explained
in \cite{Alexandrov:2008nk}, from which the
variation of the metric follows. While the integral formulae in  \cite{Alexandrov:2008nk}
are rather involved, they generalize the well-known Penrose integral formula
\be
\label{penroseint}
 \sum_j \int_{C_j} \frac{\de\varpi}{\varpi}\, e^{\Phi\ui{j}(t)}\,  \hHij{ij} \left(\xii{i}^\Lambda(t),
\txii{j}_\Lambda(t),\ai{j}(t)\right)
\ee
 which
maps a holomorphic section of $H^1(\cZ_\cM,\cO(-2))$, which we again represent
by a local holomorphic function $\hHij{ij}$, into a function  on $\cM$
which satisfies a certain set of second order differential equations
determined by the \qk structure on $\cM$ \cite{Neitzke:2007ke}.
In Section \ref{sec_nonlin}, we shall use \eqref{penroseint} to construct
a scalar-valued fivebrane partition function on $\cM$.

Finally, we recall that via the
moment map construction, continuous isometries of $\cM$ are in one-to-one correspondence
with real elements of $H^0(\cZ_\cM,\cO(2))$, i.e. with global holomorphic sections of the
line bundle $\cO(2)$ over $\cZ_\cM$, invariant under the antipodal map.
It follows that any continuous isometry of $\cM$
can be lifted to a holomorphic isometry of $\cZ_\cM$. Moreover, one can always
choose the local Darboux coordinates such that the holomorphic Killing
vector is $\pa_{\ai{i}}$ on each patch $\cU_i$. In that case,
the generating function $\hHij{ij}$ becomes independent of $\ai{i}$, the contact
one-form $\CX\ui{i}$
becomes globally defined, and $\hHij{ij}$ can be viewed as a generating
function for symplectomorphisms with respect to the complex symplectic form
$\de\CX$ on the quotient of $\cZ_\cM$ by the vector field $\pa_{\ai{i}}$. Thus, any
\qk manifold with a Killing vector can be mapped locally to a \hk manifold of the same
dimension.\footnote{We are grateful to A. Neitzke for discussions on this construction.}

\subsection{Twistorial description of $\cM$ \label{sec_twimh}}

The twistorial description of the HM moduli space was worked out at tree-level
in \cite{Neitzke:2007ke}, the one-loop correction was included in
\cite{Alexandrov:2007ec,Alexandrov:2008nk} and D-instantons
were incorporated in \cite{Alexandrov:2008gh,Alexandrov:2009zh}.
The structure of the D-instanton corrected twistor space is closely
related to that of the twistor space of the
\hk manifold governing the moduli space of $\cN=2$ gauge theories
on $\IR^3\times S^1$ \cite{Gaiotto:2008cd}, in line with the remark at
the end of the last subsection.

\subsubsection{Perturbative twistor space}

In the absence of non-perturbative effects, i.e. for the one-loop corrected
metric \eqref{hypmetone}, the twistor space may be described by three patches
$\cU_+,\cU_-,\cU_0$ around the north pole $(t=0)$, south pole $(t=\infty)$ and
equator of the $\CP$ fiber over a fixed point on $\cM$. The transition functions
from $\cU_0$ to $\cU_\pm$ are governed by the prepotential $F$ on $\cS\cK$,
viewed as function of Darboux coordinates $\xi^\Lambda$,
as explained in \cite{Alexandrov:2008nk,Alexandrov:2008gh}. The one-loop
correction is incorporated by a non-zero ``anomalous dimension", i.e. as a logarithmic
singularity around the north and south pole. In the patch $\cU_0$, the ``Darboux
coordinates" may be chosen as
\be
\label{gentwi}
\begin{array}{rcl}
\xi^\Lambda &=& \zeta^\Lambda + \frac{\tau_2}{2}
\left( \varpi^{-1} z^{\Lambda} -\varpi \,\bz^{\Lambda}  \right)\, ,
\\
\txi_\Lambda &=& \tzeta_\Lambda + \frac{\tau_2}{2}
\left( \varpi^{-1} F_\Lambda(z)-\varpi \,\bF_\Lambda(\bz) \right)\, ,
\\
\tilde\alpha&=& \sigma + \frac{\tau_2}{2}
\left(\varpi^{-1} W(z)-\varpi \,\bar W(\bz) \right) +\frac{\I\chi(\cX)}{24\pi} \,\log \varpi \, ,
\end{array}
\ee
where
$W(z) \equiv  F_\Lambda(z) \zeta^\Lambda - z^\Lambda \tzeta_\Lambda $.
As in \eqref{shze}, we define
\be
\label{defrhop}
\txi'_\Lambda=\txi_\Lambda-A_{\Lambda\Sigma}\xi^\Sigma\, ,
\qquad
\alpha'= \alpha +\frac{1}{2}\, A_{\Lambda\Sigma}\xi^\Lambda \xi^\Sigma\, ,
\ee
such that, for any charge vector $\gamma\in\Gamma$,
$q_\Lambda \xi^\Lambda-p^\Lambda \txi_\Lambda
=q'_\Lambda \xi^\Lambda-p^\Lambda \txi'_\Lambda$.

Under monodromies $M$ in $\cS\cK$, $(\xi^\Lambda,\txi_\Lambda)$ transforms
linearly under $\rho(M)$. On the other hand, $\tilde\alpha$ is shifted by $\kappa(M)$ from \eqref{sigmon},
whereas the middle term is canceling against the variation of the logarithmic term in \eqref{gentwi} under the
R-symmetry rotation $t\mapsto t \, e^{\I \Im f_M}$. In particular,
under a monodromy $b^a\mapsto b^a+\epsilon^a$ around the large volume point,
the Darboux coordinates transform as
\be
\label{bjac}
\begin{split}
\xi^0\mapsto & \xi^0\, ,
\qquad
\xi^a\mapsto \xi^a + \epsilon^a \xi^0\, ,
\qquad
\txi'_a\mapsto \txi'_a -\kappa_{abc}\xi^b \epsilon^c-\frac12\,\kappa_{abc} \epsilon^b \epsilon^c \xi^0\, ,
\\
&\txi'_0\mapsto \txi'_0 -\txi'_a \epsilon^a+\frac12\, \kappa_{abc}\xi^a \epsilon^b \epsilon^c
+\frac16\,\kappa_{abc} \epsilon^a \epsilon^b \epsilon^c \xi^0\, ,
\qquad
\tilde\alpha\mapsto\tilde\alpha+2\kappa_a\epsilon^a\, ,
\end{split}
\ee
where we used \eqref{gentwi}, \eqref{bjacr}, and took into account the fact that $\kappa(M)$
must be linear in $\epsilon^a$, $\kappa(M)=\kappa_a \epsilon^a$. The constants
$\kappa_a$ should be computable along the lines suggested in the discussion
below Eq. \eqref{ansig},
but we do not know their values at this stage.

Similarly, the Heisenberg action \eqref{heisext} lifts to a holomorphic action on $\cZ_\cM$ given by
\be
\label{heisalgz}
\begin{split}
T'_{(H,\kappa)}\ :  &\left(\xi^\Lambda,\ \txi_\Lambda,\ \tilde\alpha\right)\mapsto
\left(\xi^\Lambda + \eta^\Lambda ,\ \txi_\Lambda+ \tleta_\Lambda, \right.\\
&\left. \qquad \qquad
\tilde\alpha +2\kappa- \tleta_\Lambda \xi^\Lambda+ \eta^\Lambda \txi_\Lambda
- \left( \eta^\Lambda \tleta_\Lambda -2 \tleta_\Lambda\theta^\Lambda+2 \eta^\Lambda\phi_\Lambda\right)
\right),
\end{split}
\ee
where $\eta^\Lambda, \tleta_\Lambda,\kappa\in\IZ$. Thus, the quotient of $\cZ_\cM$ by
translations along $\pa_{\tilde\alpha}$ defines a complexified torus $\cT_\IC$, parametrized
by the coordinates ($\xi^\Lambda, \txi_\Lambda)$ and their complex conjugates,
while $e^{\I\pi  \tilde{\alpha}}$ parametrizes the fiber of a $\IC^\times$-bundle
$ \cL_\Theta^{\mathbb{C}}$ over $\cT_\IC$. The restriction of $ \cL_\Theta^{\mathbb{C}}$
to the intermediate Jacobian torus $\cT\subset\cT_\IC$,
at a fixed point on the special K\"ahler base $\cS\cK$,
coincides with the theta line bundle $\cL_\Theta$.
It is interesting to note that the two-form $\de\cX$ endows  $\cT_\IC$ with  a complex symplectic
structure, and therefore a natural \hk metric, as explained at the end of section 5.1.

\subsubsection{D-instanton corrected twistor space\label{sec_Dtwi}}

In the presence of D-instantons, the covering of $\cZ_\cM$ must be refined as
follows \cite{Alexandrov:2008gh,Gaiotto:2008cd}.
Over a fixed point in $\cM$, the
$\CP$ fiber of $\cZ_\cM$ is divided into angular sectors extending between two
BPS rays
\be
\ell_\gamma=\{ t:\  Z_\gamma(z^a)/t\in \I\IR^-\}.
\label{lgam}
\ee
Across such
a BPS ray, the Darboux coordinates $(\xi^\Lambda,\txi_\Lambda)$ are related
by a complex symplectomorphism generated by\footnote{Eq. \eqref{lgam}
holds in the one-instanton approximation only, and must be
amended to take into account the difference between Darboux coordinates in different
angular sectors. The exact generating function
was derived \cite{Alexandrov:2009zh}, and reproduces
the symplectomorphism appearing in \cite{Gaiotto:2008cd}, up to the quadratic refinement
$\sigma_{\text{D}}(\gamma)$ which must be added by hand.}
\be
\label{elcon}
H_\gamma
 = \frac{\hnkl }{(2\pi)^2}\, \Li_2\left[\sigma_{\text{D}}(\gamma) \, \expe{- \Xi_\gamma}  \right]\, ,
\ee
where $\Xi_\gamma= q_\Lambda\xi^\Lambda-p^\Lambda \txi_\Lambda$, the factor
$\hnkl$ is the generalized Donaldson-Thomas (DT) invariant for the charge
vector $\gamma$, $\Li_2(x)\equiv\sum_{n=1}^{\infty} x^n/n^2$ is the dilogarithm
function, taking care of multi-covering effects,
and $\sigma_{\text{D}}(\gamma)$ is a quadratic refinement \eqref{quadraticrefinementpq}
of the intersection form on the charge lattice $\Gamma$, which plays a crucial role in ensuring consistency
across walls of marginal stability  \cite{Gaiotto:2008cd}.  As discussed in Section \ref{subsec_topolinst},
although $\sigma_{\text{D}}(\gamma)$ should be related to the quadratic refinement which governs the
fivebrane partition function and the NS-axion, and moreover they are expected to coincide,
we refrain from identifying them at this point.

The gluing conditions \eqref{xitrafo} for the Darboux coordinates $\xi^\Lambda,\txi_\Lambda$
can be summarized by the following integral equation
\be
\Xi_\gamma = \Theta_\gamma +
 W_\gamma/t - \bar W_\gamma t -\frac{1}{8\pi^2}
\sum_{\gamma'} \Omega(\gamma')\, \langle \gamma,\gamma'\rangle
\int_{l_{\gamma'} }\frac{\de t'}{t'}\, \frac{t+t'}{t-t'}\,
\Li_1\left[\sigma_{\text{D}}(\gamma') \, \expe{- \Xi_{\gamma'}(t')}  \right] ,
\label{eqTBA}
\ee
where $\Theta_\gamma= q_\Lambda\zeta^\Lambda-p^\Lambda \tzeta_\Lambda$,
and we recall that $\Li_1(x)=-\log(1-x)=\pa_x \Li_2(x)$.
It is worth noting that Eq. \eqref{eqTBA} is
identical to the integral equation found in  \cite{Gaiotto:2008cd} in the context
of \hk geometry. Moreover,
it has the form of a Thermodynamic Bethe
Ansatz equation \cite{Gaiotto:2008cd} with an $S$-matrix satisfying all axioms of integrable field theories
\cite{Alexandrov:2010pp}, hinting at a possible integrable structure.

Once the Darboux coordinates $(\xi^\Lambda,\txi_\Lambda)$ have been
determined, the Darboux coordinate $\tilde\alpha$ follows from the
contour integral
\beq
\tilde\alpha&=& \sigma
+\frac{\tau_2}{2} \(\varpi^{-1} W-\varpi \,\bar W\) +\frac{\I\chi(\cX)}{24\pi} \log \varpi
-\frac{1}{4\pi^2}\sum_\gamma \Omega(\gamma)\(\varpi^{-1}\Wg{}+\varpi\bWg{} \)\Igg{}(0)
\nn\\
&& + \frac{1}{8\pi^2}\sum_\gamma \Omega(\gamma) \[\frac{\I}{\pi }
\int_{\ellg{\gamma}}\frac{\d \varpi'}{\varpi'}\,
\frac{\varpi+\varpi'}{\varpi-\varpi'}\,
\Li_2\[\sigma_{\text{D}}(\gamma) \expe{-\Xigi{}(\varpi')}\]
+\(\Thg{}+\varpi^{-1}\Wg{}-\varpi\bWg{}\)\Igg{}(\varpi) \]
\nn \\
&& -\frac{1}{64\pi^2}\sum_{\gamma\ne \gamma'} \Omega(\gamma)\Omega(\gamma')\langle \gamma,\gamma'\rangle
\int_{\ellg{\gamma}}\frac{\d \varpi'}{\varpi'}\,
\frac{\varpi+\varpi'}{\varpi-\varpi'}\,
\Li_1\left[\sigma_{\text{D}}(\gamma) \, \expe{- \Xigi{}(\varpi')}  \right]
\cJ_{\gamma'}(\varpi'),
\label{manyalpha}
\eeq
where
\be
\Igg{}(\varpi)=-\int_{\ellg{\gamma}}\frac{\d \varpi'}{\varpi'}\,
\frac{\varpi+\varpi'}{\varpi-\varpi'}\,
\Li_1\left[\sigma_{\text{D}}(\gamma) \, \expe{- \Xigi{}(\varpi')}  \right].
\label{newfun}
\ee
The contact potential is similarly given by
\be
e^{\Phi} = \frac{\tau_2^2}{8}\, e^{-\cK(z,\bz)} +\frac{\chi(\cX)}{96\pi}
+\frac{ \I}{16\pi^2}\sum\limits_{\gamma} \Omega(\gamma)
\int_{\ellg{\gamma}}\frac{\d \varpi}{\varpi}\,
\( \varpi^{-1}\Wg{} -\varpi\bWg{}\)
\Li_1\left[\sigma_{\text{D}}(\gamma) \, \expe{- \Xigi{}(\varpi)}  \right].
\label{phiinstmany}
\ee
Note that it is constant on $\CP$, as required by the continuous shift isometry along $\sigma$,
and may be chosen to define the ``instanton corrected 4D string coupling" $e^\phi=e^\Phi$.

Eqs. \eqref{eqTBA} (and subsequently \eqref{manyalpha}, \eqref{phiinstmany})
may be solved recursively by first
substituting $\Xi_\gamma(t)$ on the r.h.s. by its perturbative value $\Xi_\gamma^{(0)}(t)=
\Theta_\gamma + W_\gamma/t - \bar W_\gamma t$, computing the integral
to obtain the one-instanton correction $\Xi_\gamma^{(1)}$ and iterating this procedure. This provides
an asymptotic series for the exact, D-instanton corrected Darboux coordinates,
\be
\Xi_\gamma(t) = \Theta_\gamma + W_\gamma/t - \bar W_\gamma t
+ \sum_{N=1}^{\infty} \Xi_\gamma^{(N)}\ ,
\ee
where $\Xi_\gamma^{(N)}$ involves $N$ nested contour integrals of sums of
products of $N$ DT-invariants $\Omega(\gamma_k)$, $k=1,\dots,N$. This
can be interpreted as corrections from multi-centered instanton configurations,
with $\Omega(\gamma_k)$ providing the contribution to the instanton measure  of
the $k$th center.  In the vector multiplet context, these are instead
multi-centered Euclidean black holes whose worldline winds around
the Euclidean time direction, and $\Omega(\gamma_k)$ is instead the
indexed degeneracy of the $k$th center. At fixed $N$, the sum over $\gamma_{k}$
typically has zero radius of convergence, due to the expected  exponential growth
of the generalized DT-invariants. It was argued in \cite{Pioline:2009ia} that the ambiguity of this
asymptotic series is of the same order as the corrections expected from NS5-branes.

Noting that the BPS rays $\ell_\gamma$ and $\ell_{n\gamma}$ are identical,
one may dispose of the dilogarithm in above formulas by replacing the integer-valued DT
invariants $  \hnkl  $ by the rational-valued DT invariants\footnote{The
rational invariants $\thnkl$ are also relevant for S-duality and
wall-crossing  \cite{Manschot:2010xp,MPS-in-progress}.}
\be
\thnkl   = \sum_{d\vert\gamma} \frac{1}{d^2} \, \Omega(\gamma/d)\, ,
\qquad
\hnkl   = \sum_{d\vert\gamma} \frac{1}{d^2}\, \mu(d)  \, \overline{\Omega}(\gamma/d),
\label{rationalinvariants}
\ee
where $\mu(d)$ is the M\"obius function.
In particular, one finds that the D-instanton corrected twistor space can be equally obtained
from the above construction where the holomorphic functions \eqref{elcon} are replaced by
\be
\label{HDinstp}
\tilde H_{\gamma}=\frac{\sigma_{\text{D}}(\gamma)}{(2\pi)^2}\, \thnkl     \,
\expe{p^\Lambda\txi'_\Lambda-q'_\Lambda \xi^\Lambda} \, .
\ee
Here we expressed the result in terms of the
primed charges $q'_\Lambda$ and Darboux coordinates $\txi'_\Lambda$
and used the fact that $\sigma_{\text{D}}(d\gamma)=\left(\sigma_{\text{D}}(\gamma)\right)^d$ for any positive
integer $d$.
As an illustration, the contact
potential in the one-instanton approximation is found to be
\cite{Alexandrov:2008gh}
\be
\begin{split}
e^{\Phi}
=&\frac{\tau_2^2}{8}\, e^{-\cK(z,\bz)} +\frac{\chi(\cX)}{96\pi}
+\frac{1}{4\pi^2}\sum\limits_{\gamma\in\Gamma}\sigma_{\text{D}}(\gamma)\, \thnkl
\, |\Wkl|\,   \cos\(2\pi\Thkl\)
K_1(4\pi|\Wkl|)+\dots\, .
\label{phiinstfull}
\end{split}
\ee
In this approximation, $e^\Phi$ will in general be discontinuous across
lines of marginal stability in $\cS\cK$, where $\thnkl$ may jump. However, this discontinuity
is canceled by the two- and higher-instanton contributions, provided the
jump in $\thnkl$ obeys the Kontsevich-Soibelman wall-crossing formula \cite{Gaiotto:2008cd}.

\subsubsection{S-duality in twistor space}

As mentioned in Section \ref{sec_mir}, in the infinite volume and zero coupling limit,
$\cQ_K(\hat \cX)$ admits an isometric action of $SL(2,\IR)$ given in \eqref{SL2Z}. This action
lifts to a holomorphic action on $\cZ_\cM$, given in terms of the Darboux coordinates
in the patch $\cU_0$ as \cite{Alexandrov:2008gh}
\be
\label{SL2Zxi}
\begin{split}
&
\xi^0 \mapsto \frac{a \xi^0 +b}{c \xi^0 + d} \, , \qquad
\xi^a \mapsto \frac{\xi^a}{c\xi^0+d} \, , \qquad
\txi'_a \mapsto \txi'_a +  \frac{c}{2(c \xi^0+d)} \kappa_{abc} \xi^b \xi^c- c_{2,a}\eps(\delta)\, ,
\\
&
\begin{pmatrix} \txi'_0 \\ \alpha' \end{pmatrix} \mapsto
\begin{pmatrix} d & -c \\ -b & a  \end{pmatrix}
\begin{pmatrix} \txi'_0 \\  \alpha' \end{pmatrix}
+ \frac{1}{6}\, \kappa_{abc} \xi^a\xi^b\xi^c
\begin{pmatrix}
c^2/(c \xi^0+d)\\
-[ c^2 (a\xi^0 + b)+2 c] / (c \xi^0+d)^2
\end{pmatrix}\, .
\end{split}
\ee
Indeed, substituting the Darboux coordinates \eqref{gentwi} in \eqref{SL2Zxi}
and using the classical mirror map \eqref{symptob}, one recovers the isometric action \eqref{SL2Z},
supplemented with the following $SU(2)$ action on the $\CP$ fiber,
\be
\label{SL2varpi}
\varpi \mapsto  \frac{c \tau_2 + \varpi  (c \tau_1 + d) +
\varpi |c \tau + d| }{(c \tau_1 + d) + |c \tau + d| - \varpi c \tau_2}\, .
\ee
Under the holomorphic action \eqref{SL2Zxi},  the complex
contact one-form transforms by an overall holomorphic factor
$\hCX\ui{i}\to \hCX\ui{i}/(c\xi^0+d)$, leaving the complex
contact structure invariant. Moreover, the relations
\be
\label{SL2phi}
e^\Phi \mapsto \frac{e^\Phi}{|c\tau+d|}\, ,\qquad
\frac{|\varpi|}{1+|\varpi|^2} \mapsto \frac{|\varpi|}{1+|\varpi|^2} \frac{|c \xi^0+d|}{|c\tau+d|}
\ee
ensure that the \kahler potential varies by a \kahler transformation, consistent
with the rescaling of $\hCX\ui{i}$,
\be
K_\cZ\mapsto  K_\cZ - \log(|c\xi^0+d| )\, .
\ee
Note that $e^\Phi$ transforms like $\tau_2^{1/2}$, i.e. has modular weight $(-\frac12,-\frac12)$.
The character $c_{2,a}\eps(\delta)$ in the transformation of $\tilde\xi'_a$ follows
from the shift of the real coordinate $c_a$ in \eqref{SL2Z}. It may be interpreted geometrically
by saying that $\expe{p^a \txi_a^{\, \prime}}$ transforms like the automorphy factor of a multi-variable
Jacobi form of index $m_{ab}=\frac12 \kappa_{abc} p^c$ and with multiplier system
$\expe{-c_{2a} p^a \eps(\delta)}$.

As discussed  in Section \ref{sec_mir}, worldsheet  instantons and the one-loop correction
break continuous S-duality, but an $SL(2,\IZ)$ subgroup can be preserved upon
including D1-D(-1)-brane instantons, as shown in \cite{RoblesLlana:2006is}.
The corrections to the Darboux coordinates from such instanton
configurations were computed  in \cite{Alexandrov:2009qq}. It was found
that the Darboux coordinates continue to
transform under  the tree-level transformation rules \eqref{SL2Zxi}, up to a local
contact transformation. It is possible to construct new Darboux coordinates which
transform precisely according to \eqref{SL2Zxi} (those were called $\xi^{[0_B]}$
in \cite{Alexandrov:2009qq}), although these coordinates
will only be valid in a certain open set in $\IC P^1$. If S-duality is indeed maintained at the
quantum level, this should remain true in the presence of D3-brane instantons, and ultimately,
in the presence of D5 and NS5-brane instantons. In the following,
we shall be using such adapted Darboux coordinates, though their explicit
construction in the presence of D3-D1-D(-1) instantons  is still an open problem
\cite{amp-in-progress}.

\subsection{Contact structure for fivebrane instantons\label{sec_poinca}}

In this section, we use the qualitative insights gained in the previous sections
and the twistor techniques reviewed above
to determine the form of NS5-brane instanton corrections to the HM
moduli space $\cM$, consistently with supersymmetry and S-duality.
Our aim is to determine the data governing the deformation of the complex contact structure on $\cZ_\cM$
which encode the NS5-brane corrections. Such data are provided by contours $\ell_{k,p,\hgam}$
and the associated holomorphic sections $H_{k,p,\hgam}\in H^1(\cZ_{\rm pert},\cO(2))$ describing discontinuities of
complex Darboux coordinates across these contours.
To find them in the one-instanton approximation,
we covariantize under S-duality the data responsible for the D-instanton corrections, which
are given by the holomorphic sections \eqref{HDinstp} and the contours $\ell_\gamma$ \eqref{lgam}.
Performing this covariantization, we assume that S-duality requires the invariance of all these data,
namely, that every pair $\(H_{k,p,\hgam},\ell_{k,p,\hgam}\)$ is mapped to another one so that the total structure
of the twistor space remains intact.
As apparent from the analysis of the D1-D(-1)-instanton sector
\cite{Alexandrov:2009qq}, this assumption is strictly speaking unwarranted, since S-duality
can produce contributions which can be canceled by local contact transformations. Moreover,
it leads to a dense set of mutually intersecting discontinuities on the $\CP$ fiber,
whose mathematical status is questionable.
However, we shall see that it does lead to an appealing result which matches general
expectations about the fivebrane contributions. We therefore believe that the main features
of our approach should subsist in a more sophisticated treatment.

Thus, we start with the holomorphic sections
$\tilde H_{\gamma}\in H^1(\cZ_{\rm pert},\cO(2))$ given in
\eqref{HDinstp}, which describe the D-instanton corrections to the perturbative
metric \eqref{hypmetone} in the one-instanton approximation.
In the absence of D5-brane instanton corrections $p^0=0$, and for a fixed
D3-brane charge $p^a$, the corrections to the moduli space metric
from $\tilde H_{\gamma}$ should preserve the isometric action of $SL(2,\IZ)$
(or a finite index subgroup thereof).
On the other hand, the instanton corrections with $p^0\neq 0$ break S-duality completely,
unless they are supplemented by additional contributions from NS5-branes, or more
generally $(p,k)$5-branes.
It is natural to propose that S-duality is restored by refining the contact structure on $\cZ_\cM$
by adding all images of $(\tilde H_{\gamma},\ell_{\gamma})$ under $\Gamma_{\infty}\backslash SL(2,\IZ)$,
where $\Gamma_{\infty}$ is the subgroup of strictly upper triangular matrices,
to the data describing discontinuities of Darboux coordinates.
The coset $\Gamma_{\infty}\backslash  SL(2,\IZ)$ can be parametrized by
matrices
\be
\label{Sdualde2}
\delta = \begin{pmatrix} a & b \\ c & d \end{pmatrix} \, ,
\ee
where $(c,d)$ run over coprime integers, and the integers $(a,b)$
are functions of $(c,d)$, determined up to multiples of $(c,d)$ by the condition
$ad-bc=1$.
Thus, we propose that in the one-instanton approximation the contact structure is described
by the following data
\be
\label{HtotD}
H_{k,p,\hgam} = \delta\cdot \tilde H_{\gamma}  \, ,
\qquad\qquad
\ell_{k,p,\hgam}=\delta\cdot \ell_{\gamma}\, ,
\ee
where $\delta\cdot \tilde H_{\gamma}$ and $ \delta\cdot \ell_{\gamma}$
denotes the image of \eqref{HDinstp} and \eqref{lgam}
under the $SL(2,\IZ)$ transformation \eqref{Sdualde2}
with $(c,d)=(-k/p^0,p/p^0)$ as in \eqref{Sdualde}, and we recall that
$\hat\gamma=(p^a, \hat q_a,\hat q_0)$.

We expect that $H_{k,p,\hgam}$ represents the effect on the contact structure of a $(p,k)5$-brane.
Using \eqref{SL2Zxi}, it evaluates to
\be
\label{5pqZ2}
H_{k,p,\hgam}=\frac{\sigma_{\text{D}}(\gamma) }{4\pi^2}\,
\thnkl  \,
\expe{-\frac{k}{2}\,  S_\alpha+ \frac{p^0\( k \hat q_a (\xi^a-n^a) + p^0 \hat q_0\)}{k^2(\xi^0-n^0)}
+ a\,\frac{p^0 q'_0}{k}- c_{2,a} p^a\eps(\delta)} ,
\ee
where
\be
\label{defSa}
S_\alpha\equiv \tilde\alpha + (\xi^\Lambda-2n^\Lambda)\txi'_\Lambda
+2 \frac{N(\xi^a-n^a)}{\xi^0-n^0}\, ,
\ee
$p^0$ is defined as $\gcd(p,k)$, $\gamma=(p^0,p^a,q_a,q_0)\in\Gamma$,
and we have denoted $n^0\equiv p/k$, $n^a\equiv p^a/k$, valued in $\IZ/k$.
On the other hand, the transformation \eqref{SL2varpi} can be used to find the contour
$\ell_{k,p,\hgam}$. It is easily seen to be a half-circle stretching between
the two zeros of $\xi^0-n^0$,
which, as will become clear below in \eqref{exprexp}, are the only singularities
of $H_{k,p,\hgam}$.
The direction of the contour is determined by the same condition as \eqref{lgam}
with $Z_\gamma$ replaced by $W_{k,p,\hgam}$, which was introduced in \eqref{Wkg}.
As a result, the complete set of contours is given by an infinite number of copies of
the ``melon-shaped" picture for D-instantons \cite{Gaiotto:2008cd,Alexandrov:2008gh},
rotated into each other by $SL(2,\IZ)$ transformations. For fixed $(p,k)$, the
new BPS rays extend between two antipodal points whose location
is completely determined by a sole rational number $p/k$.

A crucial requirement on the new contact structure is that
it should be invariant under the Heisenberg group \eqref{heisalgz}.
The invariance with respect to $T'_{(0,\tilde\eta),\kappa}$
 follows trivially from the invariance of the holomorphic
functions \eqref{5pqZ2}, provided the fivebrane characteristics $\theta^\Lambda$ are set to zero.
In contrast,  $H_{k,p,\hgam}$ are not individually invariant
under $T'_{(\eta,0),0}$. To be nevertheless consistent with
the Heisenberg symmetry, the set of functions \eqref{5pqZ2}
should be globally invariant under  $T'_{(\eta,0),0}$, i.e. $H_{k,p,\hgam}$ should be
mapped to $H_{k,p',\hgam'}$ for a suitably chosen map $(p,\hgam)\mapsto (p',\hgam')$,
with the contours $\ell_{k,p,\hgam}$ following the same pattern.

As we demonstrate in Appendix \ref{apsec_Heis}, under the crucial assumption that
$\thnkl$ are invariant under the spectral flow \eqref{flow},
the combined action of the Heisenberg symmetry
together with a shift $p\mapsto p+k\eta^0$ and the spectral flow action \eqref{flow}
with parameter $\epsilon^a=k \eta^a/p^0$  on the charges $p^a,q_a,q_0$
leaves the set of  functions \eqref{5pqZ2} globally invariant, up to a phase factor
$\nu(\eta)$ computed in \eqref{ratS},
\be
T'_{(\eta^\Lambda,0,0)}\cdot
H_{k,p+k \eta^0,\hgam[k\eta^a/p^0]}
=\nu(\eta)\, H_{k,p,\hgam} ,
\label{invH_heis}
\ee
where $\hgam[\epsilon]=(p^a+\epsilon^a,\hat q_a,\hat q_0)$.
Moreover, the contours $\ell_{k,p,\hgam}$ stay invariant under this combined transformation.
Indeed, their endpoints, being the zeros of $\xi^0-n^0$
are explicitly invariant and the  direction of approach is also conserved
by virtue of  the invariance of $W_{k,p,\hgam}$ \eqref{Wkg}.
Under the same assumptions, $H_{k,p,\hgam}$ are also invariant,
 up to a phase factor $\nu'(\eta)$ computed in \eqref{totnup},
under the combined action of monodromies
\eqref{bjac} and of the spectral flow \eqref{flow} with parameter $\epsilon'^a\equiv (p/p^0)\epsilon^a$,
\be
M_{\epsilon^a}\cdot
H_{k,p,\hgam[p\epsilon/p^0]}
=\nu'(\epsilon) \, H_{k,p,\hgam},
\label{invH_mon}
\ee
where $M_{\epsilon^a}$ is the monodromy operator acting on complex Darboux coordinates as in \eqref{bjac}.

Unfortunately, while the phase $\nu(\eta)$ can be set to one in the $k=1$ case
by an {\it ad hoc} choice of the characteristics, namely
$\thetad{\Lambda}=0$, $\phi_0=A_{00}/2$, $\phi_a-\phid{a}=A_{aa}$, this is
not possible for general values of $k$, and it does not lead to a cancelation of
the phase $\nu'(\epsilon)$.
Moreover, these conditions  are
not compatible with the transformation rule \eqref{sympchar} of the characteristics
under monodromies, and therefore should probably not be taken seriously.
This tension between S-duality, Heisenberg  and monodromy invariance
indicates that it may be necessary to relax some of our assumptions about the way these
symmetries are realized. In this paper, we shall ignore this difficulty
and proceed with the analysis as though the phases $\nu(\eta)$ and $\nu'(\epsilon)$ were absent, leaving
a more complete analysis for future work.

\subsection{Poincar\'e series for NS5-instantons and the topological string amplitude\label{sec_ns5top}}

It is instructive to  construct a formal section of $H^1(\cZ_{\rm pert},\cO(2))$ given by a sum of
the holomorphic functions \eqref{5pqZ2} over the integer charges $p,p^a$ and $q_\Lambda$:
\be
H_{{\rm NS5}}^{(k)}(\xi,\txi,\tilde\alpha)=\sum_{p,p^a, q_\Lambda} H_{k,p,\hgam}(\xi,\txi,\tilde\alpha).
\label{fullH}
\ee
This sum is formal since each term is attached to a different contour on $\CP$.
Nevertheless, it can be meaningfully inserted into general formulas for Darboux
coordinates and the contact potential from \cite{Alexandrov:2008nk}, provided
the sum is  performed {\it after} integration along $\CP$.

Let us now recast \eqref{fullH} in the form
\be
\label{hthxin}
H_{{\rm NS5}}^{(k)}(\xi,\txi,\tilde\alpha)=\frac{1}{4\pi^2}
\!\!\!\! \sum_{\substack{ \mu \in
(\Gamma_m/|k|)/\Gamma_m \\ n \in \Gamma_m + \mu+\theta}}\!\!\!\!
H_{{\rm NS5}}^{(k,\mu)}\( \xi^\Lambda , n^\Lambda\)\,
\expe{k n^\Lambda(\txi_\Lambda-\phi_\Lambda)
-\frac{k}{2}\,(\tilde\alpha+\xi^\Lambda\txi_\Lambda)} .
\ee
Using results of the previous subsection, we find that the characteristics $\theta^\Lambda$ vanish
and that the function $H_{{\rm NS5}}^{(k,\mu)}\( \xi^\Lambda , n^\Lambda\)$ is invariant,
up to the phase \eqref{ratS}, under simultaneous shift
of $\xi^\Lambda$ and $n^\Lambda$. Therefore, up to the same phase, it is actually a function
of only their difference,
\be
H_{{\rm NS5}}^{(k,\mu)}\( \xi^\Lambda , n^\Lambda\)=
\nu(n^\Lambda-\mu^\Lambda)\,
H_{{\rm NS5}}^{(k,\mu)}\( \xi^\Lambda - n^\Lambda\),
\label{relHH}
\ee
where the function on the right is given by
\be
\label{HNS5}
H_{{\rm NS5}}^{(k,\mu)}( \xi^\Lambda) =
\sum_{q_\Lambda} K_{k,\mu^\Lambda,q_\Lambda} \thnkl\,
\expe{- \frac{k\, N(\xi^a)}{\xi^0} +
\frac{p^0\( k \hat q_a \xi^a+ p^0 \hat q_0\)}{k^2 \xi^0}
+\frac{k}{2}\, A_{\Lambda\Sigma}\xi^\Lambda\xi^\Sigma },
\ee
and we collected various constant factors into
\be
K_{k,\mu^\Lambda,q_\Lambda}=\sigma_{\text{D}}(\gamma)\,
\expe{-\frac{k}{2}\, A_{\Lambda\Sigma}\mu^\Lambda\mu^\Sigma
+a \,\frac{p^0  q'_0}{k}-k c_{2,a} \mu^a\eps(\delta)+k\mu^\Lambda\phi_\Lambda}.
\ee
In this expression, $\gamma=(p^0,k\mu^a,q_a,q_0)$ where
$p^0=\gcd(k,k\mu^0)$,
\be
\hat q_a= q'_a+ \frac{k^2}{2p^0}\kappa_{abc} \mu^b\mu^c\ ,\qquad
\hat q_0= q'_0 + \frac{k}{p^0} \mu^a q'_a+ \frac{k^3}{3(p^0)^2}\kappa_{abc}
\mu^a\mu^b\mu^c\,  ,
\ee
and $q'_a,q'_0$ are related to $q_a,q_0$ via \eqref{chargeshift}. The vectors
$k(\mu^0,\mu^a)$ are identified as the residue class of $(p,p^a)$ modulo $k$.
Setting  the phase factor $\nu(n-\mu)$ in \eqref{relHH} to one
(which can be achieved   in the case $k=1$ by the aforementioned
{\it ad hoc}  choice of characteristics, and should result more generally from
a fully consistent  treatment of Heisenberg, S-duality and monodromy symmetries),
we find that the holomorphic section $H_{\rm NS5}^{(k)}$ \eqref{hthxin}
encoding the fivebrane corrections has the form of a (non-Gaussian) theta series.

Having recast the fivebrane corrections \eqref{fullH} into the form \eqref{hthxin}, we can
now unravel the connection between the fivebrane wave-function  $H_{\rm NS5}^{(1)}$
for $k=1$ and the A-model topological string amplitude discussed in Section \ref{sec_top}.
To this end, notice  that for $k=1$, Eq. \eqref{HNS5} simplifies dramatically to
\be
\label{HNS51}
H_{{\rm NS5}}^{(1,0)}( \xi^\Lambda) =\epsilon
\sum_{\hat q_a,\hat q_0} \thnkl \, (-1)^{\hat{q}_0}\,
\expe{ -  \frac{N(\xi^a)}{\xi^0}+
\frac{\hat q_a \xi^a+ \hat q_0}{\xi^0}+\frac12\,A_{\Lambda\Sigma}\xi^\Lambda\xi^\Sigma  } .
\ee
Here we fixed the characteristics as in Eqs. \eqref{phi0}, \eqref{phiA}
as required for Heisenberg invariance
and took into account that  $p^0=1,\mu^\Lambda=0$ for $k=1$.
The factor $\epsilon=(-1)^{A_{00}-\phid{0}/2}$ is  irrelevant and
will be omitted in what follows.
Now recall the relation \eqref{gvc2} which expresses the A-model wave function
$e^{F_{\rm hol}(z,\lambda)}$ in terms of the DT partition function.
Identifying $(Q_a,2J)=(\hat q_a+c_{2,a}/24, \hat q_0)$ as in \eqref{defQJ} (which are
integer-valued when $p^{0}=1$, as noted below \eqref{defQJ}) and using the
relation\footnote{Here we use the relation $\lambda=2\pi/(\I\xi^0)$, rather than
$\lambda=1/(\xi^0\sqrt{2\pi})$  as stated in \eqref{psiholR}. We do not understand the
origin of this normalization mismatch. Note also that the prefactor in
\eqref{PsiDT} behaves as $\left(\xi^0\right)^{1+(2\epsilon_{\text{DT}} +\epsilon_{\text{GW}})
\frac{\chi(\hat\cX)}{24}}$, which seems unnatural for the usual choices of
$\epsilon_{\text{GW}},\epsilon_{\text{DT}}  $ given in the literature.} \eqref{psiholR}
between $F_{\rm hol}(z,\lambda)$ and
the wave function $\Psi_{\IR}^{\rm top}(\xi^\Lambda)$, we obtain the twistor
space version of the real-polarized A-model topological string wave function
\be
\begin{split}
\label{PsiDT}
\Psi_{\IR}^{\rm top}(\xi^\Lambda)&=
\left(\xi^0\right)^{1+(1+\epsilon_{\text{GW}})\frac{\chi(\hat\cX)}{24}}
[M(e^{2\pi\I/\xi^0})]^{(\frac12-\epsilon_{\text{DT}}  )\chi(\hat\cX)}\,
\sum_{\hat q_0,\hat q_a} (-1)^{\hat{q}_0}N_{DT}\left(\hat q_a+\frac{1}{24} c_{2a}, \hat q_0\right)
\\
&\times
\expe{- \frac{N(\xi^a)}{\xi^0}+\frac12\,A_{\Lambda\Sigma}\xi^\Lambda\xi^\Sigma
+ \frac{\hat q_a \xi^a}{\xi^0}
+ \frac{\hat q_0}{\xi^0} }.
\end{split}
\ee

Comparing \eqref{PsiDT} with \eqref{HNS51} and identifying
the DT-invariants $N_{DT}(\hat q_a+\frac{1}{24} c_{2a}, \hat q_0)$
with the rational instanton measure $\thnkl$,
we find that the wave function $H_{{\rm NS5}}^{(1,0)}( \xi^\Lambda)$ governing NS5-brane
instantons in type IIB/$\hat\cX$ is proportional to the
 wave function of the  topological A-model on $\hat\cX$ in the real polarization,
\be
H_{{\rm NS5}}^{(1,0)}(\xi^\Lambda) =
\left(\xi^0\right)^{-1-(1+\epsilon_{\text{GW}})\frac{\chi(\hat\cX)}{24}}
\, [M(e^{2\pi\I/\xi^0})]^{(\epsilon_{\text{DT}}  -\frac12)\chi(\hat\cX)}\,
 \Psi_\IR^{\rm top}(\xi^\Lambda) \, .
\label{NS5DTrelation}
\ee
We note that the sign factor $(-1)^{2J}=e^{\pi i {\hat q}_0}$ predicted by the
GW/DT relation nicely agrees with the factor
$\expe{-\frac12 q_\Lambda p^\Lambda}$ in the quadratic refinement $\sigma_\Theta$.
The relation \eqref{NS5DTrelation} is moreover generally consistent
with the fact that both $H_{{\rm NS5}}^{(1,0)}(\xi^\Lambda)$ and $\Psi_\IR^{\rm top}(\xi^\Lambda)$
should transform under
monodromies according to the metaplectic representation,
although the powers of $\xi^0$ and $M(e^{2\pi\I/\xi^0})$ appear to spoil
these transformation properties.
Eq. \eqref{NS5DTrelation} also suggests that the proper mathematical
interpretation of the real-polarized wave function $\Psi_\IR^{\rm top}$ may be as
a section of $H^1(\cZ_\cM)$, rather than $H^0(\cM_K)$ as is commonly assumed.

By mirror symmetry, \eqref{NS5DTrelation} should also determine the wave function
governing NS5-brane corrections in type IIA string theory compactified on $\cX$,
in terms of the wave function $\Psi_\IR^{\rm top}$ of the topological B-model  on $\cX$
in the real polarization, in agreement with general expectations
expressed in \cite{Kapustin:2004jm}, and earlier in
\cite{Dijkgraaf:2002ac,Nekrasov:2004js}.
We offer further support for this assertion
in the remainder of this section.
More generally, we expect that the wave function $H_{{\rm NS5}}^{(k,\mu)}(\xi^\Lambda)$
for $k>1$ in type IIA should originate from a higher rank version of the topological B-model
on $\cX$, presumably related to the generalized invariants of Joyce and
Song \cite{Joyce:2009xv}.

\subsection{Fivebrane partition function from twistor space \label{sec_nonlin}}

Having identified candidate sections $H_{k,p,\hgam}$ in $H^1(\cZ_\cM,\cO(2))$  governing the
corrections to the contact structure on the twistor space $\cZ_\cM$ from $k$ fivebranes, we would now
like to make contact with the qualitative discussion of fivebrane instanton corrections
to the HM moduli space metric of Sections \ref{sec_pert} and \ref{sec_IIB}. To this end, we should
in principle evaluate the corresponding corrections to the Darboux coordinates
and contact potential $\Phi$ using the integral formulae in \cite{Alexandrov:2008nk,Alexandrov:2008gh},
and from them obtain the corrections to the metric on $\cM$.
This analysis is however beyond the scope of the present work.

Instead, we shall address the simpler question raised in Section \ref{subsec_fivepftwist}, namely
construct a \emph{scalar-valued function}
on $\cM$ which generalizes the Gaussian flux partition function
$\ZG{k}$ at finite coupling. To this end, let us view
\eqref{fullH} as a formal holomorphic section of
$H^{1}(\cZ_\cM,\cO(-2))$ (barring global issues) and apply
the standard Penrose transform \eqref{penroseint}. This produces
a certain function $\ZNS{k}$ on the perturbative HM moduli space, satisfying
certain second order differential equations. We shall see that this
function reduces to a close variant of the Gaussian flux partition function
$\ZG{k}$ discussed in Section \ref{sec_zns5} in the limit $g_s=0$, thereby motivating
the name ``non-Gaussian NS5-brane partition function".
This computation may also be viewed
as a warm-up for the more complicated computations involved in extracting
the corrections to the HM metric from the $H^{1}(\cZ_\cM,\cO(2))$ section \eqref{fullH}.

\subsubsection{Penrose transform\label{sec_pent}}

We thus consider the Penrose transform \eqref{penroseint} of a single term $H_{k,p,\hgam}$ in
the sum \eqref{fullH}, given in  \eqref{5pqZ2}:
\be
\label{varPhiNS5}
\int_{\ell_{k,p,\hgam}} \!\!\frac{\de\varpi}{\varpi}\,
\expe{-\frac{k}{2}\(\tilde\alpha+(\xi^\Lambda-2n^\Lambda) \txi'_\Lambda\)
- \frac{k N(\xi^a-n^a)}{\xi^0-n^0}
+\frac{k p^0 \hat q_a(\xi^a-n^a) +(p^0)^2\hat q_0}{k^2(\xi^0-n^0)}},
\ee
where the contour $\ell_{k,p,\hgam}$ interpolates between the two zeros of $\xi^0-n^0$,
and passes through the saddle point to be analyzed below.
We omitted the contact potential which appears as the overall factor since
in our approximation it is independent of $t$ and
also dropped the constant factors given by the quadratic refinement and the
last two terms in \eqref{5pqZ2}.
Note that, by construction, \eqref{varPhiNS5} is proportional to
$e^{-\I\pi k \sigma}$, and manifestly invariant under
\kahler transformations and (after summing over charges, ignoring the phases
discussed in Appendix A) under Heisenberg translations.

Using the expression for the Darboux coordinates given in \eqref{gentwi},
the integrand evaluates to
\be
\varpi^{-\frac{\chi(\hat\cX)}{24}}\,
\expe{
-k\[ \frac{1}{2} \(\sigma+(\zeta^\Lambda-2n^\Lambda)\tzeta'_\Lambda\)-\frac{\tau_2^2}{4}\,\Re(\bz^\Lambda F'_\Lambda)
+\frac{\cB+\frac{\tau_2}{2}\(t^{-1}\cA-t\bar\cA\)}
{\ttau_1+\frac{\tau_2}{2}\(t^{-1}-t\)}
\]} ,
\label{exprexp}
\ee
where we abbreviated $\ttau=\tau-n^0$ and introduced
\beq
\cA &=& \frac{\tau_2^2}{4}\, \bz^\Lambda F'_\Lambda
-\frac12\, (\zeta^\Lambda-n^\Lambda)(\zeta^\Sigma-n^\Sigma) F'_{\Lambda\Sigma}
-\frac{p^0\hat q_a}{k^2}\,z^a,
\\
\cB &=& N(\zeta-n)-\frac{\tau_2^2}{2}\( \ttau_1\Re N(z)+(\zeta^a-n^a)\kappa_{abc}t^b t^c\)
-\frac{p^0\hat q_a}{k^2}\(\zeta^a-n^a\)-\frac{(p^0)^2 \hat q_0}{k^3}.
\nonumber
\eeq
It is useful to note that
\be
\frac{k\tau_2}{2|\ttau|^2}\(\ttau_1\Re\cA-\cB-\I|\ttau|\Im\cA\)=W_{k,p,\hgam}
\label{ABW}
\ee
reproduces the quantity introduced in \eqref{Wkg}.
The integral over $t$ is dominated in the weak coupling limit by a saddle point at
\be
t_s
=\I\,\frac{-\tau_2\Im\cA-\sqrt{(\cB-\ttau_1\Re\cA)^2+|\ttau|^2(\Im\cA)^2}}{\cB-\ttau_1\bar\cA}.
\label{fullsaddle}
\ee
Substituting into \eqref{exprexp}, we find that \eqref{varPhiNS5} is given, in the semi-classical
approximation, by the expression
\be
 J\,\varpi^{-\frac{\chi(\hat\cX)}{24}}_s\,
e^{-S(t_s)},
\ee
where the classical action is given by
\be
\begin{split}
S(t_s) = &\frac{2\pi k \tau_2}{|\ttau|^2}\, \sqrt{(\cB-\ttau_1\Re\cA)^2+|\ttau|^2(\Im\cA)^2}
\\
&+2\pi \I k\[
\frac{1}{2}\, \(\sigma+(\zeta^\Lambda-2n^\Lambda)\tzeta'_\Lambda\)
-\frac{\tau_2^2}{4}\,\Re(\bz^\Lambda F'_\Lambda)
+|\ttau|^{-2}\(\tau_2^2\Re\cA+\ttau_1\cB \)
\]
\label{exprexp2}
\end{split}
\ee
and $J$ is defined as
\be
J= \frac{p-k\xi^0(t_s)}{|p-k\tau|}|\, W_{k,p,\hgam}|^{-1/2}
=\frac{|p-k\tau|\, |W_{k,p,\hgam}|^{1/2}}{(p-k\tau_1)|W_{k,p,\hgam}|-\I k \tau_2 \Re W_{k,p,\hgam}}\, .
\ee
It is straightforward to check that \eqref{exprexp2} (plus the omitted constant terms) reproduces
the action $S_{k,p.\hgam}$ in \eqref{ImSpk} obtained by  S-duality
transformation of the standard D-instanton action \eqref{SD}.

It is also instructive to compute \eqref{varPhiNS5} directly in the weak coupling limit.
Assuming that the integral is controlled by a saddle point
with $t_s\sim 1/\tau_2$ as $\tau_2\to\infty$ (which is indeed the case of \eqref{fullsaddle}),
we can redefine  $t=t'/\tau_2$ and expand the exponent in the limit $\tau_2\to\infty$,
keeping $t'$ finite.  Neglecting the logarithmic contribution,
we obtain that the argument of the exponential in \eqref{varPhiNS5} reduces to
\beq
S&\approx & \pi \I k \(\sigma+(\zeta^\Lambda-2n^\Lambda) \tzeta'_\Lambda\)
+4\pi k e^\phi -2\pi \I\,\frac{p^0}{k}\, \hat q_a z^a
\label{amexp}\\
&&
-\pi k\(   \I\tau_{\Lambda\Sigma}
(\zeta^\Lambda-n^\Lambda)  (\zeta^\Sigma-n^\Sigma)
+2  \tau_2 t (\zeta^\Lambda-n^\Lambda) \Im\tau_{\Lambda\Sigma} \bar z^\Sigma
-\frac12\, \tau_2^2 t^2 \bar z^\Sigma \Im\tau_{\Lambda\Sigma} \bar z^\Sigma\).
\nonumber
\eeq
The saddle point for the $t$ integral lies at
\be
\label{saddlealp}
t_s=\frac{2(\zeta^\Lambda-n^\Lambda) \Im\tau_{\Lambda\Sigma} \bar z^\Sigma}
{\tau_2\, \bar z^\Lambda
\Im\tau_{\Lambda\Sigma} \bar z^\Sigma}=
4 \, e^\cK \tau_2^{-1} \, (\zeta^\Lambda-n^\Lambda) \Im\cN_{\Lambda\Sigma} z^\Sigma\, ,
\ee
consistently with the weak coupling limit of \eqref{fullsaddle},
so that  the semi-classical approximation to \eqref{varPhiNS5}
becomes
\be
  \tau_2^{-1}\, \varpi_s^{-1-\frac{\chi(\hat\cX)}{24}} \,
(\bar z^\Lambda
\Im\tau_{\Lambda\Sigma} \bar z^\Sigma)^{-1/2}\, e^{-S(t_s)}
\ee
with
\be
\label{SpenG}
S(t_s)= 4\pi k e^\phi +\pi\I k \left(  \sigma + (\zeta^\Lambda-2n^\Lambda) \tzeta_\Lambda
- \bar\cN_{\Lambda\Sigma} (\zeta^\Lambda-n^\Lambda)  (\zeta^\Sigma-n^\Sigma)\right)
-2\pi \I\,\frac{p^0}{k}\,\hat q_a z^a.
\ee
The classical action \eqref{SpenG} indeed reproduces the Gaussian fivebrane action \eqref{NS5instact},
after using identifications \eqref{defcarge}. While this agreement was guaranteed given the
fact that \eqref{exprexp2} equals $S_{k,p.\hgam}$ which reduces to \eqref{NS5instact}
at weak coupling, this computation allows us to make contact with the analysis
in \cite{Dijkgraaf:2002ac}, where an auxiliary variable $t$ was introduced as a way
to turn the Gaussian partition function in the Weil polarization into an indefinite Gaussian
partition function in the Griffiths polarization: this auxiliary variable $t$ is just the coordinate
on the $\CP$ fiber over $\cM$, after a simple rescaling.

Eq. \eqref{amexp} also indicates that
the non-Gaussian theta series \eqref{HNS5} is formally divergent, as it reduces to an indefinite
Gaussian sum in the regime of weak coupling. However, it should be kept in mind that
each term in \eqref{HNS5} is integrated on a different contour in $\CP$. Provided one first carries
out the contour integral in $\CP$ and then the sum over charges, the end result will involve
a sum of exponentially suppressed instantonic corrections, though each of them  might be
multiplied by an exponentially growing summation measure as in the D-instanton
sector \cite{Pioline:2009ia}.

\subsubsection{Non-Gaussian fivebrane partition function}

We define the non-Gaussian fivebrane partition function by summing
the Penrose transform evaluated in \ref{sec_pent} over charges,
\be
\ZNS{k,\mu}\equiv e^\phi \sum_{p,p^a, \hat q_a,\hat q_0} \int_{\ell_{k,p,\hgam}}  \frac{\de\varpi}{\varpi}\,
H_{k,p,\hgam}(\xi(\varpi),\txi(\varpi),\tilde\alpha(\varpi))\, .
\ee
The fact that the  Gaussian action \eqref{SpenG}  resulting from the integral and
the omitted constant prefactor, both depend on charges $\hat q_\Lambda$ linearly,
allows us to carry out the sum
over electric charges explicitly.
We now evaluate this sum for $k=1$ in terms of the partition function of DT invariants.
Setting $n^\Lambda=0$ by Heisenberg invariance,
we have
$p^0=1$, $\hat q_a=q_a-\frac{c_{2,a}}{24}$ and so only one term in \eqref{SpenG}
contributes. Reinstating the summation measure $\thnkl$ and the quadratic refinement
with the characteristics fixed as above,
the sum over electric charges leads formally to
\be
\sum_{q_a,q_0} \thnkl\, \expe{ \hat q_a z^a-\frac{1}{2}\, \hat q_0}
= \expe{- \frac{c_{2a} z^a}{24}}\, Z_{\rm DT}(z^a, \lambda)\Big|_{\lambda\to 0},
\ee
where $Z_{\text{DT}} $ is the DT-partition function \eqref{ZDTdef}.
Using the DT/GW relation \eqref{gvc2}, this may be further rewritten as
\be
\lambda^{\frac{\chi(\hat\cX)}{24}\epsilon_{\text{GW}}}\,
[M(e^{-\lambda})]^{(\epsilon_{\text{DT}}-\frac12)\chi(\hat\cX)}\,
e^{F_{\rm hol} +\frac{(2\pi \I)^3}{\lambda^2}
\left( N(z^a) -\frac12 A_{\Lambda\Sigma}{z}^{\Lambda} {z}^{\Sigma} \right)}\Big|_{\lambda\to 0}.
\label{sumpenrtr}
\ee
In the limit $\lambda\to 0$, only genus 0 and genus 1 contributions to $F_{\rm hol}$
remain.
For $\epsilon_{\text{DT}}  =0$  the power of the Mac-Mahon function
cancels the degenerate Gromov-Witten contributions in $F_{\rm hol}$, while the total
remaining power of $\lambda$ cancels for $\epsilon_{\text{GW}}=1$. Still, the non-degenerate
genus zero Gromov-Witten contributions seem to make the limit $\lambda\to 0$
singular. It is plausible that these singular contributions may be canceled when
$\alpha'$ and D(-1)-instanton
corrections to the Darboux coordinates are included, and we shall ignore them in what follows.
Thus, we conclude that the expression \eqref{sumpenrtr} can be replaced by  $e^{f_1(z)}$
where $f_1(z)$ is the holomorphic part of the one-loop vacuum amplitude $F_1$ \eqref{F1}.

As a result, we find that the weak coupling approximation of the fivebrane partition function is given by
\be
\ZNS{1} \sim
\tau_2\,  e^{ f_1-\cK}\, (\bar z^\Lambda
\Im\tau_{\Lambda\Sigma} \bar z^\Sigma)^{-1/2}
\sum_{n\in\Gamma_m+\theta}
\varpi_s^{-1-\frac{\chi(\hat\cX)}{24}} \,
e^{-2\pi\I n^\Lambda\phi_\Lambda-S'(t_s)} \, ,
\label{Ptotres}
\ee
where $S'(t_s)$ is the action \eqref{SpenG} without the last term.
Extracting the factor $e^{-4 \pi e^{\phi}-\I\pi   \sigma}$, as in \eqref{ns5quali},
we obtain the Gaussian NS5-partition function
\be
\ZNSG{1}
\equiv \sum_{n\in\Gamma_m+\theta}
\cF\, \expe{\frac{1}{2} (\zeta^\Lambda-n^{\Lambda})\bar \cN_{\Lambda\Sigma}
(\zeta^\Sigma-n^{\Sigma})
+ n^{\Lambda}(\tzeta_{\Lambda}-\phi_\Lambda)
-\frac{1}{2} \zeta^{\Lambda}\tilde\zeta_{\Lambda}},
\label{NS5partitionfunctioncorrect}
\ee
where the prefactor $\cF$ is given by
\be
\cF(n;\cN, z, \phi, \zeta)=
\frac{ \(e^{-\cK}\tau_2\)^{2+\frac{\chi(\hat\cX)}{24}}}{\sqrt{\bar{z}^{\Lambda}\text{Im}\,
\tau_{\Lambda\Sigma}\bar{z}^{\Sigma}}} \left[(\zeta^{\Lambda}-n^{\Lambda})\text{Im}\,
\cN_{\Lambda\Sigma} z^{\Sigma}\right]^{-1-\frac{\chi(\hat\cX)}{24}}\, e^{f_1(z)}.
\label{normalizationfactor}
\ee

Comparing the result \eqref{NS5partitionfunctioncorrect} with the analysis
in Section \ref{sec_factorize} shows that the dependence on
the $C$-field is given by a generalized Siegel theta series,
with an insertion of a power of $t_s$ (given in Eq. \eqref{saddlealp})
in the sum.\footnote{It is interesting
to note that  insertions of powers of $t_s$ in the sum do not spoil the modular
properties of the theta series, at least in the semi-classical approximation.
This is obvious for transformations of type \eqref{cAtrans} and
\eqref{vartheB}. For the inversion $\bar\cN\mapsto -\bar\cN^{-1}$, it suffices to
evaluate the Fourier transform
$$
\label{intze}
\int \de\zeta^\Lambda\,
(z^\Lambda \Im\cN_{\Lambda\Sigma} \zeta^\Sigma)^{-1-\frac{\chi(\hat\cX)}{24}}
\, e^{-\I\pi \zeta^\Lambda \bar\cN_{\Lambda\Sigma}\zeta^\Sigma+
2\pi\I k \zeta^\Lambda \tzeta_\Lambda}
$$
in the saddle point approximation. At the extremum $\zeta=\bar\cN^{-1}\tzeta$, we have
$ z^\Lambda \Im\cN_{\Lambda\Sigma}  \zeta^\Sigma =
\frac{1}{2\I}  z^\Lambda {[\cN (\bar\cN^{-1}-\cN^{-1})]_{\Lambda}}^\Sigma
\tzeta_\Sigma = - F_\Lambda  [\Im(\cN^{-1})]^{\Lambda\Sigma} \tzeta_\Sigma $,
where we used the identity $\cN_{\Lambda\Sigma} z^\Sigma = F_\Lambda$, so that
the integral becomes
$$
(\det \bar\cN)^{-1/2}\, (- F_\Lambda  [\Im(\cN^{-1})]^{\Lambda\Sigma}
\tzeta_\Sigma)^{-1-\frac{\chi(\hat\cX)}{24}}\,
e^{\I\pi \tzeta_\Lambda (\bar\cN^{-1})^{\Lambda\Sigma}\tzeta_\Sigma}\, ,
$$
consistently with the transformations properties of the various quantities involved.}
Thus, the assumption made in Section \ref{sec_factorize} that
the normalization factor $\cF$ did not depend on the flux $H$ was erroneous.
In retrospect, such flux-dependent insertions into the fivebrane theta series were already
present in the automorphic studies \cite{Bao:2009fg,Pioline:2009qt}.

The fact that the holomorphic part ${f_1(z)}$ of the B-model one loop amplitude
${F_1}$ \eqref{F1} appears in the prefactor \eqref{normalizationfactor}
implies that the one-loop determinant of the non-chiral fivebrane partition
function in the flux sector $H$ is proportional to  $e^{F_1}$, as anticipated
in \cite{Dijkgraaf:2002ac}. The latter
is a product of analytic Ray-Singer torsions \cite{Bershadsky:1994cx},
\be
e^{F_1} = \prod_{0\leq p,q\leq 3} \left[\det' \Delta_{\bar\partial}^{p,q}\right]^{\frac12 pq (-1)^{p+q}}
=
 \frac{\big(\det' \Delta_{\bar\partial}^{0,0}\, \big)^{9/2}\big(\det' \Delta_{\bar\partial}^{1,1}\, \big)^{1/2} }
{\big(\det' \Delta_{\bar\partial}^{1,0}\, \big)^{3}}\, ,
\ee
where $\Delta_{\bar\partial}^{p,q}$ is the Laplacian on $p$-forms valued in $\Lambda^q T\cX$,
and $\det'$ is the determinant with zero-modes removed. In the second equality we used
the standard identities $\det'\Delta_{\bar\partial}^{p,q}=\det'\Delta_{\bar\partial}^{3-p,q}
=\det'\Delta_{\bar\partial}^{q,p}$ (see e.g. \cite{Harvey:1996ts}). The fact that this normalization
factor differs from the one computed in \cite{Belov:2006jd,Monnier} should come as no surprise,
since these authors considered the  partition function of the chiral two-form, whereas our
result applies to  the partition function of the (2,0)  supersymmetric field theory
on the NS5-brane with an insertion of $(-1)^{2J_3} (2J_3)^2$, as required for computing
instanton corrections to the two-derivative low energy effective action. It would be
very interesting to perform the one-loop determinant computation explicitly.

To summarize, we have found that the saddle point approximation of the Penrose transform \eqref{varPhiNS5}
produces a refinement of the chiral Gaussian partition function constructed
in Section \ref{sec_factorize}. It is therefore natural to consider
the original Penrose transform as an extension of the Gaussian fivebrane partition
function \eqref{NS5partitionfunctioncorrect} in the regime
where the energy stored in the three-form flux is of the same order or larger than the
energy in the fivebrane itself. Building upon the discussion at the end of
Section \ref{sec_theta}, we thus define the full non-linear partition
function of $k$ chiral NS5-branes as the integrated matrix element
\be
\label{matelp}
\ZNS{k,\mu}
=e^\phi \int \frac{\de t}{t}\,
e^{-\I\pi k\tilde\alpha} \,
\langle \Psi^{\Gamma_m,k,\mu} |
\, e^{- \I(\xi^\Lambda \tilde T_\Lambda - \txi_\Lambda T^\Lambda)} |
H_{\rm NS5}^{(k,\mu)} \rangle\, ,
\ee
where $\Psi^{\Gamma_m,k,\mu}$ is the same distribution as in \eqref{PsiRNS5G}, and
$|H_{\rm NS5}^{(k,\mu)} \rangle$ is the state whose wave-function in the real polarization is
given by \eqref{HNS5}. Note that we are here including the dependence on the NS-axion $\sigma$
in the definition of the non-Gaussian partition function.
This is quite natural given the observation at the end of
Section \ref{subsubsec_monodr} that $\sigma$ is in fact not well defined by itself.

Finally, we note that the functional space where the generators $\tilde T_\Lambda$ and
$T^\Lambda$ operate is recognized as a space of (local) holomorphic sections
on $\cZ_\cM$. For supergravity theories with a symmetric moduli space, this same space
is the habitat of the quaternionic discrete series representations, whose relevance
to the issue of instanton corrections to the HM moduli space has been advocated
previously \cite{Gunaydin:2005mx,Bao:2009fg,Pioline:2009qt}.

\section{Discussion\label{sec-dis}}

In this work, we have taken  steps towards understanding
NS5-brane instanton corrections to the hypermultiplet moduli space $\cM$ in type IIA
string theory compactified on a Calabi-Yau threefold $\cX$, and other moduli
spaces related to it by T-duality and mirror symmetry
(see Table \ref{dico} on page \pageref{dico}).

Our first main result, announced previously in \cite{Alexandrov:2010np},
is the identification of the topology of $\cM$ around the weak coupling limit:
$\cM$ is a foliated by hypersurfaces $\cC(r)$, $r\in\IR^+$, each of which being
a circle bundle $(\cL_\Theta\otimes \cL^{\frac{\chi(\cX)}{24}})^{\circ}$ over the intermediate Jacobian $\cJ_c(\cX)$.
Recall from \eqref{intJac} that $\cJ_c(\cX)$ is the total space of the torus bundle over the complex
structure moduli space $\cM_c(\cX)$ with fiber $\cT=H^3(\cX,\IR)/H^3(\cX,\IZ)$.
Physically, $r\sim 1/g_{(4)}^2$ denotes the 4D string coupling, $\cT$ parametrizes
the harmonic $C$-field on $\cX$, and the circle fiber $S^1_\sigma$ of $\cC(r)$ parametrizes
the Neveu-Schwarz axion. The curvature of $\cC(r)$ \eqref{c1C} has two
components:  i) the K\"ahler class $\omega_\cT$ reflects the fact that translations on $\cT$
commute up to a translation along the circle fiber $S^1_\sigma$, ii) while $\frac{\chi(\cX)}{24}\omega_\sk$
reflects the fact that $\sigma$ shifts under phase rotations of the holomorphic
three-form $\Omega_{3,0}$ on $\cX$ as well as under monodromies in $\cM_c(\cX)$.
In the strict weak coupling limit, $g_{(4)}=0$, the metric
is invariant under continuous translations on $\cT$ and $S^1_\sigma$, but these
continuous isometries are broken to discrete identifications \eqref{heisext}
by D-instantons and NS5-brane instantons, respectively. In particular, the
identifications \eqref{heisext} and \eqref{sigmon} involve a choice of quadratic refinement
$\sigma_\Theta$ of the intersection form on $H^3(\mathcal{X},\mathbb{Z})$,
together with a unitary character $e^{2\pi\I\kappa(M)}$ of the monodromy group
which, to our knowledge,  had not been noticed
previously (the former did however appear in \cite{Bao:2010cc}, which was
developed concurrently to the present work). The character $e^{2\pi\I\kappa(M)}$  is
related to the multiplier system of the one-loop amplitude of the topological B-model
on $\cX$, and enters in the definition of the twisted line bundle $\cL^{\chi(\cX)/24}$ where
the NS-axion is valued.

The hypermultiplet moduli space in type IIB string theory compactified
on the mirror CY threefold $\hat \cX$ exhibits the same structure as in type IIA, as required
by quantum mirror symmetry. The hypersurfaces $\cC(r)$ are now circle bundles
over the ``symplectic Jacobian" $\cJ_K(\hat \cX)$ \eqref{intJac2},
which is the total space of a torus bundle over the \kahler moduli space
$\cM_K(\hat\cX)$ with fiber $H^{\rm even}(\hat \cX,\IR)/K(\hat\cX)$.
In particular, we have clarified the map from
the K-theory lattice $K(\hat \cX)$ to the D-brane charge lattice $H^{\rm even}(\hat \cX,\IZ)$,
by showing explicitly that it is given by a modification \eqref{modifmukai}
of the standard generalized Mukai map. This modified Mukai map
 crucially involves the quadratic real matrix $A_{\Lambda\Sigma}$
appearing in the prepotential \eqref{lve}.
Due to the fractional value of the (primed) D(-1)-brane charge,
we furthermore found it necessary to modify the action of S-duality given in
\cite{Gunther:1998sc,Alexandrov:2008gh,Alexandrov:2009qq}
on the D3-brane axion $c_a$, by a shift proportional to the multiplier system
of the Dedekind eta function. Of course, the same considerations also apply for
D6-D4-D2-D0 branes in type IIA string theory compactified on the CY threefold $\hat\cX$.

Using S-duality, we inferred the qualitative form of $(p,k)$5-brane corrections to the metric,
and verified agreement with the type IIA Gaussian result in the weak coupling limit.
It should be stressed that although the type IIB fivebrane is non-chiral, its partition function
can still be written as a theta series of a similar form
as in the type IIA picture. In particular, it must
involve a quadratic refinement $\sigma_\Theta$, which can in principle be computed by
S-duality from the quadratic refinement $\sigma_{\Theta_{\rm D}}$ governing D-instantons.
It would be interesting to understand the origin of this apparent
``chirality'' on the type IIB NS5-brane worldvolume, perhaps along the lines of \cite{Witten:1999vg}.
The independence of the physical hypermultiplet metric on the choice of quadratic refinement
(as is known to be the case in $\cN=2$ field theories \cite{Gaiotto:2008cd}) further suggests
that the NS5- and D-instanton characteristics $\Theta$ and $\Theta_{\rm D}$ should be equal.
On the other hand, our naive implementation of Heisenberg and S-duality indicates a different answer,
Eq. \eqref{phi0} and \eqref{phiA}. It is an important open problem to resolve this discrepancy.

To implement NS5-brane instanton corrections consistently with supersymmetry, it is necessary
to reformulate them in terms of deformations of the complex contact structure on the twistor
space $\cZ_\cM$ over $\cM$. At the perturbative level and over a fixed point in $\cM_c(\cX)$,
$\cZ_\cM $ can be obtained as the quotient
of $H^3(\cX,\IC)\times \IC$ by the discrete Heisenberg identifications \eqref{heisalgz}.
Corrections to the perturbative metric are then encoded in holomorphic sections of $H^1(\cZ_\cM,\cO(2))$.
As alluded to above, by applying S-duality to the holomorphic sections describing D5-D3-D1-D(-1)
instanton corrections in type IIB, we constructed a candidate section $H_{k,p,\hgam}$
in Eq. \eqref{5pqZ2} encoding the contribution of a charge $k$ NS5-brane instanton.
The resulting deformation of the contact structure is formally invariant
under Heisenberg shifts and under monodromies around the large volume
point, up to phases $\nu(\eta)$ and $\nu'(\epsilon)$ computed in Appendix A.
The most conservative explanation of this clash is probably that our assumption of
the invariance of the transition functions is too restrictive, and one must allow for
local compensating contact transformations, as already observed in \cite{Alexandrov:2009qq}.
It is also conceivable that one (or more) of the symmetries must give in.
For instance, one might expect that S-duality
is broken to a finite index subgroup, as is sometimes the case of electric-magnetic
duality in $\cN=2$ field theories.

In spite of the problems mentioned above, our results nonetheless indicate that our approach is reasonable.
In particular, we find that the single fivebrane wave-function $H_{\text{NS5}}^{(1,0)}$ \eqref{HNS5}
is proportional to the wave-function $\Psi_\IR^{\rm top}$ of the topological
B-model in the real polarization, up to certain factors which deserve further study (see  \eqref{NS5DTrelation}).
Moreover, its Penrose transform reproduces a variant of the Gaussian flux partition function
with a normalization factor proportional to the one-loop B-model amplitude,
consistent with the topology of the one-loop corrected HM moduli space.
It would be interesting to verify this result
by a direct computation of the twisted partition function of the (2,0) field theory
on the fivebrane worldvolume, and elucidate the origin of the flux-dependent insertion in the Gaussian sum.

While we feel that the above results constitute significant progress towards understanding
fivebrane instanton corrections, they are still far from providing the exact metric on the HM
moduli space. In particular, our prescription for ``summing" over images under $SL(2,\IZ)$
formally leads to a dense set of mutually intersecting ``BPS rays" on the twistor fiber at
fixed values of the moduli, whose mathematical status is unclear.
While the twistorial construction of NS5-brane instantons presented in this paper
is adapted to type IIB S-duality, it is natural to wonder if there exists an alternative
construction, more suitable for type IIA,  which would make symplectic invariance manifest,
and hopefully remove the dense set of contours mentioned above,
as in  the D1-D(-1) sector considered in \cite{Alexandrov:2009qq}.

It is also important to understand how our prescription is consistent with wall-crossing.
Indeed, one of the hints in
uncovering the structure of the D-instanton corrections
\cite{Alexandrov:2008gh,Alexandrov:2009zh,Alexandrov:2009qq}
was the Kontsevich-Soibelman wall-crossing formula  \cite{ks},
which involves a product of symplectomorphisms: as shown
in \cite{Gaiotto:2008cd}, the KS formula finds a natural interpretation in the twistorial description
of the \hk moduli space of $\cN=2$ gauge theories on $\IR^3\times S^1$.
In the presence of NS5-brane instantons we require a similar formula
now involving a product of contact transformations,
perhaps arising from a suitable limit of the motivic wall-crossing formula of \cite{ks}.
We hope to return to these issues in future work.

\acknowledgments

We are especially grateful to G. Moore for numerous clarifying discussions.
We also thank M. Douglas, J. Evslin, D. Jatkar, A. Kleinschmidt, J. Manschot, R. Minasian, S. Monnier,
A. Neitzke, B. Nilsson, C. Petersson, Y. Soibelman, S. Vandoren, P. Vanhove,
A. Wijns and D. Zagier
for  valuable discussions and correspondence.
S.A. is grateful to Perimeter Institute for Theoretical Physics for the kind hospitality
and the financial support. D.P. is grateful to LPTHE
for hospitality where part of this work was carried out.
D.P. and B.P. acknowledge the hospitality of the Chalmers University of Technology, G\"oteborg University and the Marine Biological Laboratory in Tj\"arn\" o during the final
stage of this project.

\appendix

\section{Action of  Heisenberg shifts  and monodromies on $H_{k,p,\hgam}$
\label{apsec_Heis}}

In this appendix we derive how the transition functions \eqref{5pqZ2}
encoding NS5-brane corrections transform under the Heisenberg action and monodromies
around the large volume point in $\cM_K(\hat\cX)$.

\subsection{Heisenberg symmetry}

Let us start from the transition function \eqref{5pqZ2} obtained by S-duality on the D-instanton series.
It can be presented in the following form
\be
\label{5pqZ2new}
\begin{split}
H_{k,p,\hgam}=&\, \frac{1}{4\pi^2}\, \thnkl  \,
\expe{-\frac{k}{2}\(\tilde\alpha +\xi^\Lambda\txi_\Lambda\)+kn^\Lambda\(\txi_\Lambda -\phi_\Lambda\)
-k\, \frac{N(\xi^a-n^a)}{\xi^0-n^0}
\right. \\
& \ \ +\frac{k}{2}\, A_{\Lambda\Sigma}(\xi^\Lambda-n^\Lambda)(\xi^\Sigma-n^\Sigma)
 +\frac{p^0\( k \hat q_a (\xi^a-n^a) + p^0 \hat q_0\)}{k^2(\xi^0-n^0)}
\\
& \ \ \left.
+ a\,\frac{p^0 q'_0}{k}- c_{2,a} p^a\eps(\delta)
-\frac{k}{2}\, A_{\Lambda\Sigma}n^\Lambda n^\Sigma
-\frac12\, q_\Lambda p^\Lambda+q_\Lambda \thetad{\Lambda}
-p^\Lambda\phid{\Lambda}+kn^\Lambda\phi_\Lambda } ,
\end{split}
\ee
where we distinguished between characteristics appearing in the D-instanton series, which we denoted by
$(\thetad{\Lambda},\phid{\Lambda})$, and the characteristics appearing in the Heisenberg
symmetry transformation \eqref{heisalgz}, $(\theta^\Lambda,\phi_\Lambda)$, which are relevant for NS5-branes.
Note that the two terms proportional to the matrix $A_{\Lambda\Sigma}$ come
from the difference between $\txi'_\Lambda$ appearing in $S_\alpha$, Eq. \eqref{defSa},
and $\txi_\Lambda$ appearing in the Heisenberg action \eqref{heisalgz}.

It is easy to see that invariance of $H_{k,p,\hgam}$ under Heisenberg shifts \eqref{heisalgz} with
$\eta^\Lambda=0$ is ensured provided the characteristics $\theta^\Lambda$ are set to zero.
Thus, it remains to consider Heisenberg shifts with $\tilde\eta_\Lambda=\kappa=0$.
It is clear that the shift $\xi^\Lambda\mapsto\xi^\Lambda+\eta^\Lambda$
should be compensated by a similar shift of the charges $n^\Lambda$.
The shift  $n^0\mapsto n^0+\eta^0$ corresponds to a S-duality transformation changing
$p\mapsto p+k \eta^0$, whereas
the shift $n^a\mapsto n^a+\eta^a$ amounts to a spectral flow
transformation \eqref{flow} on the magnetic charges $p^a$, with flow
parameter $\epsilon^a=k \eta^a/p^0$.
Applying the rest of the
spectral flow transformation \eqref{flow} to the charges
$q_a,q_0$, and using the invariance of  $\hat q_a, \hat q_0$,
one observes that the first two lines in \eqref{5pqZ2new}
are explicitly invariant under this combined action provided
$\thnkl$ is invariant under the spectral flow,
$\overline{\Omega}(\gamma[\epsilon])=\thnkl$.
As a result, under this crucial assumption,
we need to evaluate the transformation of only the constant terms in the last line
and the full variation of $H_{k,p,\hgam}$ takes the form
\be
\label{relHHH}
H_{k,p+k\eta^0,\hgam[\epsilon]}
\( \xi^\Lambda +\eta^\Lambda, \txi_\Lambda,\tilde \alpha +\eta^\Lambda(\txi_\Lambda-2\phi_\Lambda) \)
=\nu(\eta)H_{k,p,\hgam}
\( \xi^\Lambda, \txi_\Lambda,\tilde \alpha \),
\ee
where $\nu(\eta)$ is the phase factor to be found.

We split the phase $\nu(\eta)$ into three contributions.
The first one is given by the variation of the quadratic refinement
\be
\label{ratsig}
\begin{split}
\nu_1(\epsilon)\equiv \frac{\sigma_{\text{D}}(\gamma[\epsilon])}{\sigma_{\text{D}}(\gamma)}
= &\,\expe{\frac12\, \kappa_{abc}p^ap^b\epsilon^c-\frac32\,(p^0)^2L_0(\epsilon)
+\frac12\,(p^0)^2 L_a(\epsilon)\epsilon^a-p^0\epsilon^a\phid{a}
\right.
\\
& \left.
-\thetad{a} \(\kappa_{abc}p^b\epsilon^c+p^0 L_a(\epsilon)\)
-\thetad{0}\(\epsilon^a q_a+p^a L_a(\epsilon)-L_0(\epsilon)-(p\epsilon\epsilon)\)
},
\end{split}
\ee
where $L_0, L_a$ are the integer valued functions defined in \eqref{defL0La}.
Second, the variation of the first two terms in the last line of \eqref{5pqZ2new}
is given by
\beq
\nonumber
\nu_2(\epsilon,\eta^0)&\equiv &
\expe{- c_{2,a}\( p^0\epsilon^a \eps(\delta[\eta^0])+p^a(\epsilon(\delta[\eta^0])-\epsilon(\delta)\)
+a\, \frac{p^0  (q'_0[\epsilon]-q_0')}{k}}
\\
&=&\expe{ p^0 c_{2,a}\epsilon^a
\left( \frac{p}{24 k}+\frac12\, s\left(\frac{p}{p^0},-\frac{k}{p^0}\right)+\frac18\right)
\right.
\label{rata}\\
&& \left.\qquad
+\frac{ap^0}{k}\(p^\Lambda L_\Lambda(\epsilon)-q_a\epsilon^a+2A_{ab}\epsilon^a p^b \)
+(p^a+p^0\epsilon^a)\,\frac{c_{2,a}}{24}\,\eta^0 },
\nonumber
\eeq
where $q'_0[\epsilon]\equiv q_0[\epsilon]-A_{0\Lambda} p^\Lambda[\epsilon]$, $\delta[\eta^0]$
is the $SL(2,\IZ)$ matrix \eqref{Sdualde2} with $d$ replaced by $d-c\eta^0$, and
we used the fact that the Dedekind sum \eqref{Dedekindsum}
satisfies $s(d-\eta^0 c,c)=s(d,c)$.
Finally, the third term in the same line generates the following contribution
\beq
\label{ratA}
\nu_3(\epsilon,\eta^0)&\equiv & \expe{-\frac{k}{2} (\eta^\Lambda+2n^\Lambda)A_{\Lambda\Sigma}\eta^\Sigma}
\\
& = &
\expe{-\frac{p^0}{2k}\, A_{ab}\(p^0\epsilon^a+2p^a\)\epsilon^b
-\frac{c_{2,a}}{24}\(\frac{ p\, p^0  }{k}\,\epsilon^a+p^a\eta^0+p^0\epsilon^a\eta^0\)
-\frac{k}2\,A_{00}(\eta^0)^2 } .
\nonumber
\eeq
As a result, the total variation of the transition function \eqref{5pqZ2new}
is described by the following phase factor
\be
\label{ratS}
\begin{split}
\nu(\eta)  &\, \equiv \expe{k\eta^\Lambda\phi_\Lambda}\nu_1(\epsilon)\nu_2(\epsilon,\eta^0)\nu_3(\epsilon,\eta^0)
\\
&\, =\expe{k\eta^0\phi_0-\frac{k}2\,A_{00}(\eta^0)^2}
\expe{ k\eta^a(\phi_a-\phid{a})
+\frac{k}{2p^0}\, (pp\eta)+\frac{ap^0}{k}\,p^\Lambda L_\Lambda(\epsilon)
\right. \\
& \qquad\left.
+\frac{k}{2}\,c_{2,a}\eta^a \(s\(\frac{p}{p^0},-\frac{k}{p^0}\)+\frac{1}{4}\(1-p^0\)\)
-\frac{k(k+1)}{2}A_{ab}\eta^a\eta^b
-\eta^a A_{ab} p^b
\right.
\\
& \qquad\left.
-\thetad{a} \(\kappa_{abc}p^b\epsilon^c+p^0 L_a(\epsilon)\)
-\thetad{0}\(\epsilon^a q_a+p^a L_a(\epsilon)-L_0(\epsilon)-(p\epsilon\epsilon)\)
}.
\end{split}
\ee

One may try to cancel this phase by a suitable choice of characteristics.
E.g. the first factor, which carries the dependence on $\eta^0$ and
the only $q$-dependent term can be canceled
by choosing, in a {\it ad hoc} fashion,
\be
\thetad{0}=0\ ,\qquad \phi_0=\half\, A_{00}\, .
\label{phi0}
\ee
Note that vanishing of $\theta^\Lambda$ is required by the Heisenberg invariance
and equivalently results from casting  \eqref{fullH} into \eqref{hthxin}.
For $k=1$,  the phase simplifies in this case drastically into
 \be
\nu_{k=1}(\eta)=\expe{\eta^a(\phi_a-\phid{a})-A_{ab}\eta^a\eta^b
-\thetad{a} \(\kappa_{abc}p^b\eta^c+p^0 L_a(\eta)\)},
\ee
where we took into account that $p^0=1, s(p,-1)=0$. It may therefore
be canceled completely by further choosing
\be
\thetad{a}=0\ ,\qquad \phi_a-\phid{a}=A_{aa} \ .
\label{phiA}
\ee
Thus, provided one takes $\theta^\Lambda=\thetad{\Lambda}=0$ and \eqref{phi0} together
with \eqref{phiA}, for $k=1$, the Heisenberg symmetry can be kept unbroken.
Unfortunately, the choices \eqref{phi0}, \eqref{phiA} are inconsistent with the
transformation properties of the characteristics under monodromies, as we now discuss.

\subsection{Monodromy transformations}

Next we discuss the  transformation of the transition functions \eqref{5pqZ2}
under monodromies $M$ around the large volume point in $\cM_K(\hat\cX)$
which acts holomorphically on the twistor space by \eqref{bjac}.
It is clear that to compensate this transformation, one should supplement it by a spectral flow
transformation of charges \eqref{flow} where the transformation parameter is
taken to be $\epsilon'^a=(p/p^0)\epsilon^a$. Under this simultaneous variation,
the quantity $S_\alpha$ defined in \eqref{defSa} simply varies by a constant
\be
S_\alpha\mapsto S_\alpha +
\kappa_{abc} \epsilon^a \left[ n^b n^c + n^0 n^b \epsilon^c+ \frac13 (n^0)^2\epsilon^b \epsilon^c \right].
\ee
In addition one should take into account that the monodromy induces a transformation of characteristics
given in \eqref{sympchar}.
It is explicit form can be obtained using the matrix $\rho(M)$
given in \eqref{monosymp}.
Then taking into account that $\theta^\Lambda=0$, the characteristics undergo the following transformation
\be
\phi_0\mapsto \phi_0-\epsilon^a\phi_a-\frac12\(L_0(\epsilon)-\epsilon^a L_a(\epsilon)-(\epsilon\epsilon\epsilon)\),
\qquad
\phi_a \mapsto \phi_a+\frac12\,\kappa_{aac}\epsilon^c,
\label{transchar}
\ee
where we took into account that
$\frac12 \(\kappa_{aac}-\epsilon^b\kappa_{abc}\)\epsilon^a\epsilon^c\in \IZ$ and can be neglected.

As a result, assuming again that the invariants $\thnkl$ are not affected by the spectral flow,
we can write the total variation as in \eqref{invH_mon}
with the phase factor $\nu'(\epsilon)$ given by
\be
\begin{split}
\nu'(\epsilon)&\,=\expe{-k\kappa_a\epsilon^a-\frac{k}{2}\,\delta_\epsilon S_\alpha
+\frac{p^0}{k}\,\hat q_a \epsilon^a+p^0\epsilon^a\phid{a}
+\frac{p^0}{2}\(L_0(\epsilon)-\epsilon^a L_a(\epsilon)-(\epsilon\epsilon\epsilon)\)
\right. \\
&\qquad \left.
-\frac12 \(  (pp\epsilon)+\frac{p}{p^0}\,(\epsilon\epsilon\epsilon)\) }
\nu_1(p\epsilon^a/p^0)\nu_2(p\epsilon^a/p^0,0).
\end{split}
\label{totnup}
\ee
Unfortunately, it does not seem to be possible to dispose of this phase factor either.
For $k=p^0=1$ it can be simplified to
\be
\nu'_{k=1}(\epsilon)=\expe{\frac{c_{2,a}\epsilon^a}{24}\,(3p^2+2)
-\frac{(p-1)^2}{2}\,A_{ab}\epsilon^a\epsilon^b-\(\kappa_a+(p-1)\phid{a}\)\epsilon^a}
\ee
and does not vanish even in this particular situation.

Moreover, the transformation of the characteristics \eqref{transchar} seems to be inconsistent
with the identifications \eqref{phi0} and \eqref{phiA}. This shows that although
the Heisenberg invariance can be achieved for $k=1$, there is a tension between the conditions
ensuring this invariance and the monodromy transformations.
In particular, this is why the considerations of the previous subsection do not necessarily
imply that the D-instanton and fivebrane characteristics are not equal.

\section{List of notations\label{sec_not}}

Here we present a list of the most important notations used throughout the paper (roughly in order of appearance):

\vskip 5mm

\begin{supertabular}{clr}
$\expe{z}$ & $\exp\[2\pi\I z\]$
\\
$(\cX,\hat\cX)$ & mirror pair of Calabi-Yau threefolds
\\
$\cM$ & $\cQ_c(\cX)$ (IIA) or $\cQ_K(\hat\cX)$ (IIB), quaternion-K\"ahler manifold
\\
$\cQ_c(\cX)$ &
 hypermultiplet moduli space in type IIA on $\cX$
\\
$\cQ_K(\cX)$ &
 hypermultiplet moduli space in type IIB on $\hat\cX$
 \\
$\cS\cK$ &$\cM_c(\cX)$ (IIA) or $\cM_K(\hat\cX)$ (IIB), special K\"ahler manifold
\\
$\cM_c(\cX)$ & complex structure moduli space of $\cX$
\\
$\cM_K({\hat \cX})$ & (complexified) K\"ahler moduli space of ${\hat \cX}$
\\
$\cK$ & K\"ahler potential for the  special \kahler metric on $\cS\cK$
\\
$\omega_\sk$ & K\"ahler form on $\cS\cK$
\\
$K_c$ & canonical bundle over $\cS\cK$
\\
$\tau_{\Lambda\Sigma}$ & period matrix in Griffiths complex structure ($\Lambda,\Sigma=0, 1, \dots, h_{2,1}$)
\\
$\cN_{\Lambda\Sigma}$ & period matrix in Weil complex structure (Eq. \eqref{defcN})
\\
$\cT$ & torus $H^3(\cX,\mathbb{R})/H^{3}(\cX,\mathbb{Z})$ (IIA),
$H^{\rm even}(\hat\cX,\mathbb{R})/K(\cX)$ (IIB)
\\
$\Gamma$ &
charge lattice, $H^{3}(\cX,\mathbb{Z})$ (IIA) or $K(\cX)$ (IIB), or its Poincar\' e dual
\\
$\langle \gamma,\gamma'\rangle$ & $q_\Lambda p'^\Lambda-p^\Lambda q'_\Lambda$,
integer symplectic pairing on $\Gamma$
\\
$\Gamma_e, \Gamma_m$ &
electric and magnetic charge sublattices of $\Gamma=\Gamma_e\oplus \Gamma_m$, isotropic
\\
$\omega_\cT$ & K\"ahler form on $\cT$, (Eq. \eqref{KahlerformT})
\\
$\cJ_c(\cX)$ & intermediate Jacobian: total space of $\cT\rightarrow
\cJ_c(\cX)\rightarrow \cM_c(\cX)$
\\
$\cJ_K(\hat\cX)$ & symplectic Jacobian: total space of $\cT\rightarrow
\cJ_K(\hat\cX)\rightarrow \cM_K(\hat\cX)$
\\
$\Omega_{3,0}$ & holomorphic 3-form on $\cX$
\\
$\cL$ & Hodge line bundle $\cL\rightarrow \cM_c(\cX)$ where $\Omega_{3,0}$ is valued
\\
$\cL_\Theta$ & ``theta line bundle'' $\cL_\Theta \rightarrow \cT$, defined by periodicity (Eq. \eqref{thperiod})
\\
$\cC_\Theta$ & unit circle bundle $\cL_\Theta^{\circ}$ inside $\cL_\Theta$
\\
$\ZG{k}=\cZ^{(k)}_{\Theta,\mu}$ & Gaussian partition function of a chiral 3-form obtained by holomorphic factorization
\\
$\ZNS{k}$ & non-linear NS5-partition function  governing instanton
corrections
\\
$\ZNSG{k}$ & weak-coupling limit of $\ZNS{k}/[e^{-4\pi|k| e^\phi-\pi \I k\sigma}]$
\\
$\cF$ & metric-dependent normalization of $\cZ^{(k)}_{\Theta,\mu}$
\\
$\cC_{\text{NS5}}$ & circle bundle $\cC_{\text{NS5}} \rightarrow \cJ_c(\cX)$ of which $\ZNS{1}$ is a section
\\
$\Theta=(\theta^\Lambda, \phi_\Lambda)$ & characteristics of the theta function $\cZ^{(k)}_{\Theta,\mu}$
\\
$\sigma_\Theta$ & quadratic refinement of the intersection form on $H^{3}(\cX,\mathbb{Z})$
modulo 2
\\
$(\cA^{\Lambda}, \cB_\Lambda)$ & symplectic basis of $H_3(\cX,\mathbb{Z})$ ($\Lambda=0, 1, \dots, h_{2,1}$)
\\
$(\alpha_\Lambda, \beta^{\Lambda})$ & symplectic basis of $H^{3}(\cX,\mathbb{Z})$
\\
$C$ & RR 3-form potential in $D=10$ type IIA string theory
\\
$B$ & 2-form in $D=4$ spacetime
\\
$\cB$ & ``chiral'' 2-form on the worldvolume $\cX$ of the NS5-brane
\\
$H=d\cB$ & imaginary self-dual field strength of $\cB$
\\
$C=(\zeta^{\Lambda}, \tilde{\zeta}_\Lambda)$ &
periods of $C$ along $(\cA^{\Lambda}, \cB_\Lambda)$, ``RR-axions''
 (Eq. \eqref{defzezet})
\\
$\Omega=(X^\Lambda,F_\Lambda)$ & periods of $\Omega_{3,0}$ along $\cA^\Lambda,\cB_\Lambda$ (Eq. \eqref{defXF})
\\
$z^{a}=X^{a}/X^{0}$ & projective coordinates on $\cM_c(\cX)=\cM_K(\hat\cX)$ ($a=1,\dots, h_{2,1}(\cX)=h_{1,1}(\hat\cX)$)
\\
$z^{\Lambda}$ & $(1,z^{a})$
\\
$F(X)$ & prepotential, $F_\Lambda=\pa_{X^\Lambda}F(X)$
\\
$\sigma$ & dual of the 2-form $B$ (``NS-axion'')
\\
$H=(n^\Lambda, m_\Lambda)$ & integral periods of
$H$ along $(\cA^{\Lambda}, \cB_\Lambda)$
\\
$\chi(\cX)$ & Euler number of $\cX$
\\
$r=e^{\phi}=g_{(4)}^{-2}$ & four-dimensional dilaton $\phi$ and string coupling $g_{(4)}$
\\
$\tau_2=1/g_s^2$ & ten-dimensional type IIB string coupling, related to $\phi$
via Eq. \eqref{phipertA}
\\
$\Psi_\mathbb{R}$ &  wave function in the real polarization
\\
$F_1$ & one-loop topological string vacuum amplitude
\\
$f_1$ & holomorphic part of $F_1$ (Eq. \eqref{F1})
\\
$\kappa (M)$ & logarithm of a unitary character of the monodromy group of $\cM_c$
\\
$\gamma$ & charge vector (Mukai vector), valued in $H_3(\cX,\mathbb{Z}), H_{\text{even}}({\hat\cX},\mathbb{Z})$ for IIA, IIB
\\
$Z_\gamma$ & central charge $e^{\cK/2}  (q_\Lambda X^\Lambda-p^\Lambda
F_\Lambda)$  (stability data)
\\
$W_\gamma$ &rescaled central charge, $W_\gamma=\frac{\tau_2}{2} e^{-\cK/2} Z_\gamma$
\\
$\cC(r)$ & circle bundle $\cC(r)\rightarrow \cJ_c(\cX)$ (IIA), or
$\cC(r)\rightarrow \cJ_K(\hat\cX)$ (IIB),
with fiber $S^1_\sigma$
\\
$N(X^a)$ & $\frac16 \kappa_{abc} X^a X^b X^c$
\\
$\omega_{\hat\cX}, \omega^a, \omega_a,1$ &
symplectic basis  of $H^6(\hat\cX)\oplus H^4(\hat\cX)\oplus H^2(\hat\cX)\oplus H^0(\hat\cX)$
\\
$\hat\cX, \gamma_a, \gamma^a$, [pt] &
symplectic basis  of $H_6(\hat\cX)\oplus H_4(\hat\cX)\oplus H_2(\hat\cX)\oplus H_0(\hat\cX)$
\\
$A_{\Lambda\Sigma}$ & real symmetric matrix defined up to integer shifts (Eq. \eqref{lve})
\\
$c_{2,a}$ & $ \int_{\hat\cX} c_2(\hat\cX) \omega_a$
\\
$\gamma=(p^{\Lambda}, q_\Lambda)$ & integer magnetic and electric D2-brane charges (resp. D(-1)-D1-D3-D5 in IIB)
\\
$q_\Lambda^{\prime}$ & primed electric charges, $\mathbb{Q}$-valued (Eq. \eqref{chargeshift})
\\
$\zeta'^\Lambda$ & primed RR axions (Eq. \eqref{shze})
\\
$\hat{q}_\Lambda$ & spectral flow-invariant combination of electric charges (Eq.
\eqref{qaq0hat})
\\
$\hgam$ & reduced charge vector $(p^a,\hat q_a,\hat q_0)$
\\
$Q_a, J$ & electric charge and momentum of 5D black hole (Eq. \eqref{defQJ})
\\
$A^{\rm even} $ & type IIB RR potential
\\
$\tau^1,c^a,c_a,c_0,\psi$ &
RR and NS axions adapted to S-duality (Eq. \eqref{symptob})
\\
$k$ & integer NS5-brane charge
\\
$\cZ$ & twistor space of $\cM$, a fibration $\cZ\rightarrow \cM$ with fiber $\mathbb{C}P^1$
\\
$\cU_i$ & open covering of $\cZ$
\\
$\cX^{[i]}$ & complex contact one-form (Eq. \eqref{contact})
\\
$\Phi_{[i]}$ & contact potential (Eq. \eqref{contact})
\\
$\cK_{\cZ}^{[i]}$ & K\"ahler potential on $\cZ$ (Eq. \eqref{Knuflat})
\\
$(x^{\mu},t)$ & generic coordinates on $\cM\times \mathbb{C}P^1$
\\
$(\xi_{[i]}^{\Lambda}, \tilde{\xi}^{[i]}_\Lambda, \alpha^{[i]})$ & complex Darboux coordinates on
$\cU_i\subset \cZ$,  $\CX\ui{i}= \de\ai{i} + \xii{i}^\Lambda \, \de \txii{i}_\Lambda$
\\
$\tilde\alpha$ & $-2\alpha-\xi^\Lambda \txi_\Lambda$, symplectic invariant Darboux coordinate
\\
$H^{[ij]}$ & holomorphic section of $H^1(\cZ,\cO(2))$, controlling deformations of $\cZ$
\\
$H_\gamma$ & complex symplectomorphism encoding deformations of $\cZ$ due to D-branes
\\
$\ell_\gamma$ & ``BPS ray'', defined as $\{ t:\  Z_\gamma(z^a)/t\in \I\IR^-\}$
\\
$\Omega(\gamma)$ & generalized Donaldson-Thomas invariants ($\mathbb{Z}$-valued)
\\
$\thnkl$ & generalized Donaldson-Thomas invariants ($\mathbb{Q}$-valued)
(Eq. \eqref{rationalinvariants})
\\
$N_{\text{DT}}$ & standard (rank 1) Donaldson-Thomas invariants ($\mathbb{Z}$-valued)
\\
$Z_{\text{DT}}$ & Donaldson-Thomas partition function
\\
$Z_{\text{GW}}$ & Gromov-Witten partition function
\\
$\delta$ & $SL(2,\mathbb{Z})$-transformation (Eq. \eqref{Sdualde})
\\
$\eps(\delta)$ & multiplier system of the Dedekind $\eta$-function (Eq. \eqref{defvareps})
\\
$s(d,c)$& Dedekind sum (Eq. \eqref{Dedekindsum})
\\
$W_{k,p,\hgam}$ & image of D-instanton central charge $W_\gamma$ under $\delta$
\\
$\ell_{k,p,\hgam}$ & image of the BPS ray $\ell_\gamma$ under $\delta$
\\
$H_{k,p,\hgam}$ & image of D-instanton transition function $H_\gamma$ under $\delta$
\\
$H_{\text{NS5}}^{(k)}$ & formal sum of $H_{k,p,\hgam}$ over all $p,\hgam$
\\
\end{supertabular}

\begin{table}[h]
$$
\begin{array}{|c|c|}
\Psi_{\rm BCOV} &  \cL^{1-\frac{\chi(\cX)}{24}} \\
e^{f_1} & \cL^{1-\frac{\chi(\cX)}{24}+ \frac{b_3}{4}} \otimes K_{c}^{\half} \\
e^{\cK} & \cL \otimes \overline{\cL} \\
\Psi_{\rm G} &\cL^{\frac{b_3}{4}} \otimes K_{c}^{\half}  \\
\Psi_{\rm W} & \cL^{\frac{b_3}{4}-1} \otimes K_{c}^{\half}  \\
J_{\rm G} & \cL^{-b_3/2}\otimes K_c^{-1}\otimes
\det_{\rm G}^{-1} \\
J_{\rm W} & \cL^{1-\frac{b_3}{2}} \otimes \bar\cL^{-1} \otimes K_c^{-1}
\otimes \overline{\det_{\rm W}}^{-1} \\
\det(-\Im\cN) & \det_{\rm W}^{-1} \otimes \overline{\det_{\rm W}}^{-1}  \\
\det(\Im \tau) & \det_{\rm G}^{-1} \otimes \overline{\det_{\rm G}}^{-1} \\
X^\Lambda \Im \tau_{\Lambda\Sigma} X^\Sigma &
\cL^2\otimes \det_{\rm W} \otimes \overline{\det_{\rm G}}^{-1}\\
X^\Lambda \Im \cN_{\Lambda\Sigma} X^\Sigma &
\cL^2\otimes \det_{\rm G} \otimes \overline{\det_{\rm W}}^{-1}\\
e^{-\frac{\chi(\cX)}{48} \cK+\I \pi \sigma} & \cL_\Theta\otimes \cL^{\chi/24} \\
\end{array}
$$
\caption{Transformation properties of various quantities under $\cL\otimes K_c\otimes
\det_{\rm G}\otimes \det_{\rm W}$.\label{tabtrans}}
\end{table}

%


\providecommand{\href}[2]{#2}\begingroup\raggedright\endgroup

\end{document}